\documentclass{emulateapj}
\usepackage{hyperref}
\usepackage{natbib}
\usepackage[caption=false]{subfig}

\begin{document}
\title{On the Morphology and Chemical Composition of the HR 4796A Debris Disk\footnotemark[*]}
\footnotetext[*]{This paper includes data obtained at the 6.5 m Magellan Telescopes located at Las Campanas Observatory, Chile.}
\author{Timothy J. Rodigas\altaffilmark{1,8}, Christopher C. Stark\altaffilmark{2}, Alycia Weinberger\altaffilmark{1}, John H. Debes\altaffilmark{3}, Philip M. Hinz\altaffilmark{4}, Laird Close\altaffilmark{4}, Christine Chen\altaffilmark{3}, Paul S. Smith\altaffilmark{4}, Jared R. Males\altaffilmark{4,6}, Andrew J. Skemer\altaffilmark{4,7}, Alfio Puglisi\altaffilmark{5}, Katherine B. Follette\altaffilmark{4}, Katie Morzinski\altaffilmark{4,6}, Ya-Lin Wu\altaffilmark{4}, Runa Briguglio\altaffilmark{5}, Simone Esposito\altaffilmark{5}, Enrico Pinna\altaffilmark{5}, Armando Riccardi\altaffilmark{5}, Glenn Schneider\altaffilmark{4}, Marco Xompero\altaffilmark{5}}

\altaffiltext{1}{Department of Terrestrial Magnetism, Carnegie Institute of Washington, 5241 Broad Branch Road, NW, Washington, DC 20015, USA; email: trodigas@carnegiescience.edu}
\altaffiltext{2}{NASA Goddard Space Flight Center, Exoplanets $\&$ Stellar Astrophysics Laboratory, Code 667, Green- belt, MD 20771}
\altaffiltext{3}{Space Telescope Science Institute, Baltimore, MD 21218, USA}
\altaffiltext{4}{Steward Observatory, The University of Arizona, 933 N. Cherry Ave., Tucson, AZ 85721, USA}
\altaffiltext{5}{INAF - Osservatorio Astrofisico di Arcetri, Largo E. Fermi 5, I-50125, Firenze, Italy}
\altaffiltext{6}{NASA Sagan Fellow}
\altaffiltext{7}{Hubble Fellow}
\altaffiltext{8}{Carnegie Postdoctoral Fellow}

\newcommand{\about}{$\sim$~}
\newcommand{\mj}{M$_{J}$}
\newcommand{\degrees}{$^{\circ}$}
\newcommand{\arcseconds}{$^{\prime \prime}$}
\newcommand{\asec}{$\arcsec$}
\newcommand{\fasec}{$\farcs$}
\newcommand{\lprime}{$L^{\prime}$}
\newcommand{\ks}{$Ks$~}
\newcommand{\mjyasec}{mJy/arcsecond$^{2}$}
\newcommand{\microns}{$\mu$m}

\shortauthors{Rodigas et al.}

\begin{abstract}
We present resolved images of the HR 4796A debris disk using the Magellan adaptive optics system paired with Clio-2 and VisAO. We detect the disk at 0.77 \microns, 0.91 \microns, 0.99 \microns, 2.15 \microns, 3.1 \microns, 3.3 \microns, and 3.8 \microns. We find that the deprojected center of the ring is offset from the star by 4.76$\pm$1.6 AU and that the deprojected eccentricity is 0.06$\pm$0.02, in general agreement with previous studies. We find that the average width of the ring is 14$^{+3}_{-2}\%$, also comparable to previous measurements. Such a narrow ring precludes the existence of shepherding planets more massive than \about 4 \mj, comparable to hot-start planets we could have detected beyond \about 60 AU in projected separation. Combining our new scattered light data with archival HST/STIS and HST/NICMOS data at \about 0.5-2 \microns, along with previously unpublished Spitzer/MIPS thermal emission data and all other literature thermal data, we set out to constrain the chemical composition of the dust grains. After testing 19 individual root compositions and more than 8,400 unique mixtures of these compositions, we find that good fits to the scattered light alone and thermal emission alone are discrepant, suggesting that caution should be exercised if fitting to only one or the other. When we fit to both the scattered light and thermal emission simultaneously, we find mediocre fits (reduced chi-square \about 2), preventing us from reporting a single good-matching model composition. To circumvent this problem, we take a more general approach and observe which compositions are most often preferred over others. We find that silicates and organics are generally the most favored, and that water ice is usually not favored. These results suggest that the common constituents of both interstellar dust and solar system comets also may reside around HR 4796A, though improved modeling is necessary to place better constraints on the exact chemical composition of the dust.
\end{abstract}

\section{Introduction}
HR 4796A is a young (\about 10 Myr old; \citealt{hr4796age}) A star located 72.8 pc away from Earth \citep{updatedhip}. The star is encircled by a ring of dusty debris at \about 80 AU that has been imaged at many wavelengths spanning the visible \citep{hr4796schneider}, the near-infrared (NIR; \citealt{4796organics,thalmannhr4796,hinkley4796,lagrange4796,perrinGPI4796,milli4796}), and mid-infrared \citep{wahhaj4796old,koerner4796,telesco4796,moerchen4796}. The spectral energy distribution (SED) of the star $+$ disk system has also largely been filled out from \about 30-850 \microns ~\citep{hr4796low,hr4796akari,hr4796herschel,hr4796greaves,hr4796sheret,hr4796jura95,hr4796nilsson}. The disk is cleared of material inside the ring, with sharp inner and outer edges and a small (\about few percent) offset from the star. These morphological features suggest the presence of one or more planets (either interior to the ring, exterior, or both; \citealt{collisiondynamics,lagrange4796}). 

Previous works have modeled the ring's thermal emission to constrain the dust's chemical composition, preferring porous mixtures of silicates, organics, and some water ice \citep{hr4796augereau,hr4796li,milli4796}. Recently, \cite{4796organics} modeled the ring's scattered light from 0.5-2 \microns, finding \textit{complex} organic materials provided a good match to the data, though \cite{4796noorganics} showed that mixtures including simple organics also fit the scattered light data.

This interesting system would benefit from further study to 1) confirm the ring offset from the star, especially along the disk's minor axis; 2) constrain the ring's width, since this property can be used to place a dynamical upper limit on the mass of an interior disk-shepherding planet \citep{medynamics,chiang}; 3) search for self-luminous exoplanets interior and exterior to the ring; and 4) determine the composition of the dust grains to constrain the fractional abundance of organic materials ans water ice.

\begin{table*}[t]
\centering
\caption{Observations $\&$ Data Reduction Summary}
\begin{tabular}{c | c c c c c}
\hline
\hline
Wavelength & Instrument & Date Observed (UT) & Total Exposure (min) & Sky Rotation (\degrees) & PCA modes \\
\hline
3.8 \microns ~(\lprime) & Clio-2 & 7 April 2013 & 150 & 150.73 & 28/455 \\
3.3 \microns ~($Ls$) & Clio-2 & 8 April 2013 & 80 & 93.5 & 13/120 \\
3.1 \microns ~($Ice$) & Clio-2 & 9 April 2013 & 87 & 115.4 & 21/349 \\
2.15 \microns ~($Ks$) & Clio-2 & 10 April 2014 & 80 & 94 & 26/240 \\
0.99 \microns ~($Ys$) & VisAO & 8 April 2013 & 95 & 93.6 & 23/285 \\
0.91 \microns ~($z'$) & VisAO & 7 April 2013 & 115 & 96.7 & 28/303 \\
0.77 \microns ~($i'$) & VisAO & 9 April 2013 & 114 & 119.4 & 90/342 \\
\hline
\end{tabular}  
\label{tab:obs}
\end{table*} 

These four goals can be accomplished with high-resolution imaging in the visible-NIR (0.5-4 \microns) from the ground with adaptive optics (AO). Imaging at these wavelengths results in high Strehl ratios, increasing sensitivity to faint sources close to the star. This spectral window also contains strong absorption features for water ice (at \about 1.6, 2, and 3.1 \microns; \citealt{inoue}, and references therein) and tholin-like organics (3.1 \microns; \citealt{phoebe}). Therefore obtaining high signal-to-noise (S/N) narrow and broadband images of the disk at these wavelengths can constrain the fractional abundance of these materials in the dust. Finally, since young exoplanets are bright at 1-5 \microns ~\citep{burrows,baraffe}, we can simultaneously detect self-luminous planets in addition to the scattering dust.

We have obtained high S/N images of the HR 4796A debris disk using the Magellan AO system (MagAO; \citealt{magao}) paired with the Clio-2 1-5 \microns ~camera \citep{suresh} and the VisAO camera \citep{visao}. We detect the disk at 0.77 \microns, 0.91 \microns, 0.99 \microns, 2.15 \microns, 3.1 \microns, 3.3 \microns, and 3.8 \microns, resolving the ring in front of and behind the star in several of these images. In Section \ref{sec:obs}, we describe our observations and data reduction. In Section \ref{sec:results}, we present our results on the disk's morphology, photometry (including reanalysis of archival Hubble Space Telescope (HST) STIS and NICMOS data first presented in \citealt{4796organics}), and limits on planets in the system. In Section \ref{sec:modeling}, we present the results of our modeling to constrain the composition of the dust, which includes all available thermal emission data and previously unpublished Spitzer/MIPS data. In Section \ref{sec:discussion}, we discuss the implications of our results for the morphology and chemical composition of the disk. In Section \ref{sec:summary}, we summarize and conclude.

\section{Observations and Data Reduction}
\label{sec:obs}
\subsection{Observations}
\subsubsection{Clio-2}
\begin{figure*}[t]
\centering
\subfloat[]{\label{fig:L}\includegraphics[width=0.34\textwidth]{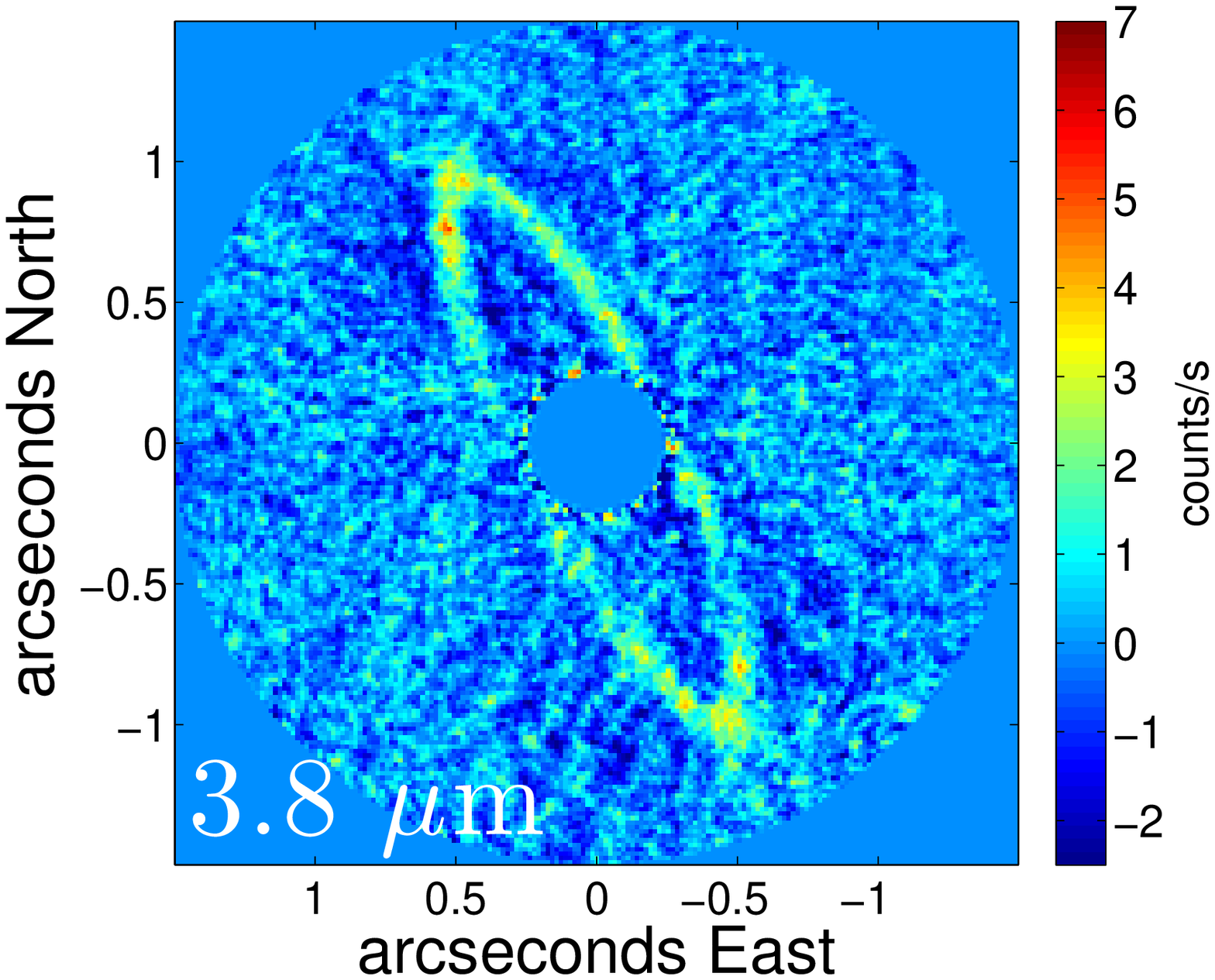}} 
\subfloat[]{\label{fig:Ls}\includegraphics[width=0.34\textwidth]{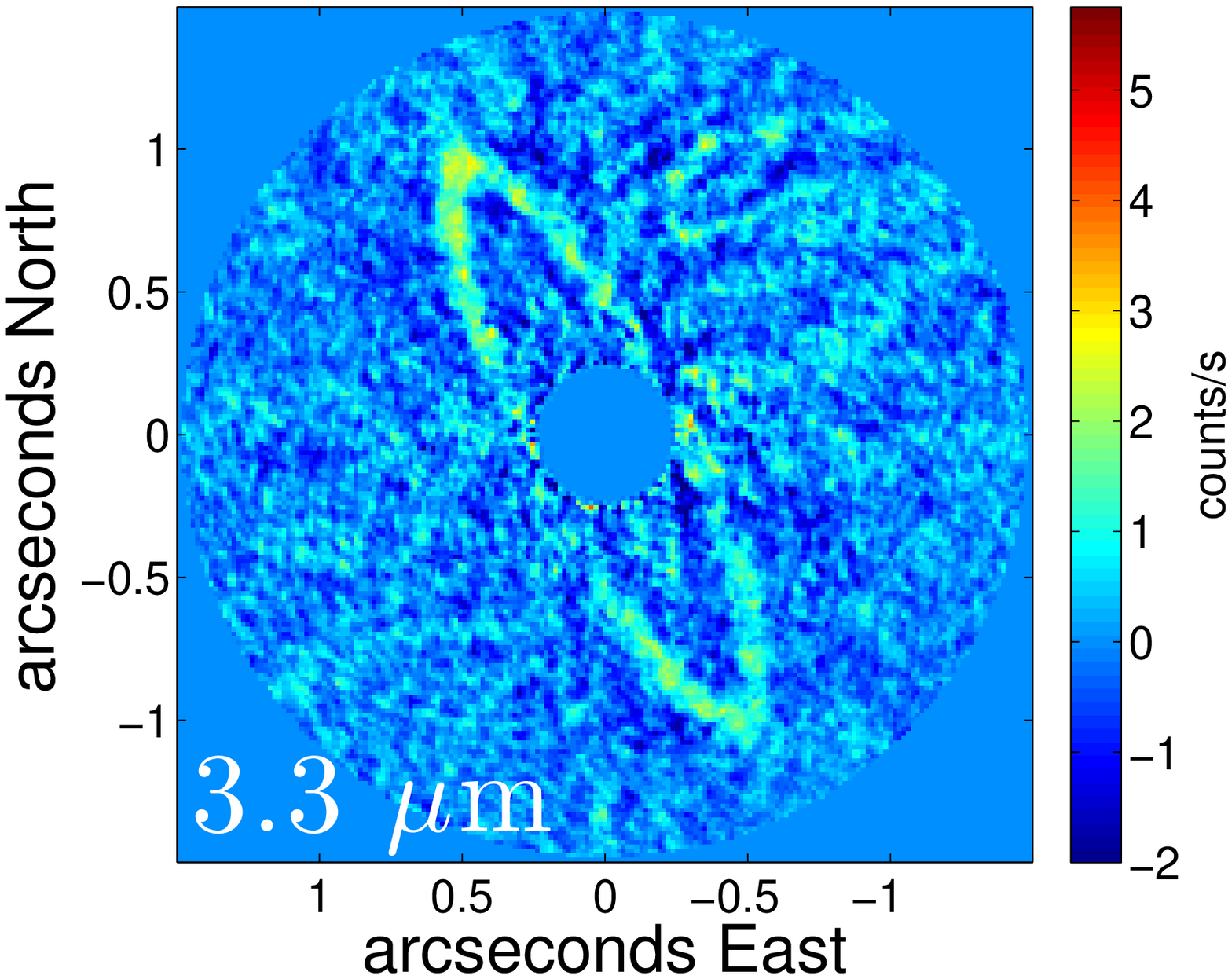}} \\
\subfloat[]{\label{fig:Ice}\includegraphics[width=0.34\textwidth]{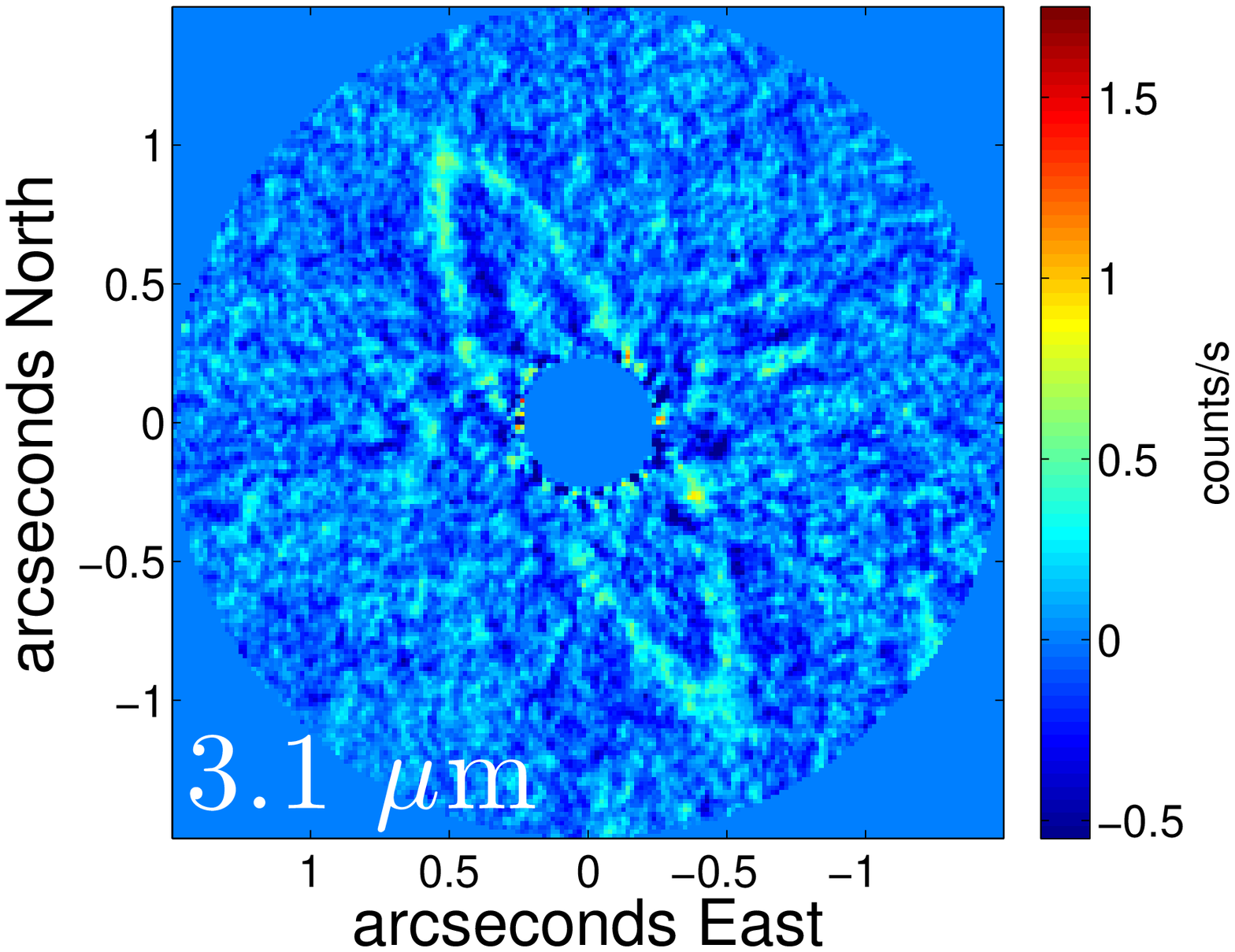}} 
\subfloat[]{\label{fig:Ks}\includegraphics[width=0.34\textwidth]{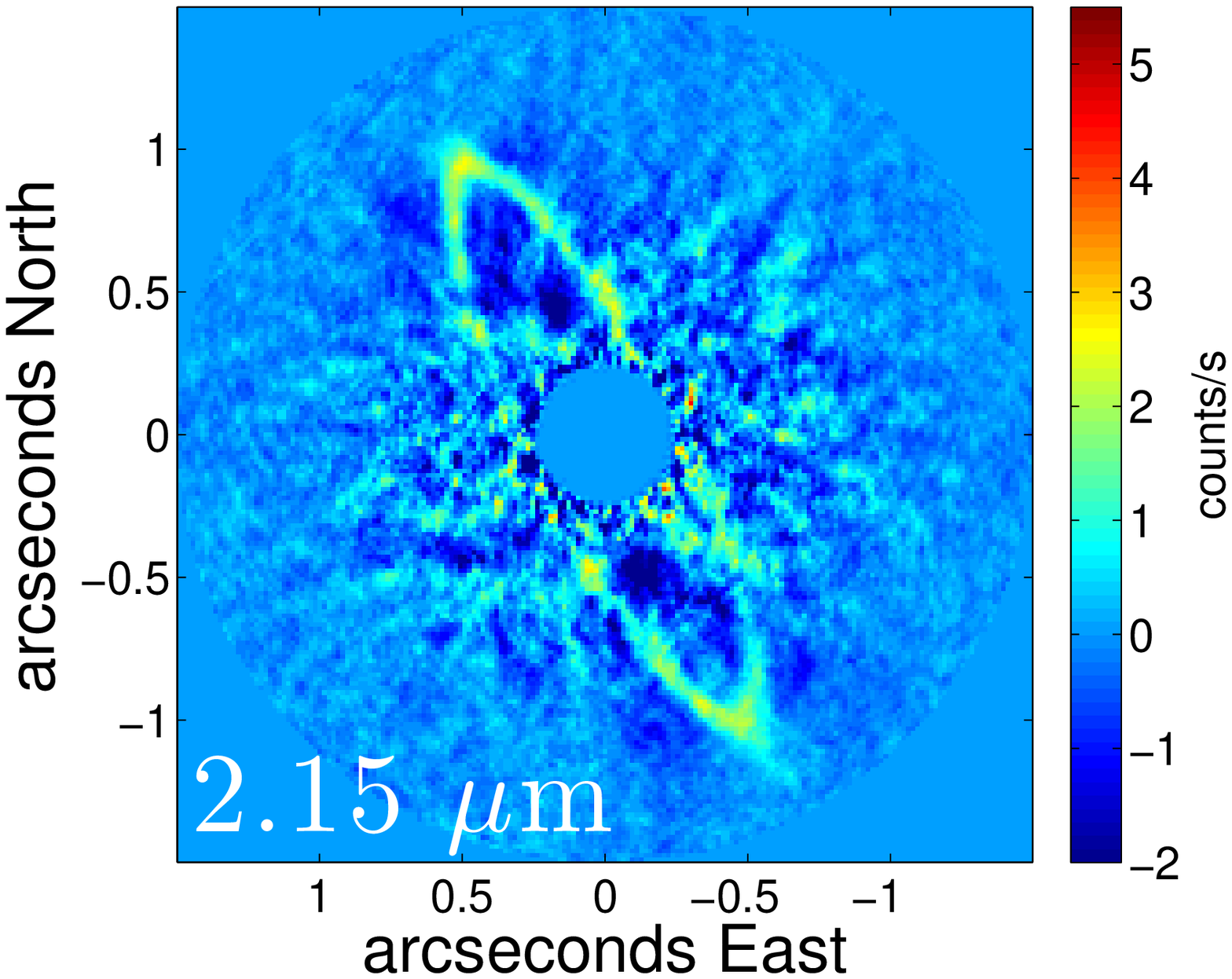}} \\
\subfloat[]{\label{fig:visaoy}\includegraphics[width=0.34\textwidth]{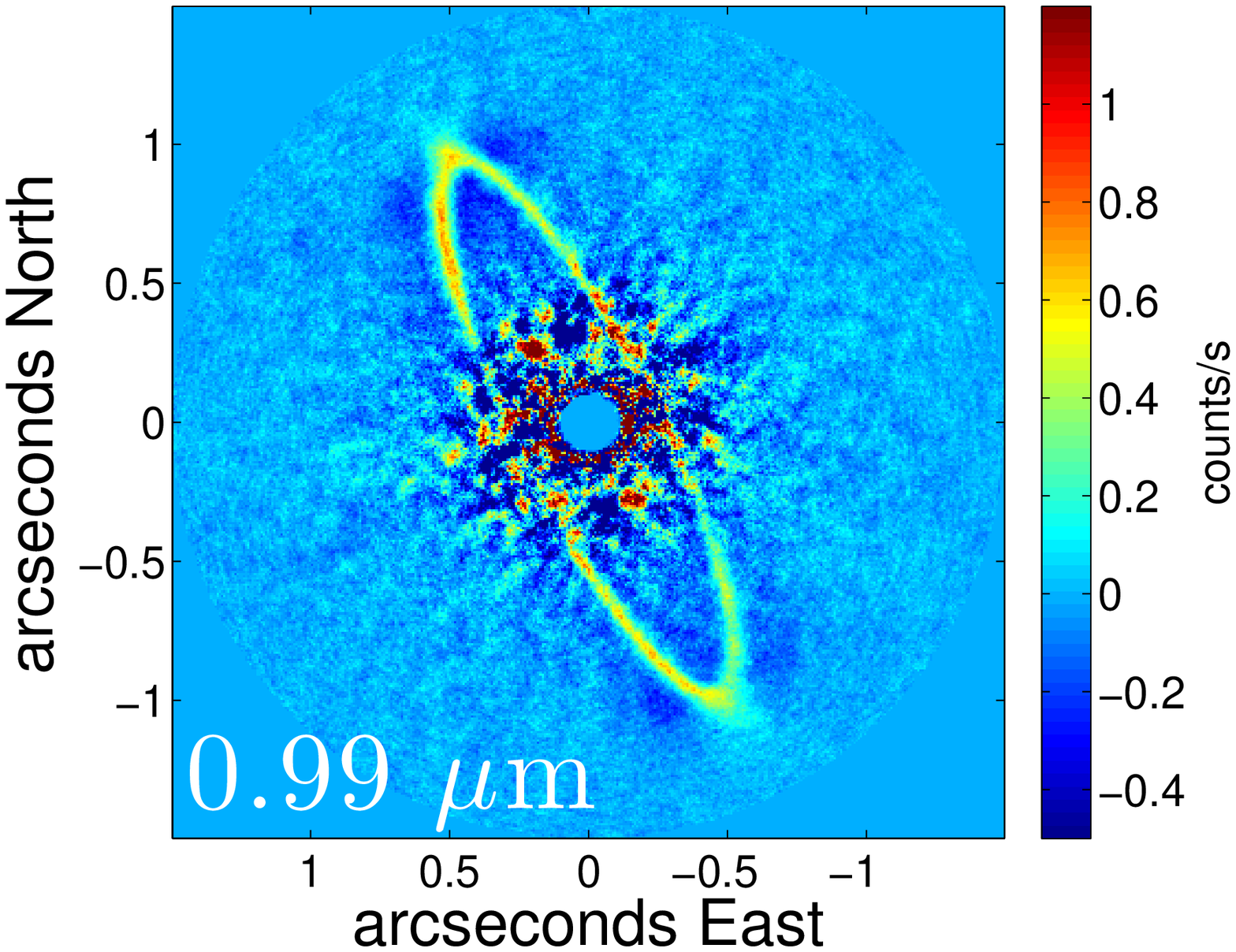}} 
\subfloat[]{\label{fig:visaoz}\includegraphics[width=0.34\textwidth]{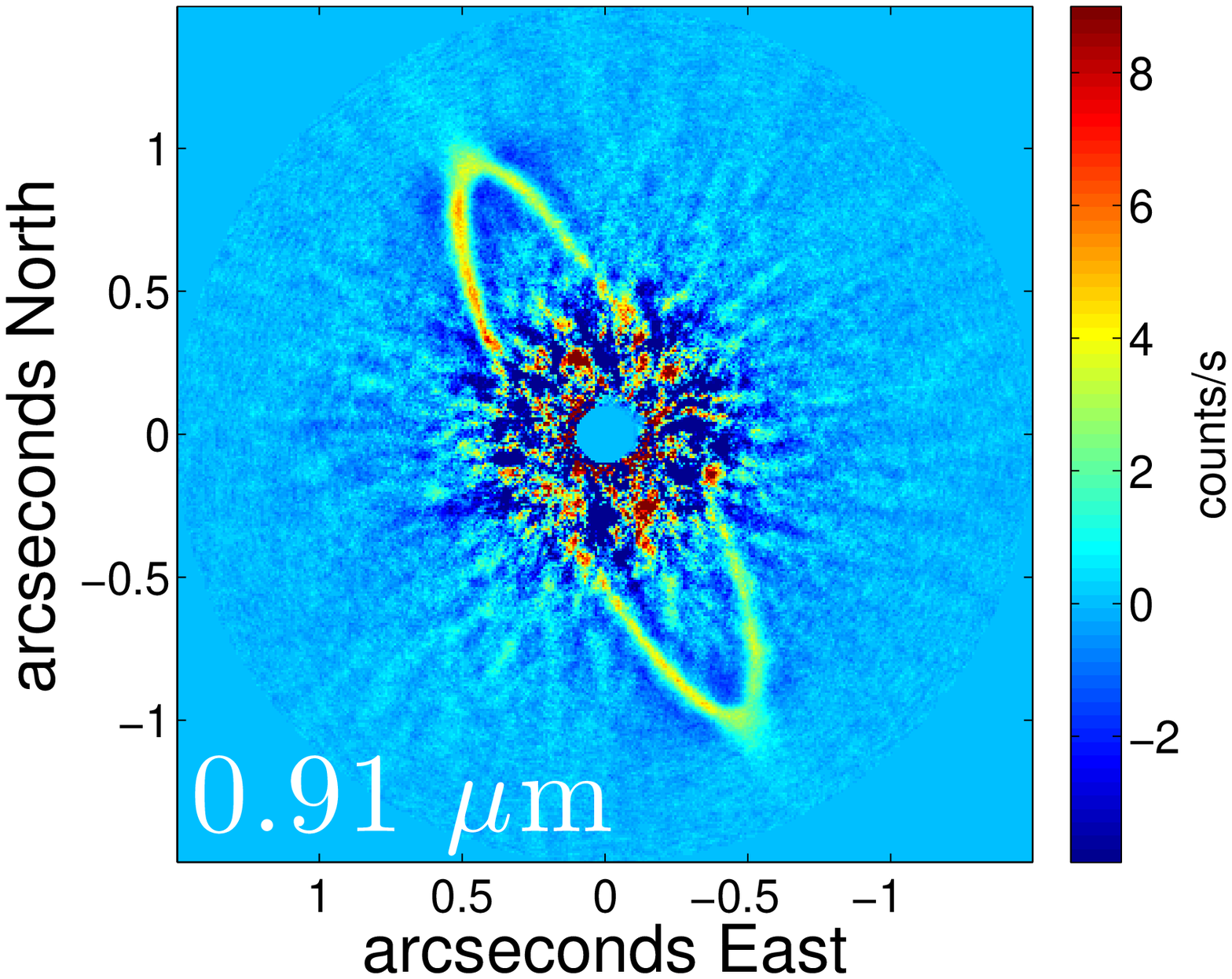}} 
\subfloat[]{\label{fig:visaoi}\includegraphics[width=0.34\textwidth]{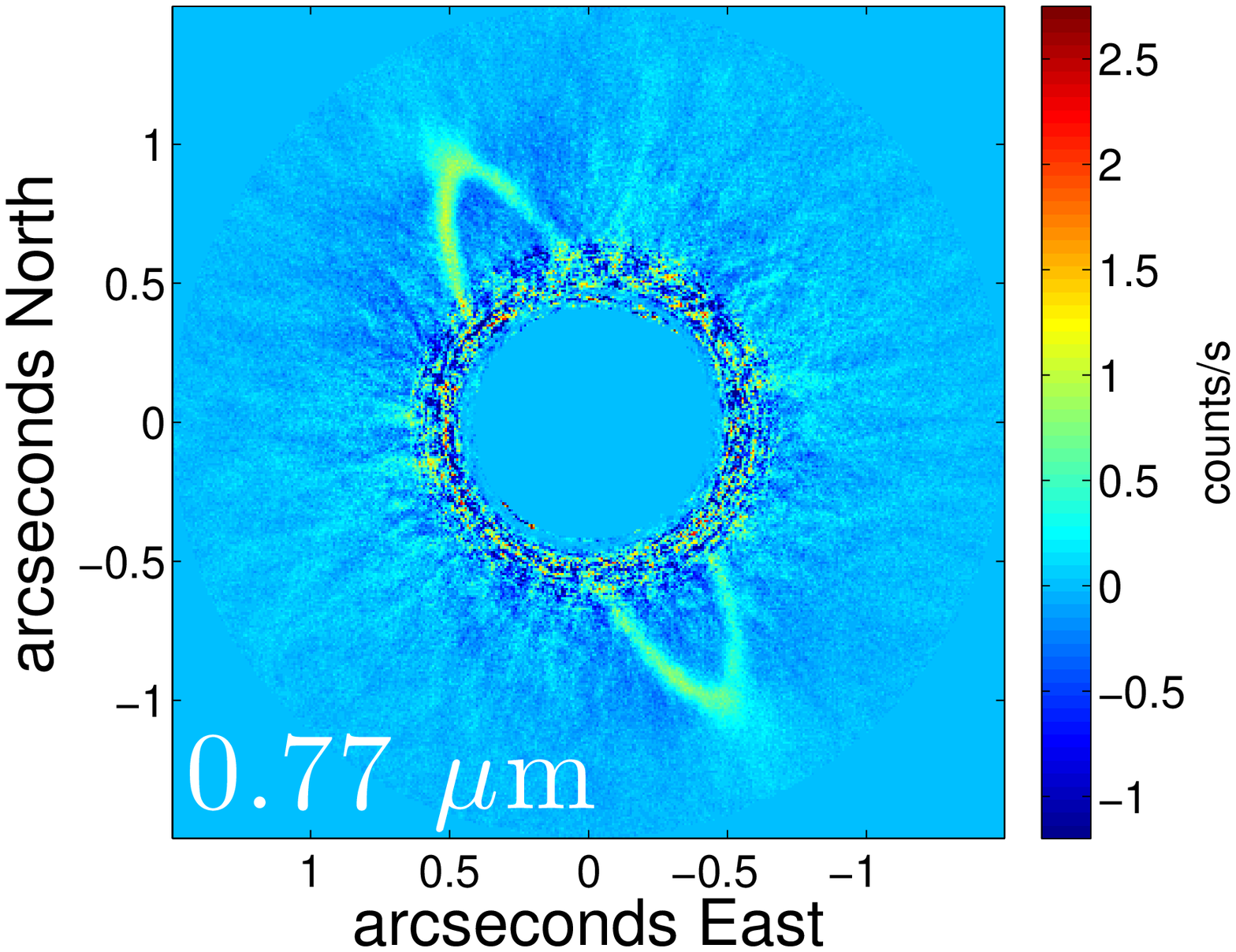}} \\
\caption{Resolved images of the HR 4796A debris disk, with North-up, East-left. Digital masks have been added for display purposes in all images.(a-d): MagAO/Clio-2 images at \lprime, $Ls$, $Ice$ band, and \ks band, respectively. (e-g): MagAO/VisAO images at $Ys$, $z'$, and $i'$ band, respectively. See Fig. \ref{fig:snremaps} for the corresponding SNRE maps.}
\label{fig:images}
\end{figure*}

\begin{figure*}[t]
\centering
\subfloat[]{\label{fig:LSN}\includegraphics[width=0.34\textwidth]{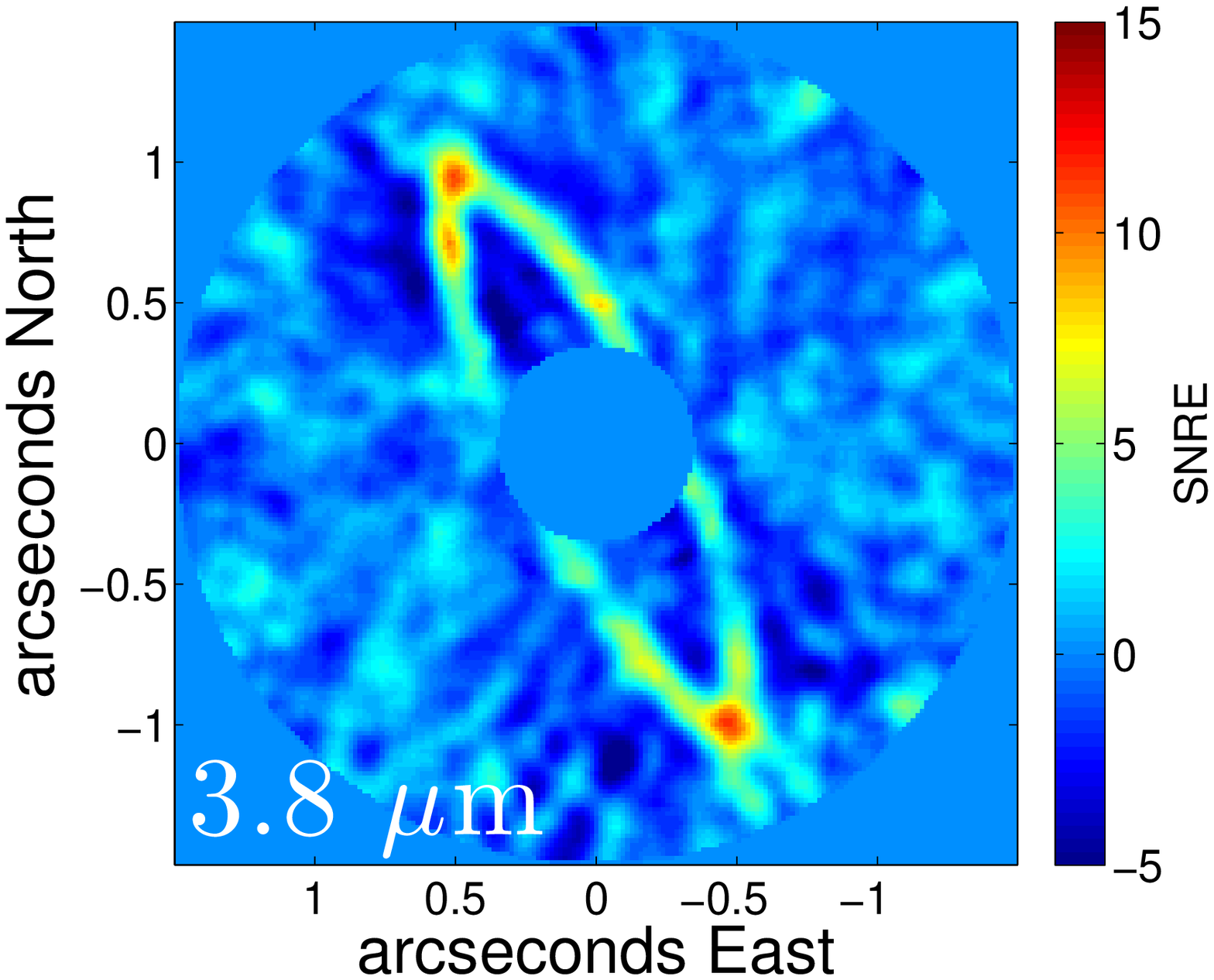}} 
\subfloat[]{\label{fig:LsSN}\includegraphics[width=0.34\textwidth]{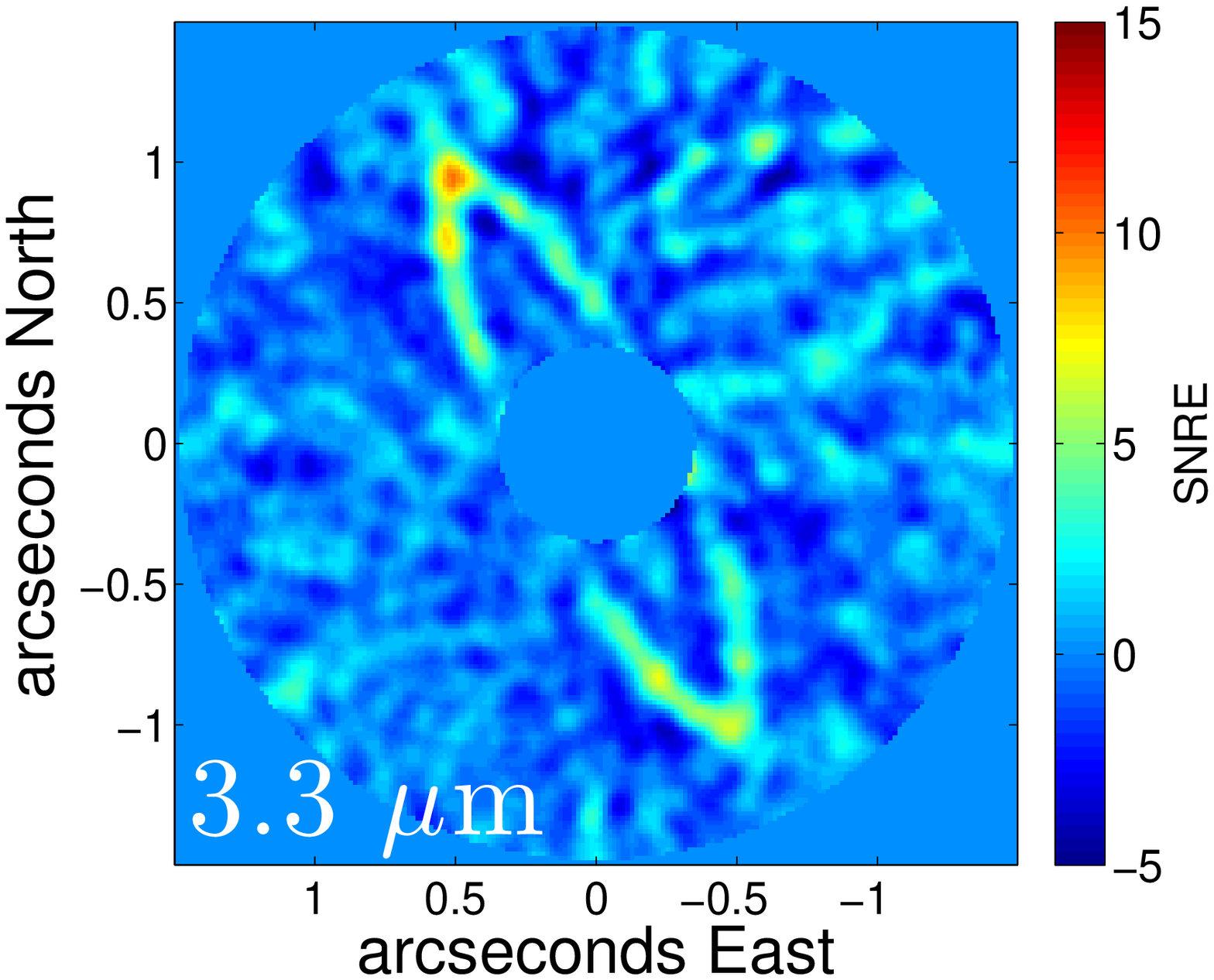}} \\
\subfloat[]{\label{fig:IceSN}\includegraphics[width=0.34\textwidth]{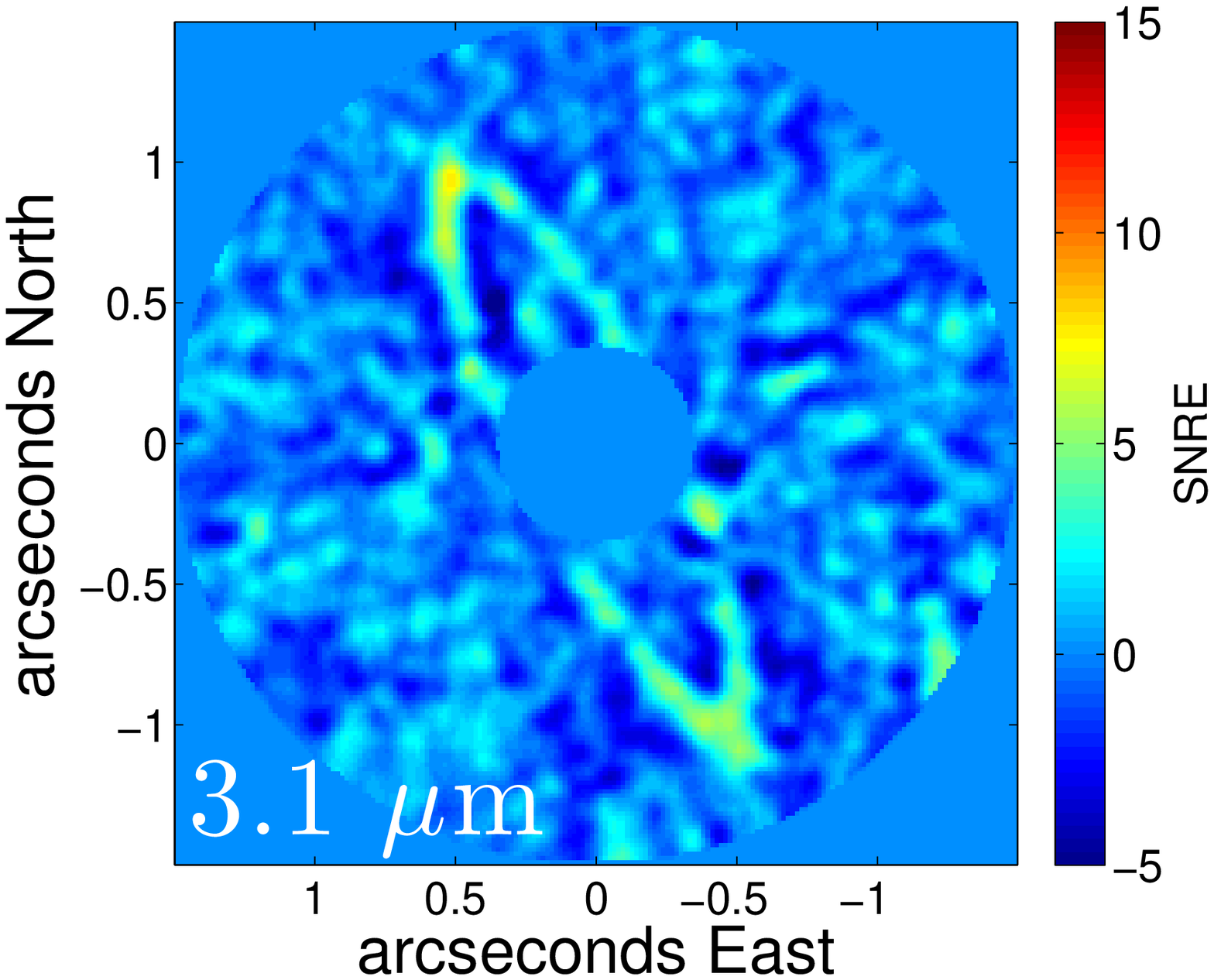}}
\subfloat[]{\label{fig:KsSN}\includegraphics[width=0.34\textwidth]{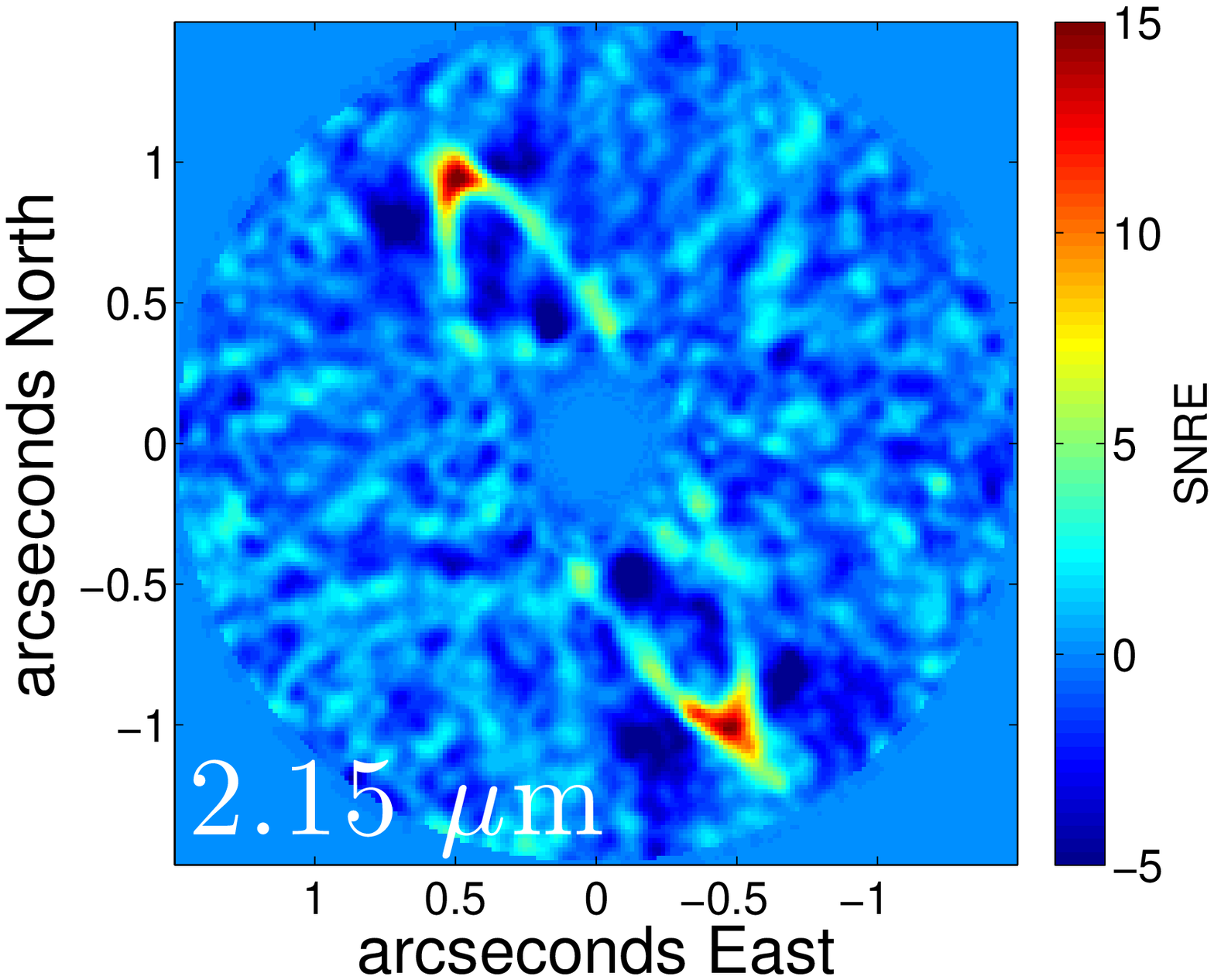}} \\
\subfloat[]{\label{fig:visaoysn}\includegraphics[width=0.34\textwidth]{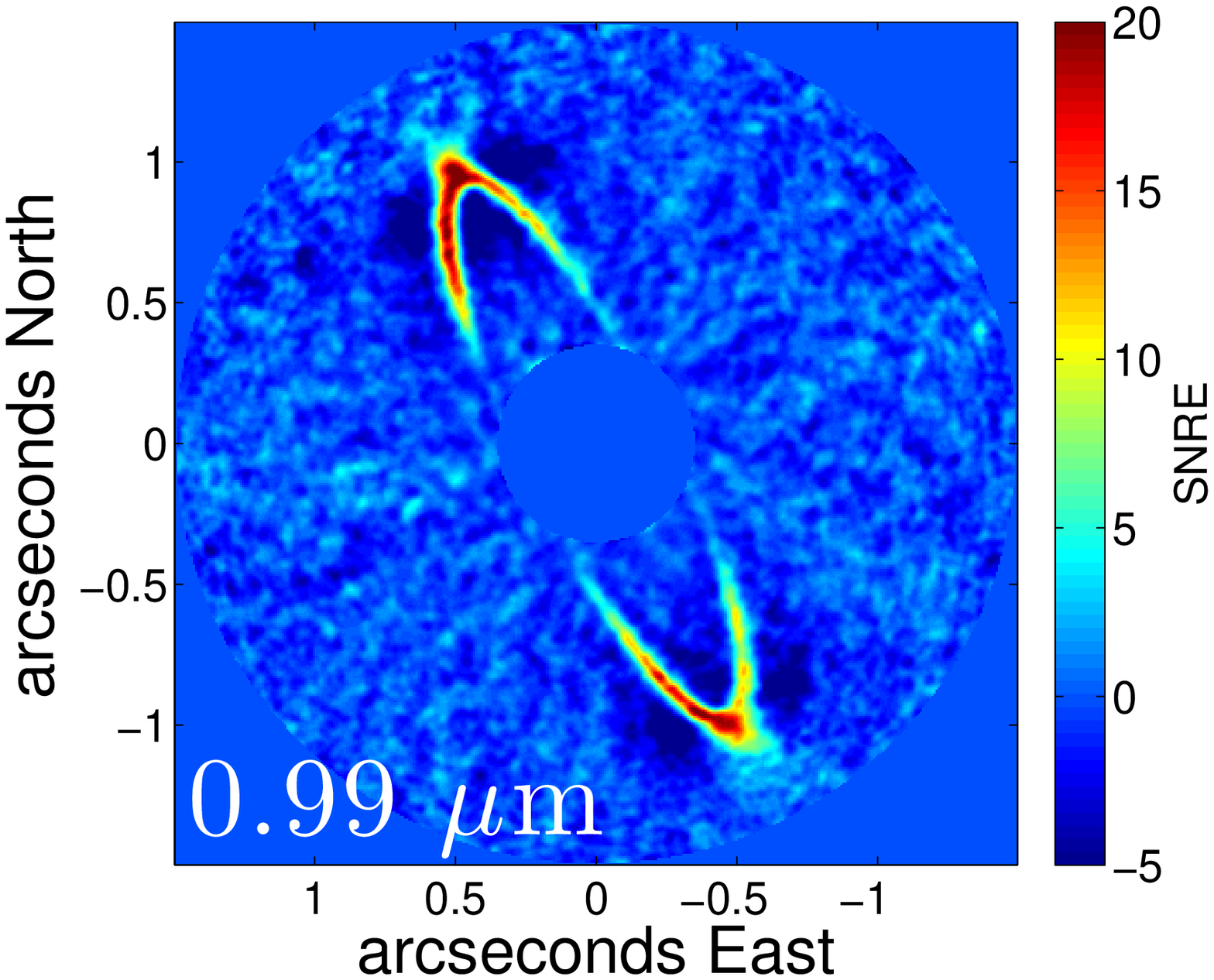}} 
\subfloat[]{\label{fig:visaozsn}\includegraphics[width=0.34\textwidth]{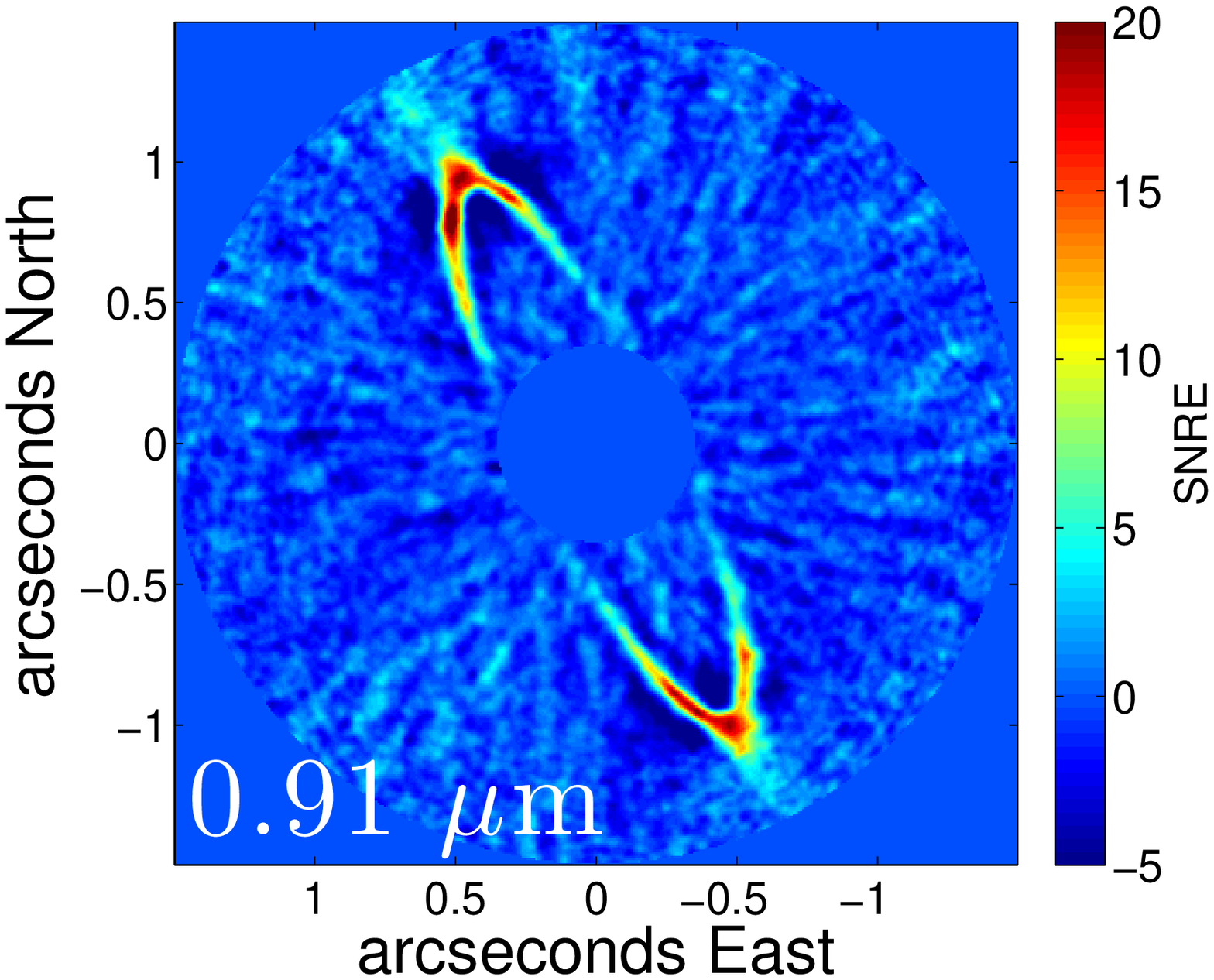}} 
\subfloat[]{\label{fig:visaoisn}\includegraphics[width=0.34\textwidth]{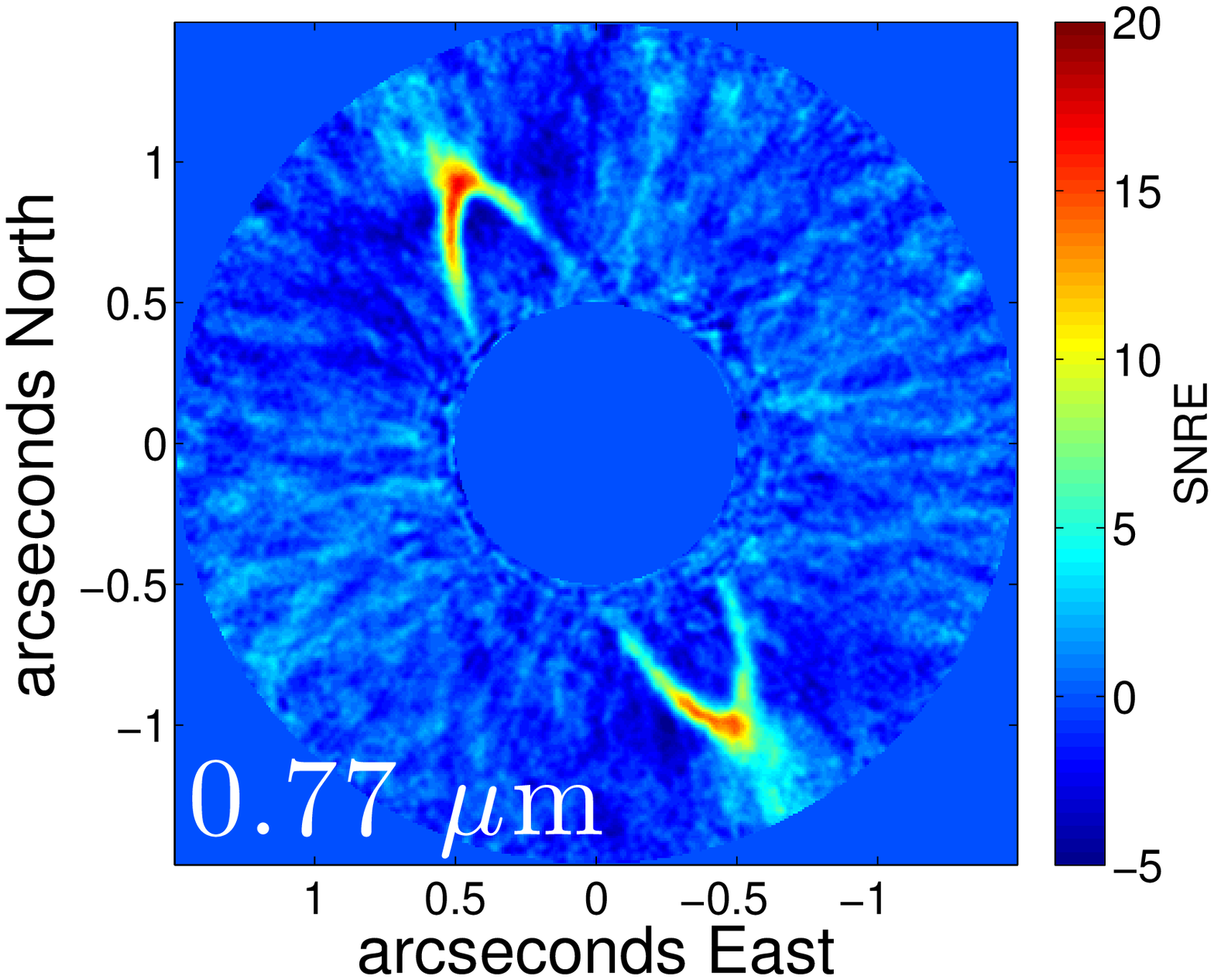}} 
\caption{SNRE maps of the HR 4796A debris disk, with North-up, East-left. Digital masks have been added for display purposes in all images. (a-d): The ring is detected at SNRE of 2-11 (\lprime), 2-10 ($Ls$), 2-8 ($Ice$ band), and 2-16 (\ks band), respectively. (e-g): The ring is detected at SNRE of 2-23 ($Ys$), 2-25 ($z'$), and 2-17 ($i'$ band), respectively.}
\label{fig:snremaps}
\end{figure*}

We observed HR 4796A at the Magellan Clay telescope at the Las Campanas Observatory (LCO) in Chile using the Clio-2 1-5 \microns ~camera and MagAO. All Clio-2 observations are summarized in Table \ref{tab:obs}. We observed the target at 3.8 \microns ~(\lprime) on UT 7 April 2013, at 3.3 \microns ~(hereafter $Ls$ for ``Lshort") on UT 8 April 2013, and at 3.1 \microns ~(hereafter $Ice$ band) and 2.15 \microns ~(\ks band) on UT 9 April 2013, all using the narrow camera (plate scale = 0\fasec 01585 per pixel\footnote{http://zero.as.arizona.edu/groups/clio2usermanual/}). Due to thermal light leakage in the $Ks$ filter, the $Ks$ band images were contaminated by excess light. This resulted in a $Ks$ band PSF that was much more ``blurred" than would have been expected for a normal $Ks$ filter. Therefore we repeated the $Ks$ band observations \about one year later, on UT 10 April 2014. Henceforth, we use only this latter dataset and ignore the original light-leaked images. The observing conditions for all nights were excellent, with seeing values ranging from 0\fasec 5-1\asec, and the AO corrected 300 modes for all observations. The observing setup and strategy was the same for each night, with the camera rotator off to facilitate angular differential imaging (ADI; \citealt{adi}), and no coronagraphs were used. We nodded the telescope by several arcseconds vertically along the detector every few minutes and dithered the star horizontally by a fraction of an arcsecond on each nod to mitigate the sky background and bad pixels/detector artifacts. Because MagAO delivers such high Strehl ratios at 1-5 \microns, nodding the telescope does not significantly change the PSF, which is critical for the PSF subtraction described in Section \ref{sec:datareduction}. At each nod position unsaturated exposures of the star were taken for photometric references. The longer exposure images were saturated out to \about 0\fasec 1.

After discarding images of poor quality (e.g., from the AO loop being open, PSF blurring due to wind shake, or improper nods), at \lprime ~we obtained 455 science exposure images of the target for a total integration of 2.5 hours; at $Ls$ we obtained 120 images for a total integration of 80 minutes; at $Ice$ band we obtained 349 images for a total integration of 87 minutes; and at $Ks$ band we obtained 240 images for a total integration of 80 minutes. The total parallactic angle rotation was 150.73\degrees, 93.5\degrees, 115.4\degrees, and 94\degrees ~for \lprime, $Ls$, $Ice$ band, and \ks band, respectively. 

\subsubsection{VisAO}
We observed HR 4796A at the Magellan Clay telescope using the VisAO camera and MagAO on UT 7 April 2013 ($z'$; $\lambda_{c}$ = 0.91 \microns), UT 8 April 2013 ($Ys$; $\lambda_{c}$ = 0.99 \microns), and UT 9 April 2013 ($i'$; $\lambda_{c}$ = 0.77 \microns), contemporaneously with the Clio-2 observations reported above. All VisAO observations are summarized in Table \ref{tab:obs}. The VisAO camera has a plate scale of 0\fasec 0079 per pixel \citep{lairdtrapmagao,jaredbetapic} and a field of view of 8\fasec 09. No coronagraph was used for any of the observations, resulting in the central 0\fasec 1-0\fasec 5 regions around the star being saturated in the images. The observing strategy was the same as for the observations conducted with Clio-2. After discarding images of poor quality, we obtained 285 images at $Ys$ for a total integration of 95 minutes; at $z'$ we obtained 3030 images for a total of 115 minutes of integration; and at $i'$ we obtained 342 images for a total integration of 114 minutes. The total parallactic angle rotation was 93.6\degrees ~at $Ys$, 96.7\degrees ~at $z'$, and 119.4\degrees at $i'$. 


\subsection{Data reduction}
\label{sec:datareduction}
\subsubsection{Clio-2}
All data reduction discussed below was performed using custom scripts in Matlab. The steps described were carried out identically for all four datasets, including the unsaturated photometric data. We divided all the images by the number of coadds, corrected for non-linearity\footnote{http://zero.as.arizona.edu/groups/clio2usermanual/}, and divided by the integration time to obtain units of detector counts/s. We then subtracted opposite-nod images from each other to remove the non-negligible sky background and corrected for bad pixels. Next we determined the sub-pixel location of the star in each of the sky-subtracted images by calculating the center of light inside a 0\fasec 5 aperture centered on the star. Based on previous imaging results \citep{mehd15115,me32297}, this is a satisfactory method as long as saturation is limited to $<$ 0\fasec 2. The images were then registered so that the star was at the center of each image. During the observing run, bright stars produced ghosts and streaks whose positions varied in unpredictable ways. To remove this unwanted noise, we calculated the standard deviation for each pixel through each star-centered datacube and masked regions where the standard deviation was high (relative to some threshold value). This filtering process significantly improved the quality of our final reduced images.

We then fed the registered, cropped, sky-subtracted images into our custom Principal Component Analysis (PCA, \citealt{pca}) pipeline. For the datasets discussed in this work, we found optimal signal-to-noise per resolution element (SNRE\footnote{SNRE is computed in the same manner as is outlined in \cite{me32297} and \cite{mehd15115}.}) detections of the disk  with ``classical" PCA (no small search areas, no rotation requirement). The number of modes that maximized the disk's average SNRE was 28 (out of 455), 13 (out of 120), 21 (out of 349), and 26 (out of 240) for \lprime, $Ls$, $Ice$ band, and \ks band, respectively. We also tested the LOCI algorithm \citep{loci} on our data but found better results with PCA, in agreement with previous studies \citep{tiffany2,me32297,hip79977,bonnefoybetapic,bocbetapic,pca}. We de-rotated all the PSF-subtracted images by their parallactic angles $+$ a small offset (-1.8\degrees\footnote{http://zero.as.arizona.edu/groups/clio2usermanual/}) to obtain North-up, East-left and combined all the images using a mean with sigma clipping.

Fig. \ref{fig:images} shows the final PCA-reduced images at each wavelength, and Fig. \ref{fig:snremaps} shows their corresponding SNRE maps. The debris ring is detected at SNRE \about 2-11 at \lprime, \about 2-10.6 at $Ls$, \about 2-8 at $Ice$ band, and \about 2-16 at $Ks$ band. The ring is detected at high SNRE outside of 0\fasec 4 (in front of an behind the star), allowing accurate characterization of the disk's geometry and morphology.

\begin{table*}[t]
\centering
\caption{HR 4796A Thermal Emission Photometry (Star + Disk)}
\begin{tabular}{c c c c c}
\hline
\hline
$\lambda$ (\microns) & $\delta\lambda$ (\microns) & $F_{\nu}$ (Jy) & Uncertainty (Jy) & Reference \\
\hline
5.46	&	--	&	0.549	&	0.029	&	1; \cite{chencatalog}	\\
5.97	&	--	&	0.454	&	0.024	&	1		\\
6.49	&	--	&	0.368	&	0.020	&	1		\\
7.00	&	--	&	0.334	&	0.018	&	1		\\
7.59	&	--	&	0.276	&	0.015	&	1		\\
7.90	&	0.87	&	0.307	&	0.044	&	2; \cite{wahhaj4796old}		\\
8.60	&	--	&	0.236	&	0.013	&	1		\\
9.63	&	--	&	0.199	&	0.011	&	1		\\
10.10	&	5.10	&	0.270	&	0.026	&	3; \cite{hr4796fajardo}		\\
10.30	&	1.01	&	0.218	&	0.024	&	2		\\
10.30	&	1.30	&	0.233	&	0.024	&	3		\\
10.66	&	--	&	0.195	&	0.010	&	1		\\
10.80	&	5.30	&	0.188	&	0.047	&	4; \cite{hr4796telesco}		\\
11.60	&	1.30	&	0.225	&	0.070	&	3		\\
11.69	&	--	&	0.211	&	0.011	&	1		\\
12.00	&	6.50	&	0.195	&	0.018	&	5; IRAS		\\
12.50	&	1.16	&	0.231	&	0.014	&	2		\\
12.50	&	1.20	&	0.253	&	0.027	&	3		\\
12.50	&	1.20	&	0.223	&	0.018	&	6; \cite{koerner4796}		\\
12.72	&	--	&	0.241	&	0.013	&	1		\\
13.78	&	--	&	0.310	&	0.017	&	1		\\
15.17	&	--	&	0.442	&	0.024	&	1		\\
16.61	&	--	&	0.680	&	0.036	&	1		\\
18.05	&	--	&	0.942	&	0.051	&	1		\\
18.10	& 1.94	& 1.106 & 0.007	& 7; \cite{moerchen4796}		   \\
18.20	&	1.70	&	0.905	&	0.130	&	4		\\
18.20	&	1.70	&	1.100	&	0.150	&	8; \cite{hr4796jura98}		\\
19.49	&	--	&	1.286	&	0.069	&	1		\\
20.00	&	9.00	&	1.860	&	0.186	&	9; \cite{hr4796jura93}		\\
20.80	&	1.00	&	1.620	&	0.160	&	2		\\
20.80	&	1.70	&	1.880	&	0.170	&	6		\\
21.77	&	--	&	2.156	&	0.116	&	1		\\
24.00	&	4.70	 &	3.030	&	0.303	&	10; \cite{hr4796low}		\\
24.50*	&	0.80 	&	2.100	&	0.170	&	2		\\
24.50*	&	0.80 	&	2.270	&	0.700	&	6		\\
24.50	& 1.92	& 3.307 &	0.047	&	 7    \\
24.65	&	--	&	3.287	&	0.176	&	1		\\
25.00*	&	11.00	&	4.518	&	0.407	&	5		\\
27.53	&	--	&	4.129	&	0.221	&	1		\\
30.41	&	--	&	4.878	&	0.261	&	1		\\
33.29	&	--	&	5.528	&	0.296	&	1		\\
54.51	&	--	&	6.414	&	0.341	&	11; Spitzer/MIPS (this work)		\\
57.91	&	--	&	5.901	&	0.277	&	11		\\
60.00*	&	40.00	&	7.835	&	0.705	&	5		\\
61.31	&	--	&	5.664	&	0.255	&	11		\\
64.71	&	--	&	5.438	&	0.251	&	11		\\
65.00	&	30.00	&	6.071	&	0.313	&	12; \cite{hr4796akari}		\\
68.11	&	--	&	5.300	&	0.259	&	11		\\
70.00	&	19.00	&	5.160	&	1.100	&	10		\\
70.00*	&	15.00	&	4.980	&	0.131	&	13; \cite{hr4796herschel}		\\
71.51	&	--	&	5.048	&	0.270	&	11		\\
74.91	&	--	&	4.806	&	0.279	&	11		\\
78.31	&	--	&	4.416	&	0.282	&	11		\\
81.71	&	--	&	4.136	&	0.284	&	11		\\
85.11	&	--	&	4.026	&	0.299	&	11		\\
88.51	&	--	&	3.879	&	0.310	&	11		\\
90.00	&	50.00	&	4.501	&	0.186	&	12		\\
91.91	&	--	&	3.716	&	0.324	&	11		\\
95.31	&	--	&	2.853	&	0.289	&	11		\\
100.00	&	37.00	&	3.854	&	0.347	&	5		\\
100.00	&	40.00	&	3.553	&	0.097	&	13		\\
160.00	&	35.00	&	1.800	&	0.360	&	10		\\
160.00	&	85.00	&	1.653	&	0.068	&	13		\\
350.00	&	40.00	&	0.160	&	0.042	&	2		\\
450.00	&	30.00	&	0.180	&	0.150	&	14; \cite{hr4796greaves}		\\
450.00	&	48.40	&	0.180	&	0.150	&	15; \cite{hr4796sheret}		\\
800.00	&	100.00	&	0.028	&	0.009	&	16; \cite{hr4796jura95}		\\
850.00	&	50.00	&	0.019	&	0.003	&	14		\\
850.00	&	96.00	&	0.019	&	0.003	&	15		\\
870.00	&	150.00	&	0.021	&	0.007	&	17; \cite{hr4796nilsson}		\\
\hline
\end{tabular} 
\\
\raggedright
$^{*}$Denotes data that are excluded from the analysis described in Section \ref{sec:modeling}. Dashes denote binned spectra. \\
\label{tab:unresolved}
\end{table*} 

\subsubsection{VisAO}
We reduced the VisAO $Ys$, $z'$, and $i'$ data in the same manner as we reduced the Clio-2 data, except for a few differences: for the $z'$ data, we manually coadded the images in sets of 10, resulting in 303 coadded images, to ease the computational effort of the PCA reduction; we did not perform any sky subtraction because the small plate scale of VisAO combined with the faint sky at 0.7-1 \microns ~renders the sky background negligible. Instead, because dark frames were periodically taken throughout the observations, we subtracted from each science image the dark frame taken closest in time to it. We also did not perform any standard deviation filtering because the detector only has one ghost whose position and brightness vary predictably. The number of PCA modes that optimized the SNRE of the disk was 23 (out of 285) at $Ys$, 28 (out of 303) at $z'$, and 90 (out of 342) at $i'$. We combined the sets of reduced images using a mean with sigma-clipping and rotated the final images by their parallactic angles $+$ a small offset (-0.59\degrees, from \citealt{jaredbetapic}) to obtain North-up, East-left. The final PCA-reduced images and their corresponding SNRE maps are shown in Fig. \ref{fig:images} and Fig. \ref{fig:snremaps}. The disk is detected at SNRE \about 2-23 beyond \about 0\fasec 4 at $Ys$, \about 2-25 beyond \about 0\fasec 4 at $z'$, and at \about 2-17 beyond \about 0\fasec 5 at $i'$.

\subsubsection{Spitzer/MIPS}
We also include here previously unpublished Spitzer/MIPS \citep{spitzer,mips} data. SED-mode observations (55.0-90.0 $\mu$m; $\lambda$/$\Delta \lambda$ $\sim$ 20) on HR 4796A (AOR key: 16169984) were obtained on 15 February 2016 using the 19.6$\arcsec$ $\times$ 157$\arcsec$ slit/grating and 5 cycles of 10 s integrations. Each observing cycle consisted of six pairs of 10 s exposures with the slit position alternated between the object and blank sky 1$\arcsec$ away. To remove the background, each sky exposure was subtracted from the immediately preceding frame containing the object. Exposures of an internal calibration source were interspersed within the cycle to track the varying response of the 70 $\mu$m detector array. The raw MIPS data were corrected for distortion, registered, mosaicked, and flat-fielded using the MIPS instrument team’s data analysis tool (DAT; \citealt{mipsreduction}). The spectrum was extracted using a 5 pixel extraction aperture and calibrated using the procedure and calibration files described in \cite{mipssed}. The processing steps included applying the dispersion solution, an aperture correction (to account for slit losses), and a flux calibration based on $\sim$20 infrared standard stars \citep{mips70um}. The Spitzer/MIPS photometry is shown in Table \ref{tab:unresolved}. 

\subsubsection{Literature far-infrared data}
We also compiled literature photometry on HR 4796A at far-infrared (FIR) wavelengths, to be used in Section \ref{sec:modeling} for the dust grain composition modeling. The data and their references are shown in Table \ref{tab:unresolved}. The observations and data reduction for the binned Spitzer/IRS data are described in \cite{chencatalog}.

\section{Results}
\label{sec:results}
\subsection{Disk Morphology}
\subsubsection{Geometrical and Orbital Parameters}
\label{sec:geometry}
\begin{table*}[t]
\centering
\caption{Geometrical Parameters of the Ring}
\begin{tabular}{c|c c c c c c c c c c}
\hline
\hline
 & $a$ ($^{\prime \prime}$) & $e$ & $\omega$ (\degrees) & $i$ (\degrees) & $\Omega^{*}$ (\degrees) & $\Delta RA$ (mas) & $\Delta Dec$ (mas) & $\Delta\tilde{a}$ (mas) & $\Delta\tilde{b}$ (mas) \\ 
 \hline
$Ys$	 & 1.08$\pm$0.01 & 0.06$\pm$0.02 & 101.40$\pm$7.00 &	76.11$\pm$0.59 & 26.70$\pm$0.16 & -7.56$\pm$4.77 & -16.86$\pm$3.91 & -11.67$\pm$4.10	 & 14.33$\pm$4.61	\\
$z'$	 & 1.08$\pm$0.01 & 0.04$\pm$0.01 & 135.18$\pm$22.88 & 77.07$\pm$0.27 & 26.22$\pm$0.29 & 5.66$\pm$7.81 & -27.61$\pm$7.08 & -27.27$\pm$7.23	 & 7.12$\pm$7.67	\\  
$Ks$	 & 1.11$\pm$0.01 & 0.08$\pm$0.02 & 108.19$\pm$7.91 &	77.75$\pm$0.64 & 26.72$\pm$0.28 & -3.92$\pm$6.8 & -29.61$\pm$8.17 & -24.69$\pm$7.92	 & 16.81$\pm$7.10	\\	  	  	  	  	  	
$Ls$	 & 1.09$\pm$0.01 & 0.09$\pm$0.01 & 100.70$\pm$4.12 &	78.02$\pm$0.64 & 26.62$\pm$0.09 & -9.43$\pm$3.64 & -24.78$\pm$6.17 & -17.93$\pm$5.75	 & 19.53$\pm$4.27	\\
\lprime & 1.08$\pm$0.01 & 0.05$\pm$0.02 & 107.58$\pm$12.26 &	77.35$\pm$0.24 & 26.55$\pm$0.11 & -3.15$\pm$6.45 & -19.19$\pm$11.31 & -15.76$\pm$10.52	 & 11.39$\pm$7.67	\\
\hline
mean & 1.09$\pm$0.01 & 0.06$\pm$0.02 & 110.61$\pm$12.67 &	76.47$\pm$0.45 & 26.56$\pm$0.20 & -3.68$\pm$6.08 & -23.61$\pm$7.72 & -19.46$\pm$7.42 & 13.84$\pm$6.44	\\
\hline
\end{tabular} 
\raggedright
\\
$^{*}$Errors in $\Omega$ are relative in that they do not include the absolute uncertainty in the position of true North on the detector (0.10\degrees ~for Clio-2 and 0.30\degrees for VisAO). \\ 
\label{tab:parameters}
\end{table*} 

The HR 4796A debris ring can be described by five sky-projected parameters: the semimajor axis, $a$, the semiminor axis, $b$, the position angle measured east of north, PA, and the ellipse center ($\Delta RA,\Delta Dec$), measured along RA and Dec, respectively. Once these parameters are measured, we can calculate the true orbital elements of the disk ($a$, $e$, $\omega$, $i$, and $\Omega$, where $\omega$ is the longitude of periastron, $i$ is the inclination from face-on viewing, and $\Omega$ is the longitude of the ascending node),  using the Kowalsky deprojection routine \citep{kowalsky,stark181327}. We set out to measure the sky-projected parameters using our high SNRE images of the disk (excluding the $Ice$ band and $i'$ images images due to the low SNRE at small inner working angles in both images). This is necessary in order to verify the results of \cite{hr4796schneider}, \cite{thalmannhr4796}, and \cite{wahhaj4796nici}, who all found that the ring is offset from the star by \about a few AU and thus might indicate the presence of one or more perturbing planetary companions \citep{wyatthr4796}.

We followed the ``maximum merit" procedure outlined in \cite{thalmannhr4796} and \cite{moth} whereby binary images of ellipses described by varying (random) parameters are multiplied with the real images of the disk until a maximum fit is obtained. The parameters were drawn from the following uniform distributions: $a \in [0.95, 1.2]$\asec, $b \in [0.2,0.3]$\asec, PA $\in [24, 28]$\degrees, $\Delta RA \in [-50, 50]$ mas, and $\Delta Dec \in [-50, 50]$ mas. These limiting values were chosen based on previous fitting results from \cite{hr4796schneider}, \cite{thalmannhr4796}, and \cite{wahhaj4796nici}. The width of the ellipse is an additional free parameter and can significantly alter the fitting results; if the ellipse is too thin, the fitting will naturally prefer the brightest parts of the ring, which may change with wavelength due to the scattering properties of the dust; if the ellipse is too wide, the fitting can be biased by residuals and noise outside the ring. We found that wide ellipses generally offered poorer fits than narrower ellipses, therefore we chose ellipses with a semimajor axis width of 0\fasec 128 (8 Clio-2 pixels, 16 VisAO pixels) and a semiminor axis width of 0\fasec 032 (2 Clio-2 pixels, 4 VisAO pixels). 

\begin{figure*}[t]
\centering
\subfloat[]{\label{fig:offseta}\includegraphics[scale=0.47]{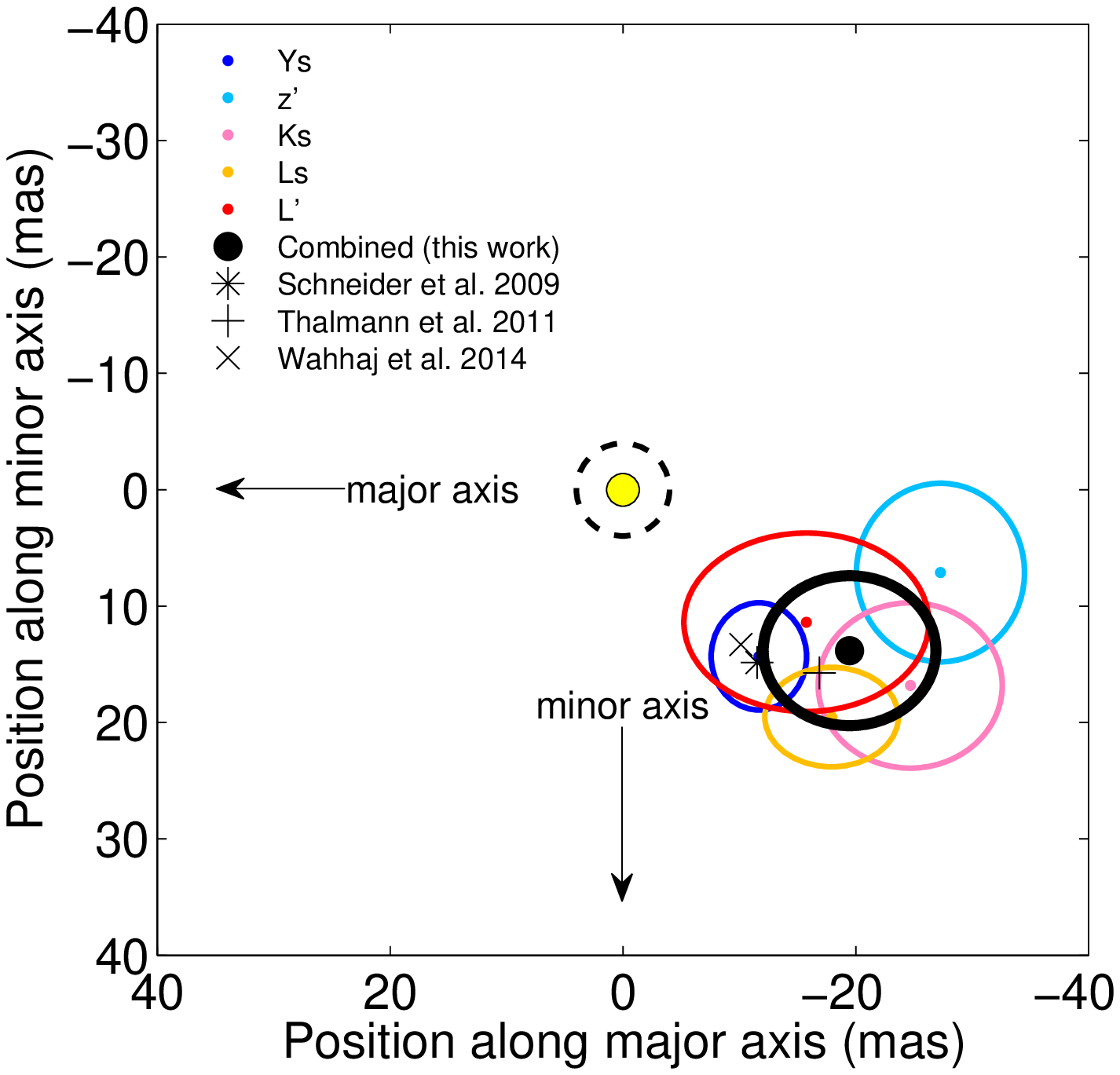}} 
\subfloat[]{\label{fig:offsetb}\includegraphics[scale=0.47]{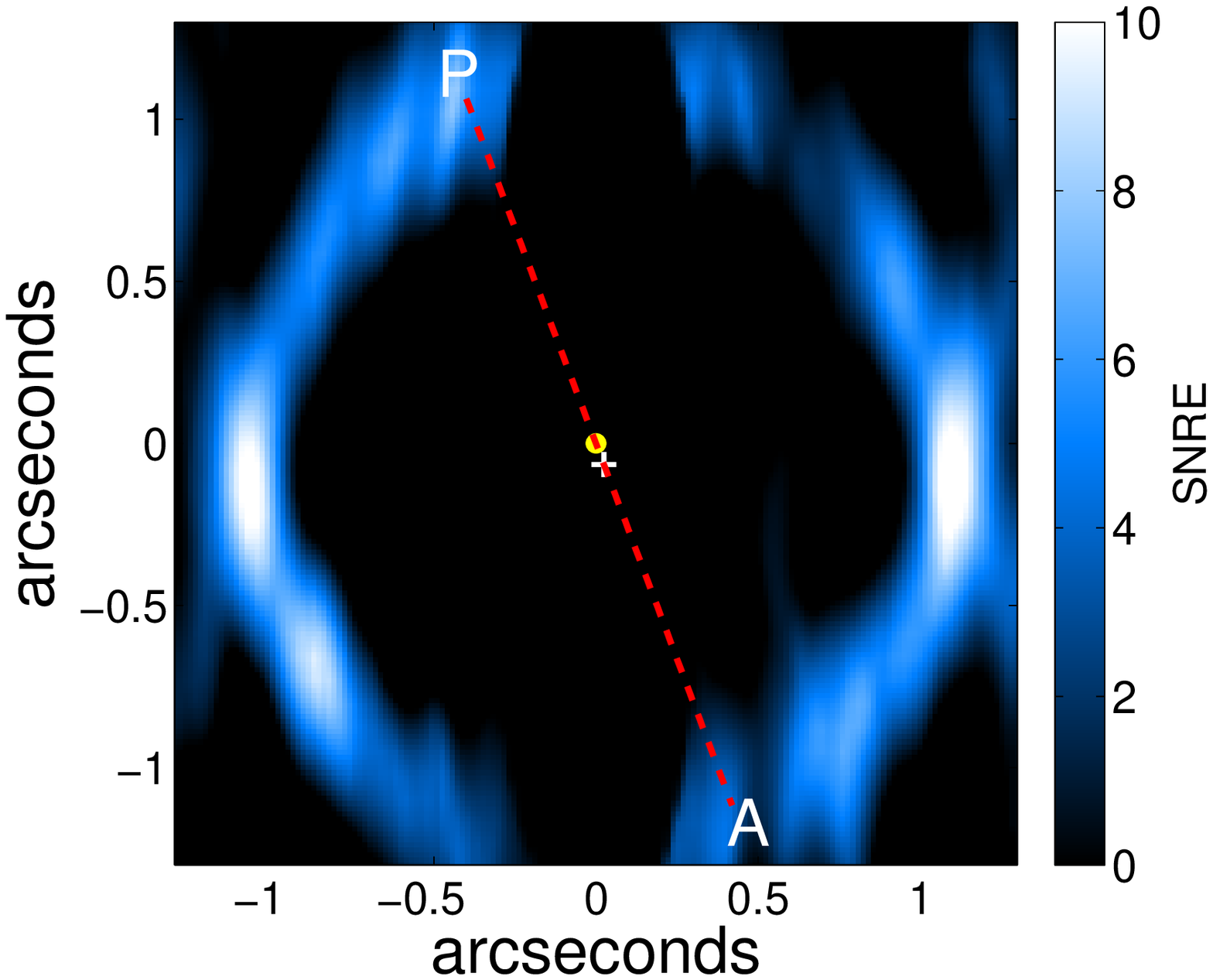}}
\caption{(a): Best-fit apparent ring center positions (colored points) and uncertainties (colored ellipses) at $Ys$, $z'$, $Ks$, $Ls$, and \lprime, relative to the star's location (yellow circle). Also plotted are the reported ring center positions from \cite{hr4796schneider}, \cite{thalmannhr4796}, and \cite{wahhaj4796nici}, without their corresponding uncertainties for visual clarity. The coordinates have been rotated counter-clockwise such that the disk's major axis is horizontal and the minor axis is vertical. North is located towards the top-left and East is located towards the bottom left. The uncertainty in the star's location (dashed circle) is estimated to be \about 4 mas. Combining all of the ring center positions and their uncertainties, the ring is offset from the star along the major axis by 2.6$\sigma$ and along the minor axis by 2.15$\sigma$. Our results generally agree with previous estimates of the ring offset. (b): Rotated and deprojected \lprime ~SNRE map of the HR 4796A debris ring, with the location of the star marked by the yellow dot and the deprojected center of the ring marked by the $+$ symbol. North and East are the same as in (a). The dashed red line defines the deprojected major axis of the disk, with ``P" and ``A" denoting the locations of the ring's periastron (110.61\degrees) and apoastron (290.61\degrees), respectively. Our fitting results indicate that the star is closer to the \textit{western} side of the ring.}
\label{fig:offset}
\end{figure*}

For each image/wavelength, we generated 50,000 binary ellipses and multiplied them with the SNRE images of the disk. As in \cite{thalmannhr4796}, we computed the merit value, defined as the average SNRE pixel value inside the ellipse, and recorded this value. We then averaged the ring parameters associated with the five highest merit values; the uncertainties on each parameter were computed as the standard deviation of the five values. We also computed the offsets along the major and minor axes, $\Delta\tilde{a}$ and $\Delta\tilde{b}$, by rotating the $\Delta RA$ and $\Delta Dec$ offsets clockwise by the fitted PA of the disk at each respective wavelength. The uncertainties in the offsets along the major and minor axes were calculated by propagating the uncertainties in $\Delta RA$ and $\Delta Dec$. Next we determined the debris ring's true orbital elements using the Kowalsky deprojection routine \citep{kowalsky,stark181327}. The debris ring's fitted sky-projected and true geometrical parameters are reported in Table \ref{tab:parameters}. 

We find similar values for all parameters to those reported by \cite{hr4796schneider}, \cite{thalmannhr4796}, and \cite{wahhaj4796nici}. In particular, we find that the ring is offset along both the major and minor axes by 2.6$\sigma$ and 2.15$\sigma$, respectively. Within the uncertainties, the center of the ring (along both RA/Dec and major/minor axis) agrees with previous results (see Fig. \ref{fig:offseta}). Most importantly, the physical deprojected center of the ring is offset from the star by a total of 4.76$\pm$1.6 AU\footnote{The error on this parameter includes the 4 mas uncertainty in the star's position.} (see Fig. \ref{fig:offsetb}). This corresponds to a physical deprojected eccentricity of 0.06$\pm$0.02, which is in general agreement with previously reported values.

Fig. \ref{fig:offsetb} shows the rotated, deprojected SNRE map of the disk at \lprime ~(since the disk is detected at the smallest inner working angles at \lprime), with the locations of the star and ring center marked. Based on our geometrical fitting, the star is closer to the \textit{western} side of the ring. 

\subsubsection{Width of the Ring}
Several numerical studies have recently shown that planets can create sharp inner and outer ring edges \citep{chiang,collisiondynamics,fomalhautalma}. \cite{medynamics} and \cite{chiang} also showed that a shepherding planet's properties can be constrained by the debris ring's intrinsic width. \cite{hr4796schneider} used HST/STIS coronagraphic images to measure the width of the ring near the ansae and found it to be 0\fasec 184 $\pm$ 0\fasec 01. Recently \cite{wahhaj4796nici} measured the normalized width of the ring from ground-based AO images obtained by NICI in the $J$, $H$, and $K$ bands to be \about 0.10, indicating a much narrower ring. Both of these studies were limited in that the ring was not detected at adequate SNRE at small inner working angles. Our VisAO and Clio-2 images resolve the ring as close as \about 0\fasec 4, which allows us to obtain more accurate measurements of the average ring width. Our VisAO images, in particular, have resolutions of \about 0\fasec 03, lessening the disk-broadening effect of PSF blurring.

\begin{figure*}[t]
\centering
\subfloat[]{\label{fig:radprofiles}\includegraphics[scale=0.47]{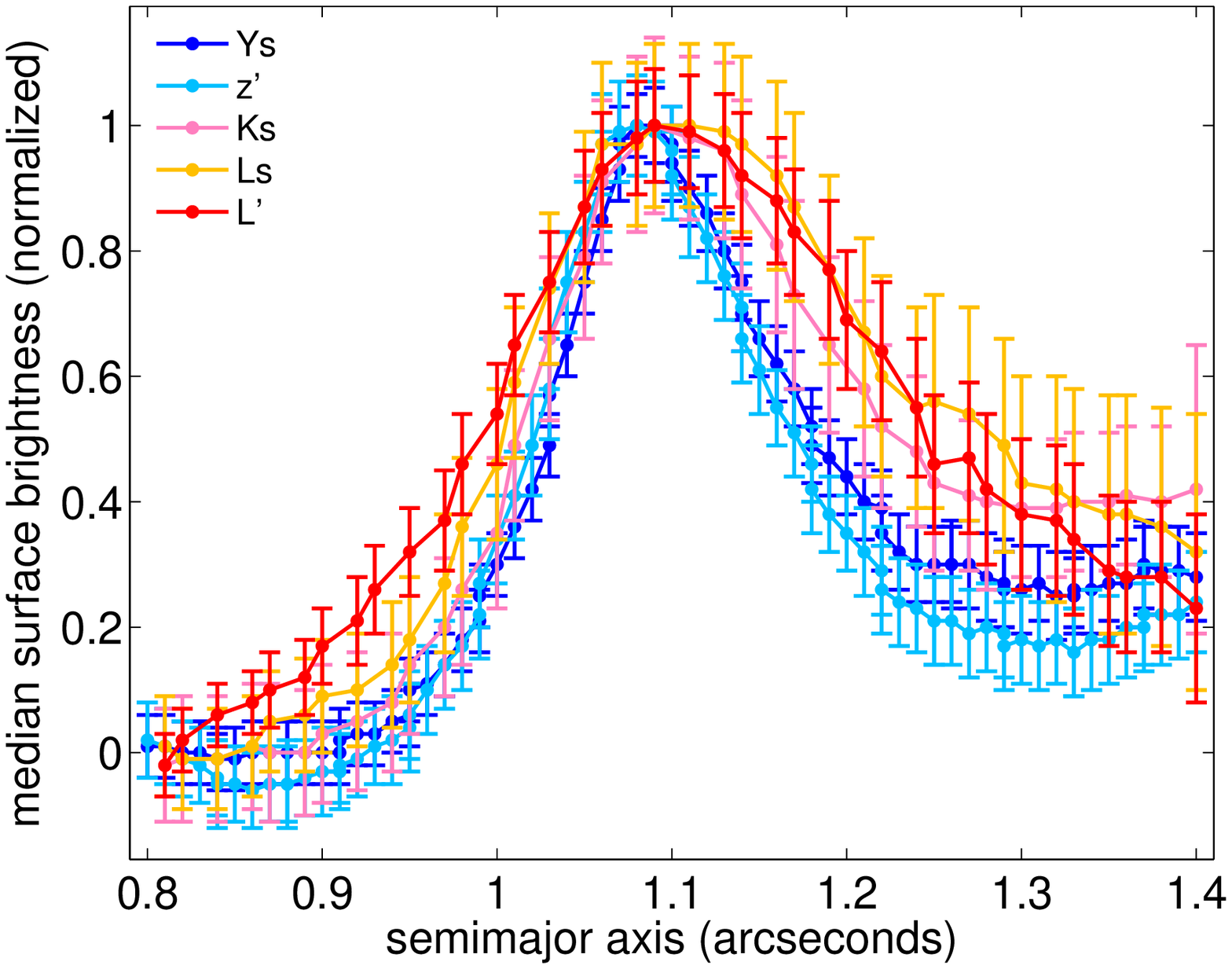}} 
\subfloat[]{\label{fig:nfwhm}\includegraphics[scale=0.47]{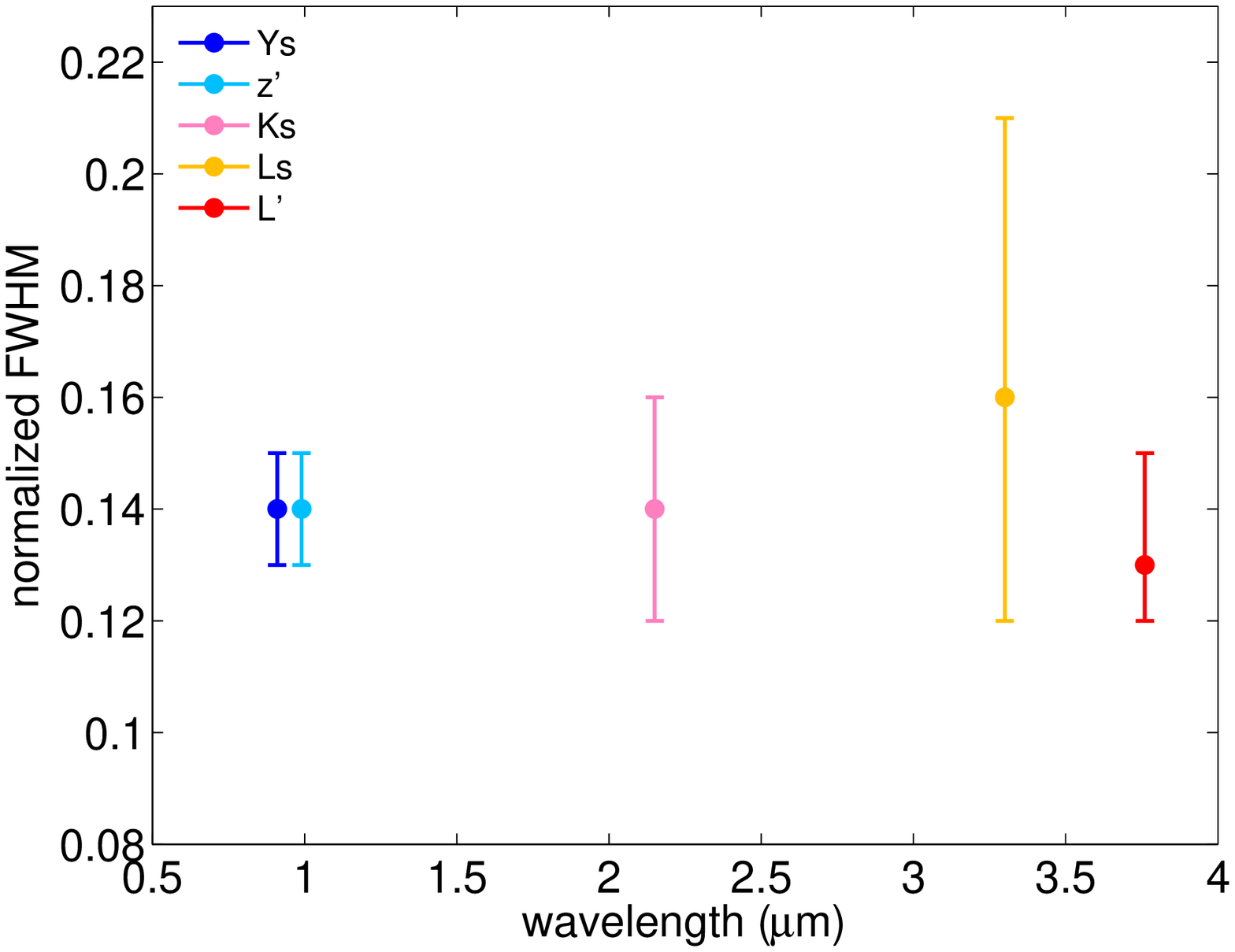}}
\caption{(a): Normalized, azimuthally-medianed radial profiles of the HR 4796A debris ring at $Ys$, $z'$, $Ls$, $Ks$, and \lprime. Because PSF broadening has not been corrected, the profiles at 2-4 \microns ~appear broader than at $< 1$ \microns. (b): The normalized FWHM (nFWHM) of the ring at each wavelength, after accounting for PSF broadening. Our measured values are equivalent within the uncertainties and are generally consistent previous measurements. We find a combined value of 0.14$^{+0.03}_{-0.02}$.}
\label{fig:FWHM}
\end{figure*}

To measure the intrinsic ring width, we followed the procedure outlined in \cite{medynamics}. As before, we considered only the $Ys$, $z'$, $Ls$, $Ks$, and \lprime ~images, since these have the highest SNRE detections of the ring. First, we rotated each image counter-clockwise by 90\degrees ~$-$ the best-fit PA (from Table \ref{tab:parameters}) so that the major axis was along the $x$-axis. Next, using the best-fit inclination, we deprojected the images. We calculated the distance of each deprojected pixel from the star (which was at the center of the images) and stored the squares of these values (for later use). We then shifted the images so that the ring center was at the center of each image. Finally we rotated the images clockwise by the corresponding argument of pericenter so that the true semimajor axis of the ring was along the $x$-axis. 

To compute the radial profiles, we generated ellipses with the same geometry as the ring. The ellipses had semiminor axis widths equal to 1 pixel (0\fasec 0079 for VisAO and 0\fasec 01585 for Clio-2). From 0\fasec 8 to 1\fasec 4, the surface brightness was computed as the median pixel value inside each annulus divided by the respective plate scale squared, and the semimajor axis of each annulus was stored. 

Images of point- and extended sources obtained using ADI typically suffer from varying degrees of self-subtraction. This can significantly shrink the apparent width of the ring. Indeed negative residuals are evident both inside and outside the ring in Fig. \ref{fig:images}. These negative residuals are retained in the radial profiles and are generally present within 0\fasec 9. To account for this bias, for each image we subtracted the average negative residual in the 0\fasec 8-0\fasec 9 region from all the values in the radial profiles. The geometric dilution of star light with distance squared was remedied by multiplying the resultant radial profiles by the previously stored squared distance values. Finally, the profiles were normalized by the respective peak surface brightness values. The uncertainties in the profiles at each semimajor axis were computed by repeating the above procedure on the noise maps used to generate the SNRE images at each wavelength. The final radial profiles are shown in Fig. \ref{fig:radprofiles}.

These profiles do not take into account the broadening of the PSF at each wavelength. At $<$ 1 \microns, where the resolution is \about 0\fasec 03, the broadening is not severe and can be neglected. But at 2-4 \microns, this is not the case. We measured the ring broadening effect at $Ks$, $Ls$, and \lprime ~by repeating the above procedure on a (qualitatively similar) convolved model disk that was inserted 90\degrees ~away from the real disk into the raw data and then recovered after ADI/PCA data reduction. The observed FWHM values (computed by measuring the distance between the half-peak locations) were compared with the FWHM of the unconvolved noiseless model disk. At $Ks$, the PSF convolution broadened the disk by a factor of 1.43; at $Ls$, by a factor of 1.57; and at \lprime, by a factor of 1.74. We then divided the observed FWHM of the real disk at $Ks$, $Ls$, and \lprime ~by their respective broadening factors.

We computed the normalized FWHM (nFWHM) at each wavelength by dividing the FWHM values by their respective peak semimajor axis values. The uncertainties in the nFWHM values were computed by repeating this procedure on the profiles $+$ and $-$ the errors at each semimajor axis. These values are shown in Fig. \ref{fig:nfwhm}. Our measured values are equivalent within the uncertainties and are generally consistent with previous measurements from \cite{hr4796schneider}, \cite{lagrange4796}, and \cite{wahhaj4796nici}. Assuming nFWHM is constant with wavelength, we find the combined nFWHM = 0.14$^{+0.03}_{-0.02}$. This corresponds to a physical width of 11.1$^{+2.4}_{-1.6}$ AU.

Inserting the nFWHM value into Eq. 5 from \cite{medynamics}, the maximum mass of an interior planet shepherding the ring would be 4.0$^{+3.0}_{-2.5}$ \mj. Using Eq. 2 from \cite{medynamics}, the planet would have a minimum semimajor axis in the 48-60 AU range and its eccentricity would be $\approx$ 0.06 (the eccentricity of the ring). Assuming the hypothetical planet is apsidally aligned with the ring such that they share the same periastron and apoastron locations (denoted in Fig. \ref{fig:offsetb}), the planet would spend more time on its orbit closer to apoastron than periastron. Specifically, at any given observation epoch it would be \about 52$\%$ more likely to be located in the bottom half of Fig. \ref{fig:offsetb} than in the top half. If the planet's orbit is more eccentric than the ring, it would be even more likely to be located near apoastron. Unfortunately, a large part of this area is inaccessible in our images (Fig. \ref{fig:images}). This means it may take several more years before the hypothetical planet is at a more favorable projected separation to be directly imaged.

\subsubsection{Streamers?}
\label{sec:streamers}


Our VisAO images (Fig. \ref{fig:images}) clearly show that excess flux resides beyond the ansae of the disk. These features are detected at SNRE \about 5-10. Their prominence appears to decrease with increasing wavelength; they are brightest at $i'$, fainter at $Ys$ and $z'$, marginally detected at $Ks$ band, and not detected at 3-4 \microns. Clearly, the multiple detections of the features over multiple nights at different wavelengths indicates they are not spurious artifacts. Indeed they have been previously detected with several different telescopes/instruments \citep{thalmannhr4796,wahhaj4796nici,perrinGPI4796}. As originally posited in \cite{thalmannhr4796}, these features likely arise from the ADI processing of a faint halo of eccentric, small grain dust, which is expected in disks around hot stars \citep{strubbe}. We therefore suggest the features be henceforth referred to as the ``halo traces." This term effectively conveys that the physical source is a halo, but we are only seeing ``traces" of it due to ADI processing. 

\subsubsection{Inner region: an inner ring?}
\begin{figure*}[t]
\centering
\subfloat[]{\label{fig:inner}\includegraphics[width=0.98\textwidth]{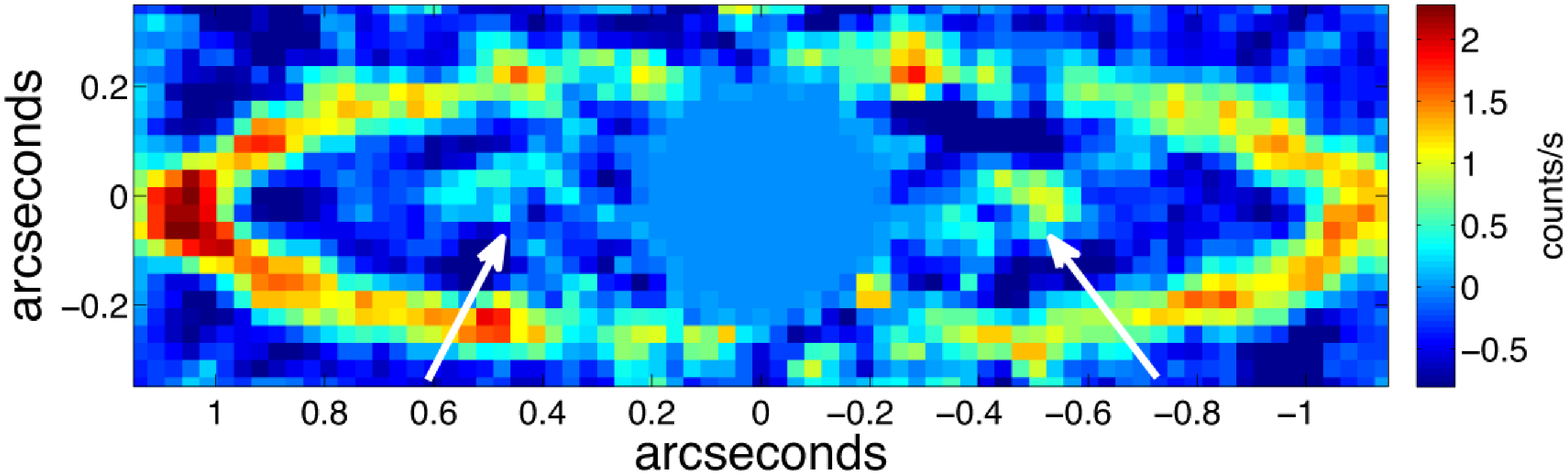}} \\
\subfloat[]{\label{fig:real}\includegraphics[scale=0.45]{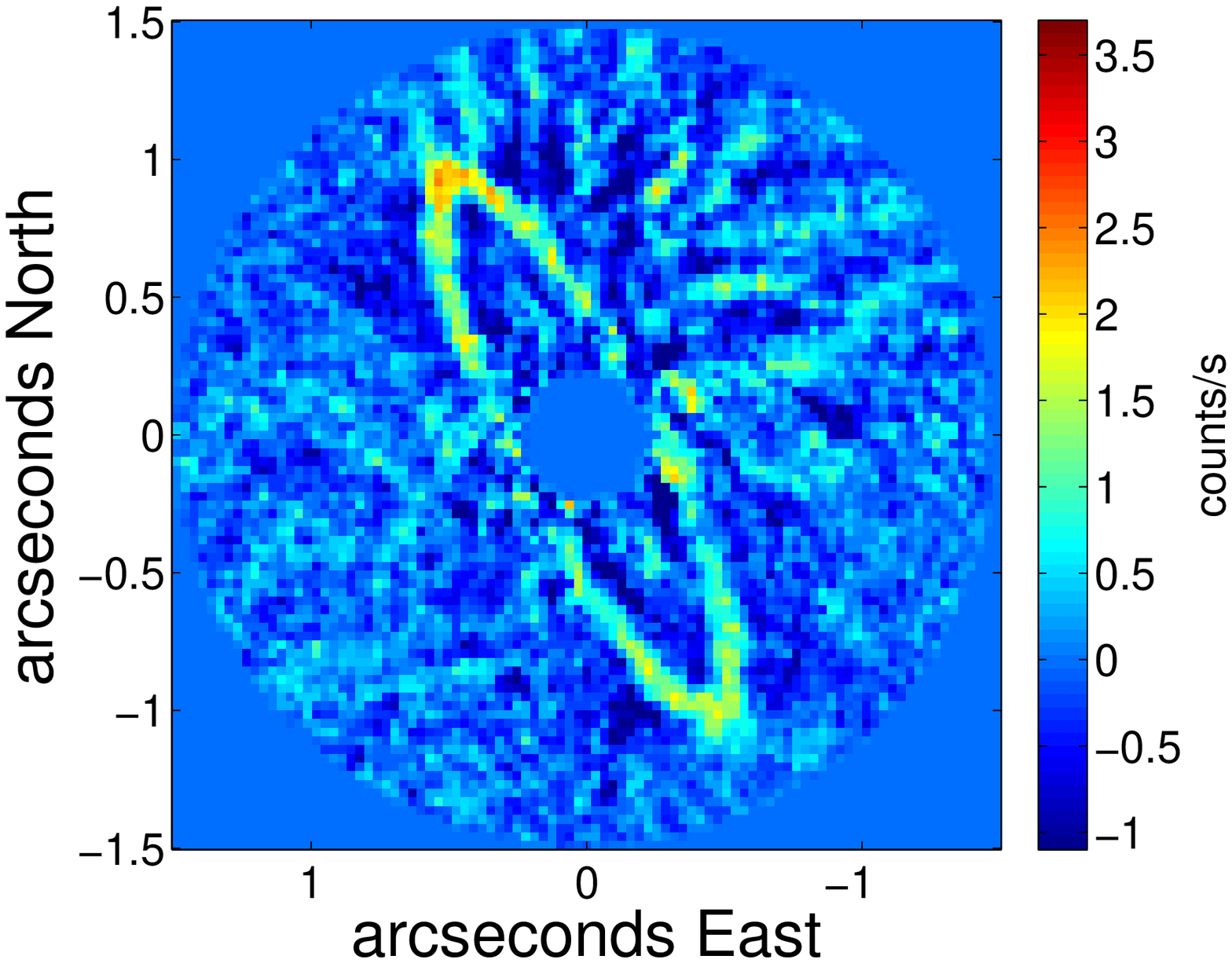}}
\subfloat[]{\label{fig:fake}\includegraphics[scale=0.45]{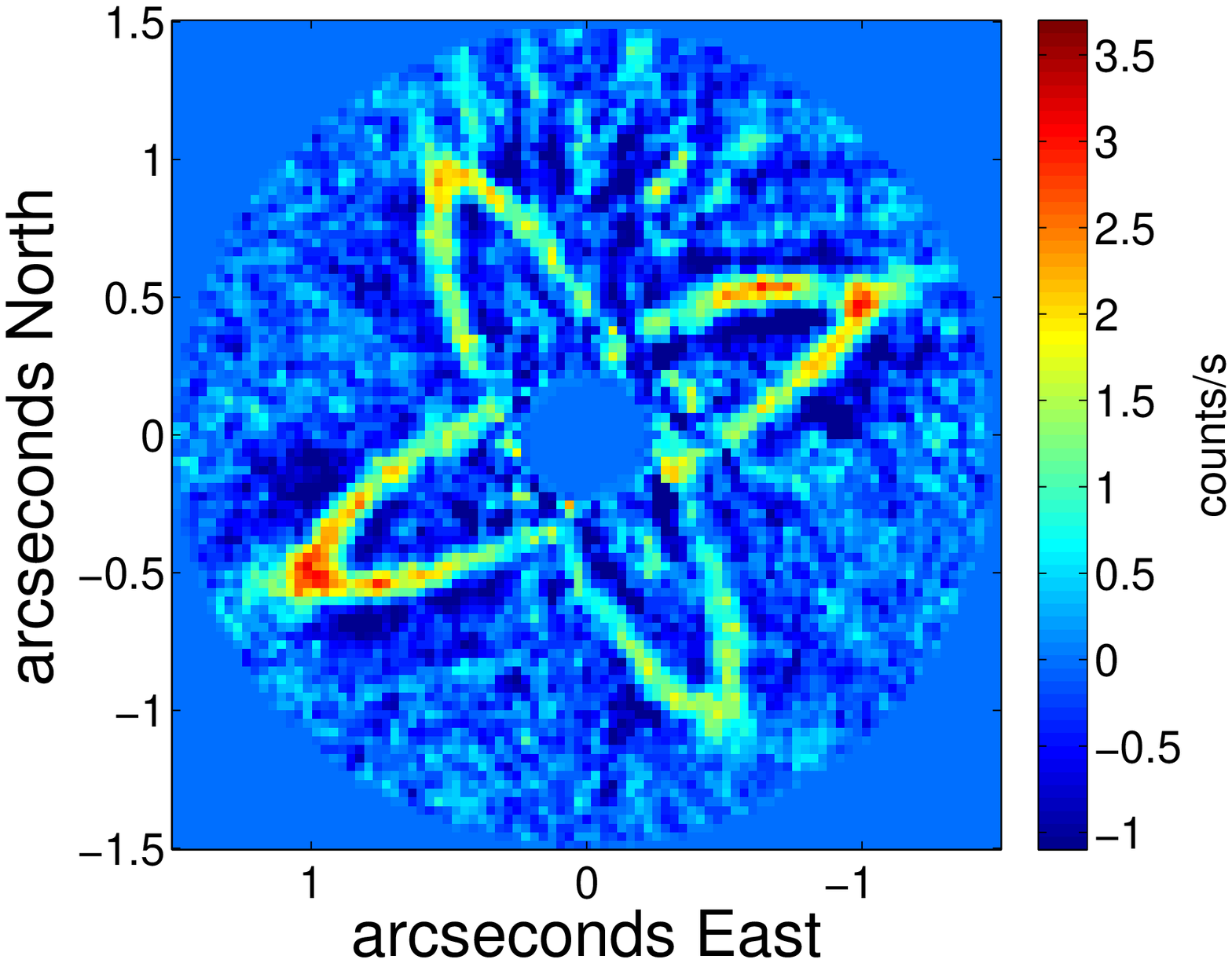}} 
\caption{(a): Binned$\times$2 image of the HR 4796A debris disk at $Ls$, rotated and zoomed-in to show the possible inner ring (denoted by the white arrows). (b): The same, but unrotated and zoomed out. (b): The same, with an artificial disk inserted and recovered 90\degrees ~away. No similar \textit{symmetric} structure is recovered inside the artificial ring, making the possibility that the residuals were ``carved out" by the ADI reduction process unlikely.}
\label{fig:innerall}
\end{figure*}

Reductions of the $Ls$ dataset consistently showed residual signal inside the ring consisting of two regions of excess flux (Fig. \ref{fig:innerall}). These features, at $<$ 0\fasec 5 and at a PA consistent with the ring major axis, are detected at marginal SNRE (\about 2-3) and thus are not assuredly real astronomical sources. To check whether the features were ``carved out" by the outer disk in ADI reduction process, we inserted and recovered an artificial disk 90\degrees ~away (Fig. \ref{fig:fake}). Residuals are present within the outer artificial disk, but there is no similar \textit{symmetric} structure as is seen in Fig. \ref{fig:real}. If real, the best explanation for the inner structure would be a ring-like inner disk located at \about 36 AU. As will be discussed in Section \ref{sec:scatphotometry}, the known/outer disk is brightest in scattered light at $Ls$, so it might make sense that the inner disk would be brightest here as well. Furthermore, the features lie along the same PA as the outer disk, which should be the case for an inner disk viewed from Earth around this star. However, it is troublesome that the Gemini Planet Imager (GPI) failed to detect any hint of these features at $Ks$ band \citep{perrinGPI4796}, though those observations were limited by $<$ 40\degrees ~of parallactic angle rotation. Furthermore, previous studies have predicted that an inner dust component should reside inside \about 10 AU \citep{hr4796augereau,wahhaj4796old,koerner4796}, which is much closer to the star than our observed features. On the other hand, an inner component so close to the star is not required to fully model the SED of the disk \citep{hr4796li}. Additional imaging at high Strehl ratio and very small inner working angles will be required to determine if the features in our images are spurious residuals or evidence for an inner ring. 

\subsection{Scattered light surface brightness}
\label{sec:scatphotometry}
We measured the surface brightness (SB) of the HR 4796A debris disk inside square apertures\footnote{We also tested circular and elliptical apertures but found that the choice of aperture shape mattered very little for the final photometry because we appropriately account for aperture size corrections (see the Appendix).} centered on the ring ansae. At these locations, the scattering phase angle is \about 90\degrees, and since the East-West asymmetry is small \citep{hr4796schneider,thalmannhr4796,hr4796augereau}, we can neglect any phase function corrections that would be necessary at other phase angles. Assuming the dust size and composition is uniform in azimuth, averaging the photometry at the ansae also allows us to decrease the overall uncertainty on the disk's photometry by a factor of $\sqrt{2}$. Photometry is reported for all wavelengths except $Ys$, $z'$, and $i'$ because we did not obtain any unsaturated images of the star in these filters.\footnote{VisAO images do contain a ghost several arcseconds away from the star that is several orders of magnitude fainter than the star. This could feasibly be used for photometry. We elected not to use the ghost for photometry because we already have similar wavelength data from HST.} We also computed photometry on archival HST/STIS and HST/NICMOS images of the disk originally reported in \cite{4796organics} at 0.5752 \microns ~(STIS 50CCD), 1.1 \microns ~(F110W), 1.6 \microns ~(F160W), 1.71 \microns ~(F171M), 1.8 \microns ~(F180M), 2.04 \microns ~(F204M), and 2.22 \microns ~(F222M).

For each wavelength, we divided each image by the respective plate scale squared (to obtain units of SB), then we placed apertures of a given size centered on the brightest pixel in each ansa and computed the median SB. The aperture size for the STIS and NICMOS images was 3 pixels on a side (0\fasec 1523 for STIS and 0\fasec 2262 for NICMOS), while for the Clio-2 images the aperture size was 7 pixels on a side (0\fasec 111). As is described in the Appendix, we ensured that the quantities measured in these apertures are appropriately corrected for the varying aperture sizes used. We also corrected for the varying PSFs of the different telescopes/instruments, and for the Clio-2 data we corrected for the disk's self-subtraction due to the ADI/PCA data reduction (see the Appendix for more details).

\begin{figure}[h]
\centering
\includegraphics[width=0.45\textwidth]{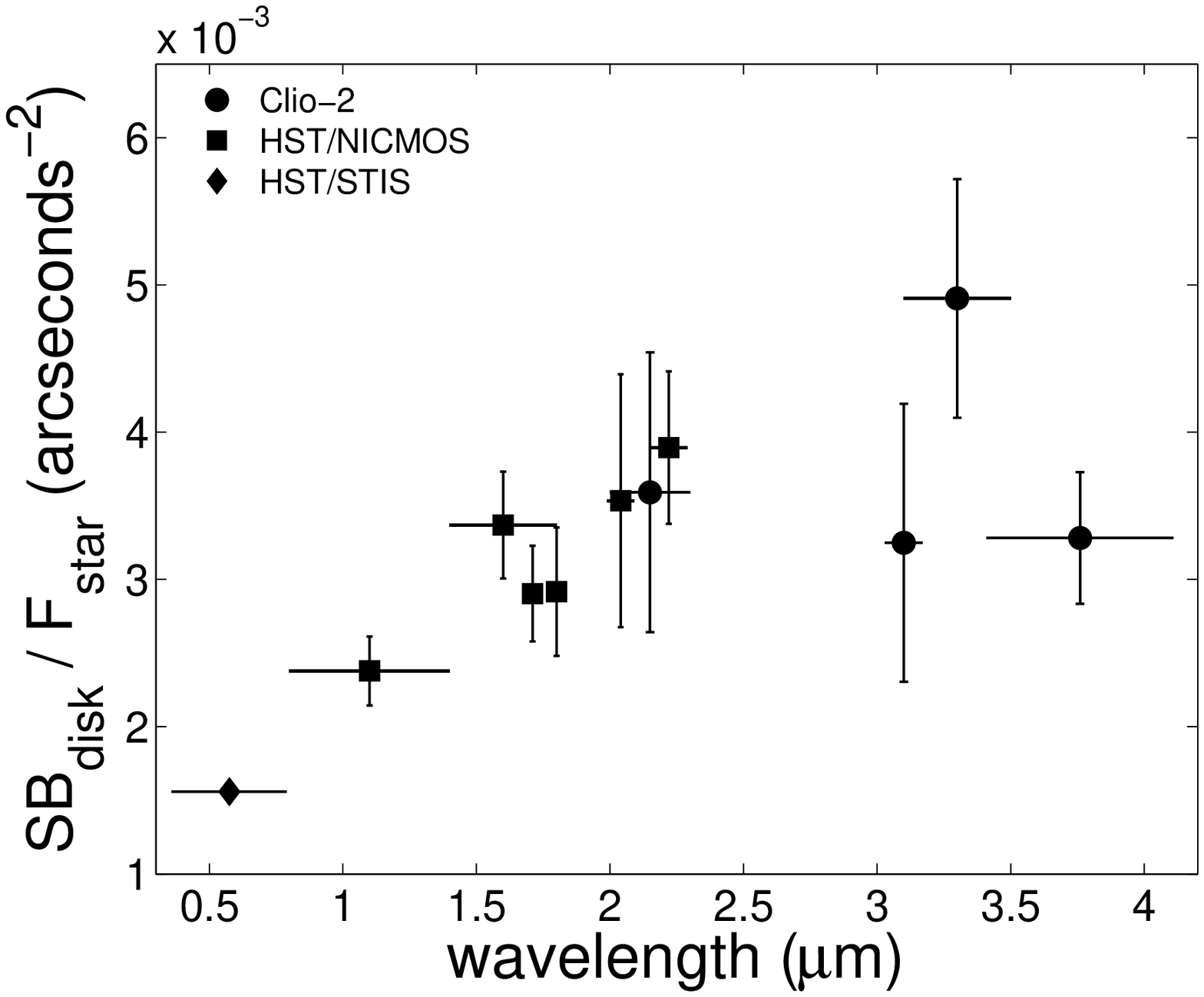}
\caption{Final calibrated SB of the HR 4796A debris ring, computed in square apertures centered on the ring ansae. The data have been normalized to the flux at 1.1 \microns ~and then averaged at the ansae to increase S/N. Horizontal bars indicate filter bandwidths. The dust is red from \about 0.5-1.5 \microns ~(as originally reported in \citealt{4796organics}) and gray/red longward of 1.5 \microns. The disk is brightest at 3.3 \microns.}
\label{fig:finalphot}
\end{figure}

Next we normalized each SB value by the star's total flux at each wavelength. This effectively removes the stellar color so the dust's scattering efficiency as a function of wavelength can be quantified. For the HST images, we used the STScI website look-up tables to determine the star's flux at each wavelength. For the Clio-2 data, we calculated the encircled flux as a function of distance from the star using the unsaturated photometric images of the star at each wavelength. We verified that the star's encircled flux leveled out beyond \about 1\fasec 5 and therefore took the sum inside a 1\fasec 5 radius aperture as the total flux. We then divided the SB measurements by these values. 

The errors for all values were computed as follows: the disk was masked out using a bar of comparable length and width to the real disk; at a given radius, the median flux was computed around the star in non-overlapping square apertures of equivalent size to those used on the real disk at each wavelength; the error was then taken as the standard deviation of these values. 

Figure \ref{fig:finalphot} shows our final corrected SB values of the HR 4796A debris ring. The disk is red in the optical-NIR, as originally reported in \cite{4796organics}, and gray/red longward of \about 2 \microns. The disk is brightest at 3.3 \microns.

\subsection{Limits on Planets}
Based on the imaging results from Clio-2 at 3-4 \microns, no point-sources were detected at high enough S/N to warrant consideration as possible sub-stellar companions. To assess the masses of planets that we could have detected, we undertook a Monte Carlo approach whereby we repeatedly inserted and recovered artificial planets with the following parameters: contrast, $9 \leq \Delta m_{L^{\prime}} \leq 12$; distance from the star, $0\fasec 35 \leq r \leq 1\fasec 45$; and position angle, 0\degrees $\leq \theta <$ 360\degrees. \cite{skemerhr8799_2} suggested that young hot planets might be more easily detected at $Ls$ rather than \lprime ~from the ground due to their observed SEDs at 3-5 \microns. However, we assessed our limits using our \lprime ~data because we did not obtain enough data at $Ls$. 

\begin{figure}[t]
\centering
\includegraphics[scale=0.45]{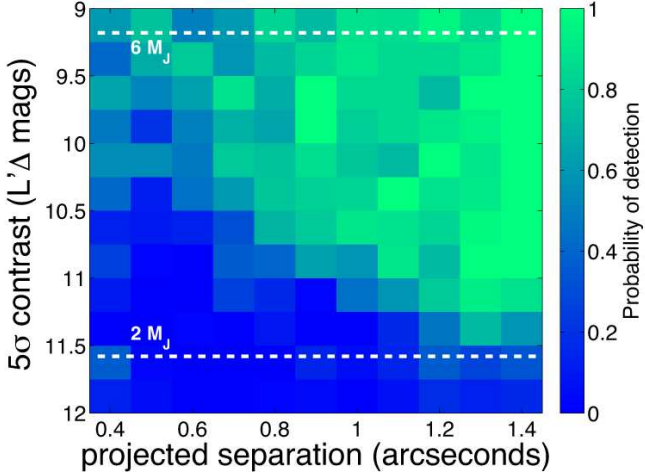}
\caption{Contrast map showing the probabilities of detecting planets of given contrasts and distances from the star. With \about 75$\%$ confidence no planets with masses $\gtrsim$ 6 \mj ~currently reside at projected separations beyond 0\fasec 4, and with $\gtrsim 90\%$ confidence no planets with masses $\gtrsim$ 3 \mj ~currently reside at projected separations beyond 1\asec. If any high-mass planets similar to those orbiting HR 8799 reside in this system, their current projected separations must be very small.}
\label{fig:planets}
\end{figure}

Artificial planets were attenuated replicas of the unsaturated PSF at \lprime. A given planet's properties were randomly chosen based on the limits above, and then the planet was inserted into the raw data. The data were reduced using the same parameters (including number of PCA modes) used to produce the highest S/N image of the disk (Fig. \ref{fig:L}). The S/N of the planet was then computed (from the SNRE map of the reduced image) and considered a successful detection if S/N $\geq$ 5. We excluded from the results any planets whose projected positions overlapped with the debris ring itself.

The insertion and recovery process was then repeated 5000 times. We determined our contrast limits by calculating the fraction of successful planet detections (S/N $\geq$ 5) in bins of size 0\fasec 1 in distance and 0.25 mags in contrast. The probability map is shown in Fig. \ref{fig:planets}. Also plotted are the expected contrasts of 2 \mj ~and 6 \mj ~planets from the COND mass-luminosity atmospheric models \citep{baraffe} for hot-start planets, which should be appropriate to use here given the young age of the host star (\about 10 Myr).\footnote{We acknowledge that our limits for ``warm-start" planets \citep{spiegel12} would likely be higher.}

Based on our contrast limits, with \about 75$\%$ confidence no planets with masses $\gtrsim$ 6 \mj ~currently reside at projected separations beyond 0\fasec 4, and with $\gtrsim 90\%$ confidence no planets with masses $\gtrsim$ 3 \mj ~currently reside at projected separations beyond 1\asec. These limits are comparable to those reported in \cite{lagrange4796}, though in that study artificial planets were only inserted along the disk major axis.

\section{Modeling}
\label{sec:modeling}
\subsection{Model Setup}
The goals of our modeling are: to more rigorously examine the relevant parameter space than has been done in previous works; to characterize generally good- and poor-fitting compositional families; and to characterize potential degeneracies involved in the fitting process. 

We modeled the scattered light and thermal emission of the dust using \textit{dustmap} \citep{dustmap}. For the scattered light, \textit{dustmap} uses Mie theory to calculate the scattering efficiencies as a function of wavelength and grain size; for the thermal emission, it calculates the absorption efficiencies and grain temperatures by balancing input and output radiation. 

\begin{table*}[t]
\centering
\caption{Root Compositions and Families}
\begin{tabular}{c c}
\hline
\hline
Family & Root Composition \\
\hline
Amorphous Olivine & astrosilicates$^{1}$, amorphous olivine$^{2}$, olivine (iron-poor)$^{3}$, \\
 & olivine (nominal iron)$^{3}$, olivine (iron-rich)$^{3}$ \\ \\
Crystalline Olivine & crystalline olivine$^{1}$ \\ \\
Pyroxene & orthopyroxene (iron-poor)$^{3}$, orthopyroxene (nominal iron)$^{3}$, \\
 & orthopyroxene (iron-rich)$^{3}$, pyroxene$^{4}$ \\ \\
Organics & amorphous carbon$^{5}$, organics (Henning)$^{3}$, organics (Li)$^{2}$ \\ \\
Tholins & Titan tholins$^{6}$ \\ \\
Water ice & water ice (Henning)$^{3}$, water ice (Li)$^{7}$, water ice (Warren)$^{8}$ \\ \\
Iron & iron$^{3}$ \\ \\
Troilite & troilite$^{3}$ \\
\hline
\end{tabular} 
\\
\raggedright
\textbf{REFERENCES:} (1) \cite{lidraine2001}; (2) \cite{ligreenberg97}; (3) Thomas Henning ($http://www.mpia-hd.mpg.de/homes/henning/Dust_opacities/Opacities/RI/new_ri.html$); (4) Aigen Li, priv. comm.; (5) \cite{zubko96}; (6) \cite{khare84}; (7) \cite{ligreenberg98betapic}; (8) \cite{warren84} 
\label{tab:compositions}
\end{table*} 

We distributed 5 million discrete points over an infinitely thin (zero scale height\footnote{Assuming zero scale height is justified for our purposes given that scale heights of debris disks are thought to be small, on the order of a \about few percent.} circular ring using the following radial number density expression:
\begin{equation}
n(r) = n_{0} \sqrt{2} \left(\left(\frac{r}{79.2 {\rm AU}}\right)^{-2\alpha_{out}} + \left(\frac{r}{79.2 {\rm AU}}\right)^{-2\alpha_{in}} \right)^{-1/2},\
\end{equation}
where $n_{0}$ is the number density at the peak semimajor axis assumed here to be 79.2 AU, and $\alpha_{out} = -6$ and $\alpha_{in} = 19.6$ (see Appendix for more details on the choice of these power-laws). We assumed an inclination angle of 77.4\degrees\footnote{This was chosen before the geometrical modeling described in Section \ref{sec:geometry} was performed. Because the real disk is likely to be optically thin, and our model disk has zero scale height, a difference of \about 1\degrees ~from the nominal value should not significantly affect our modeling results.} for the model disk and a distance from Earth = 72.8 pc. For the star, we fit its visible photometry to Kurucz stellar models \citep{newkurucz}, with the best fits being: $T_{eff} = 9250$K, $L/L_{\odot} = 22.2$, and $\log{g} = 4.5$. We then synthesized model images using the bandpasses and resolutions of STIS, NICMOS, and Clio-2. We produced individual images for 50 values of grain size, logarithmically distributed from 0.1 to 1000 \microns. For each grain size, we calculated the photometry at the wavelengths listed in Table \ref{tab:unresolved}. Thermal data with bandwidths $>20\%$ were modeled by resolving the bandpasses into three individual wavelengths. For the thermal emission data, we calculated the stellar and total disk flux within 10\arcseconds of the star, with the model disk extending out to 2\asec. For the scattered light data, we computed the aperture photometry values of the model images using the aperture sizes and locations described in Section \ref{sec:scatphotometry}. The model images were not convolved with PSFs because we account for PSF convolution when computing the real disk photometry (see the Appendix for more details). 

Models of debris disk compositions commonly treat scattered and emitted light separately, and observed albedos are often difficult to reproduce (e.g., \citealt{hd181327ice,kristhd207129}). This may be in part due to the poorly constrained scattering phase function of dust, which can act effectively as an unknown modification to the true albedo \citep{stark181327}. For our model fits, we included solutions that self-consistently treat scattered and emitted light using the scattering phase function calculated by Mie theory, which can deviate significantly from the commonly used Henyey-Greenstein scattering phase function \citep{stark181327}. 

We examined a total of 8,426 compositions, calculated from mixtures of 19 unique ``root" compositions and vacuum (acting as porosity). The 19 root compositions are listed in Table \ref{tab:compositions}. These compositions are representative of the constituents of comets, asteroids, and micrometeorites in the solar system (e.g., \citealt{hstmeteorimpacts,stardustwaterice,cometdustcarbon}). Our final library of compositions included these 19 root compositions, 361 porous mixtures of the root compositions and vacuum, 1,341 two-component mixtures, and 6,705 porous two-component mixtures. Two-component mixtures were only ever made from different ``families" as defined in Table \ref{tab:compositions} (i.e., compositions from the same family were never mixed). To create mixtures of compositions we used Bruggeman effective medium theory\footnote{Other works have used \cite{emt} to construct the mixtures. We chose \cite{bruggeman} because it uses a series implementation and is therefore generally faster to implement.} (\citealt{bruggeman}). For the porous single-composition mixtures, we examined porosities ranging from 5-95$\%$ in steps of $5\%$. For the 2-component mixtures, we examined volumetric mixing ratios ranging from 10-90$\%$ in steps of 10$\%$. For the porous 2-component mixtures, we blended each 2-component mixture with vacuum, examining porosities ranging from 10-90$\%$ in 20$\%$ steps. For each combined mixture, we calculated the best fit grain size distribution, assuming a simple power law scaling, and minimum and maximum grain sizes. We then calculated the reduced chi square values ($\chi_{\nu}^{2}$) when fit to the scattered light data alone, the thermal emission data alone, and both the scattered light and thermal emission simultaneously.

Previous works have modeled dust grain mixtures containing more than two unique root compositions (e.g., \citealt{donaldson32297,milli4796,hd181327ice}), which they accomplish by limiting the range of other relevant fitting parameters. For example, a Dohnanyi collisional cascade \cite{dohnanyi} or slight variants \citep{andras} is often assumed for the size distribution, which significantly reduces the computational workload. But combinations of certain parameters can produce very similar fits (e.g., a red slope at visible wavelengths; \citealt{4796noorganics}). Furthermore, many assumptions go into producing complicated mixtures of dust grains, and most of these still need to be physically tested (rather than just modeled). Therefore we chose to explore the relevant parameter space with as few preferences a priori as possible. Specifically, we allowed the size distribution power-law exponent to vary between -2 and -5 in steps of 0.125, and we allowed the maximum grain size to be between 100-1000 \microns ~in logarithmic steps of 0.0204. We did not use the theoretically-motivated blow-out size for the minimum grain size because the true blow-out size strongly depends on the dust's density and scattering properties (see Fig. 12 in \citealt{hd181327ice}). Therefore we set the minimum allowable grain size to 0.1 \microns. 

\subsection{Model Results}
\label{sec:modelresults}
\subsubsection{Scattered light only}
\label{sec:scattered}
\begin{figure*}[t]
\centering
\subfloat[]{\label{fig:scatteredmodels}\includegraphics[width=0.48\textwidth]{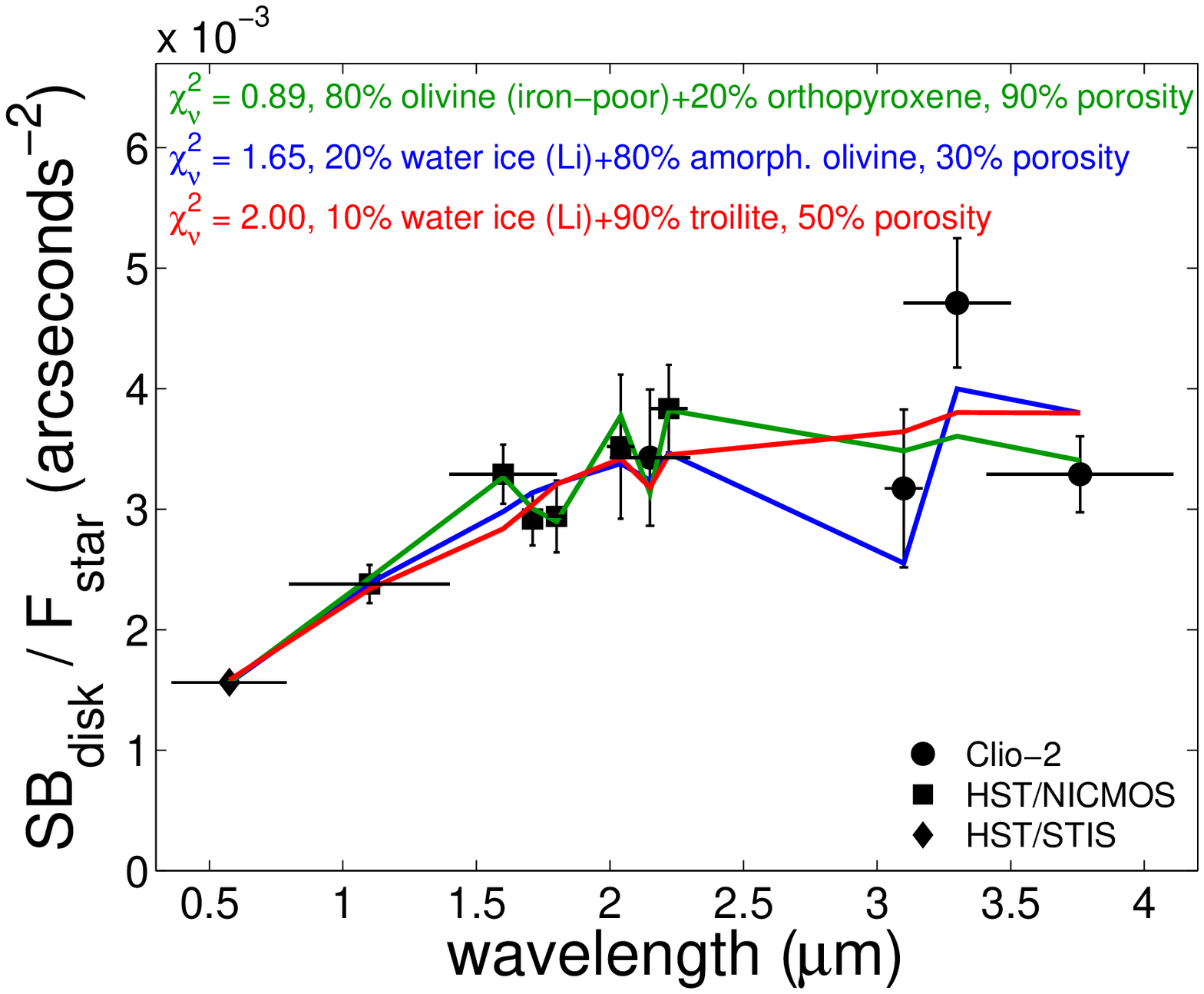}} 
\subfloat[]{\label{fig:scatteredmodelsthermal}\includegraphics[width=0.48\textwidth]{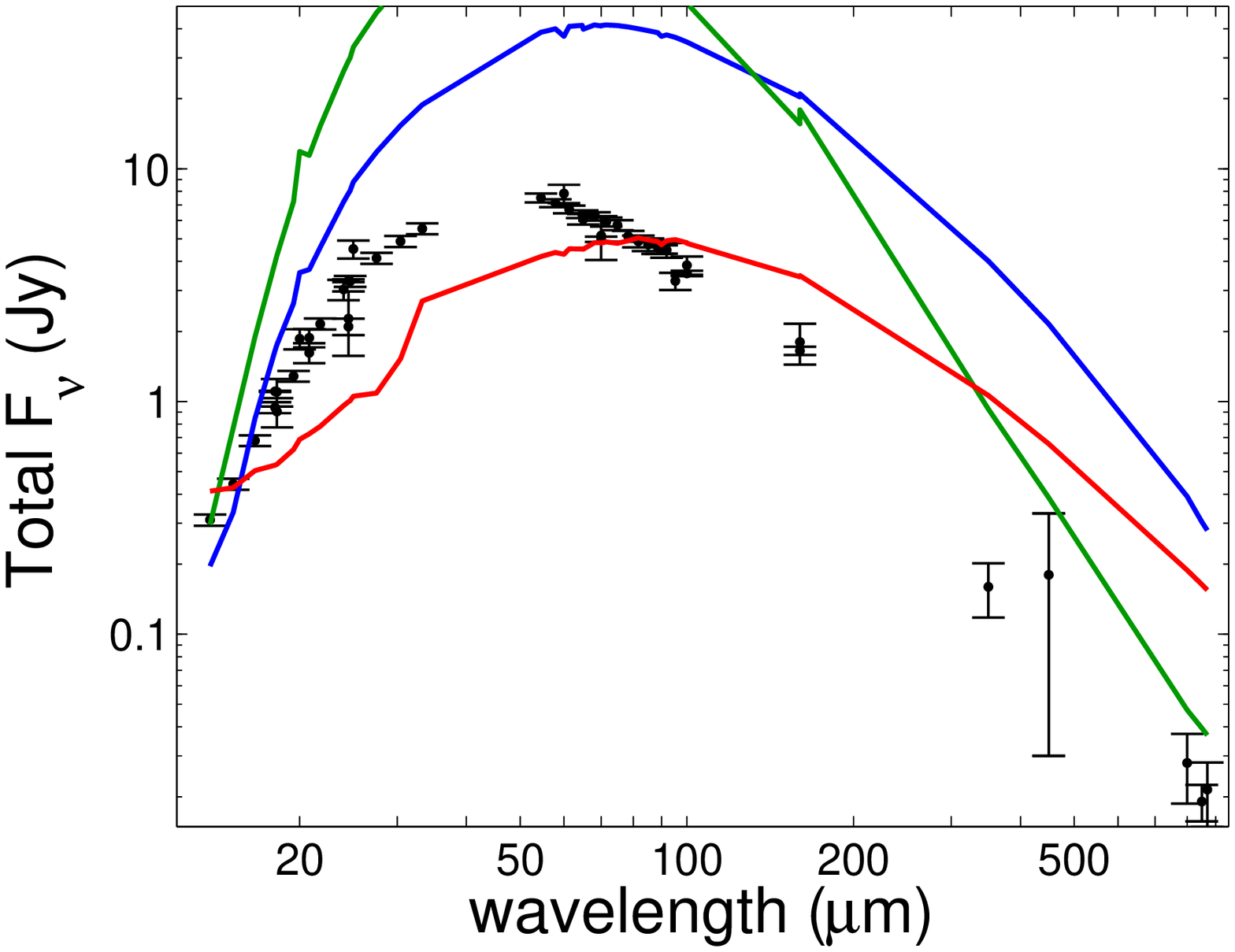}} \\
\subfloat[]{\label{fig:thermalmodelscattered}\includegraphics[width=0.48\textwidth]{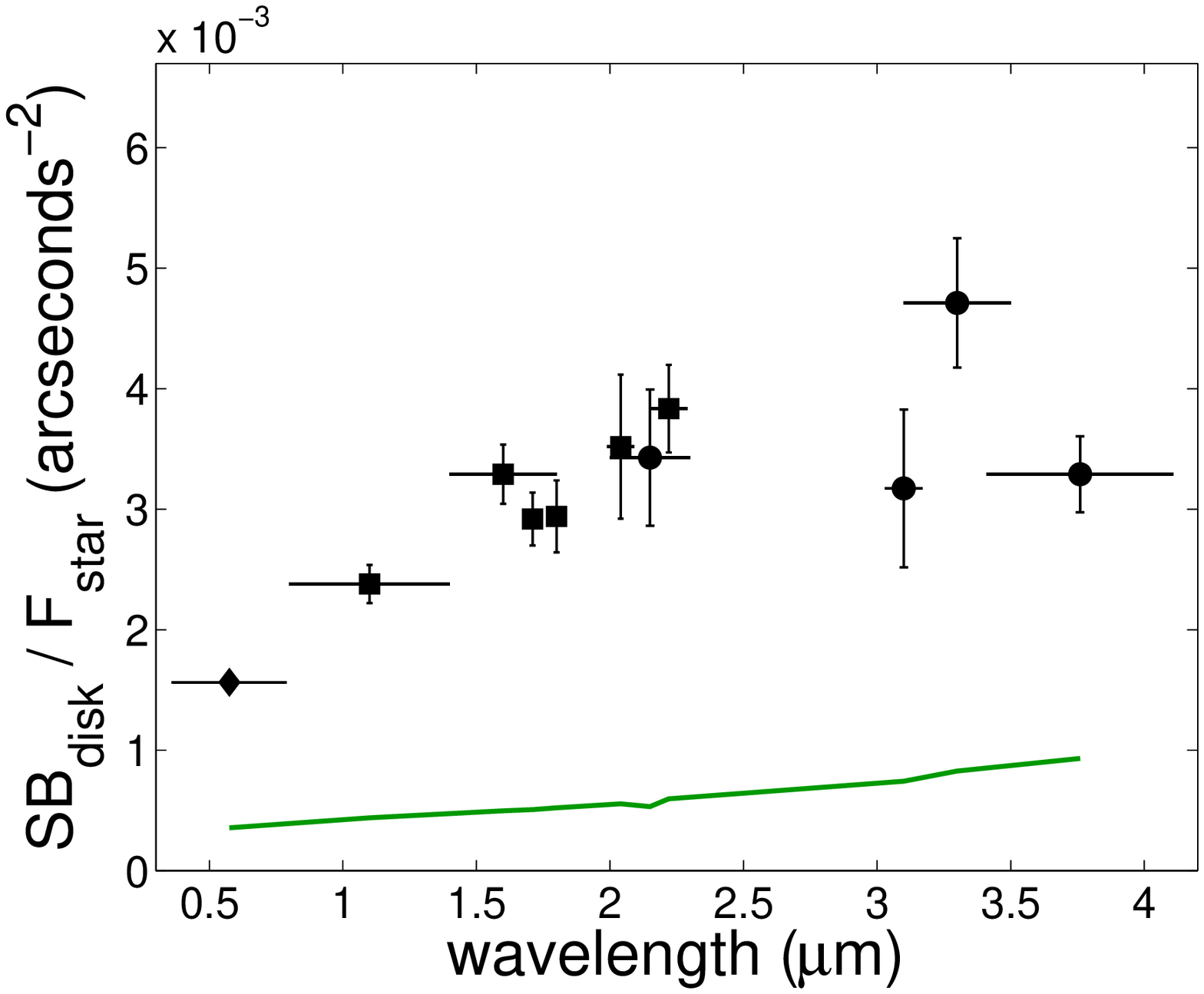}} 
\subfloat[]{\label{fig:thermalmodel}\includegraphics[width=0.48\textwidth]{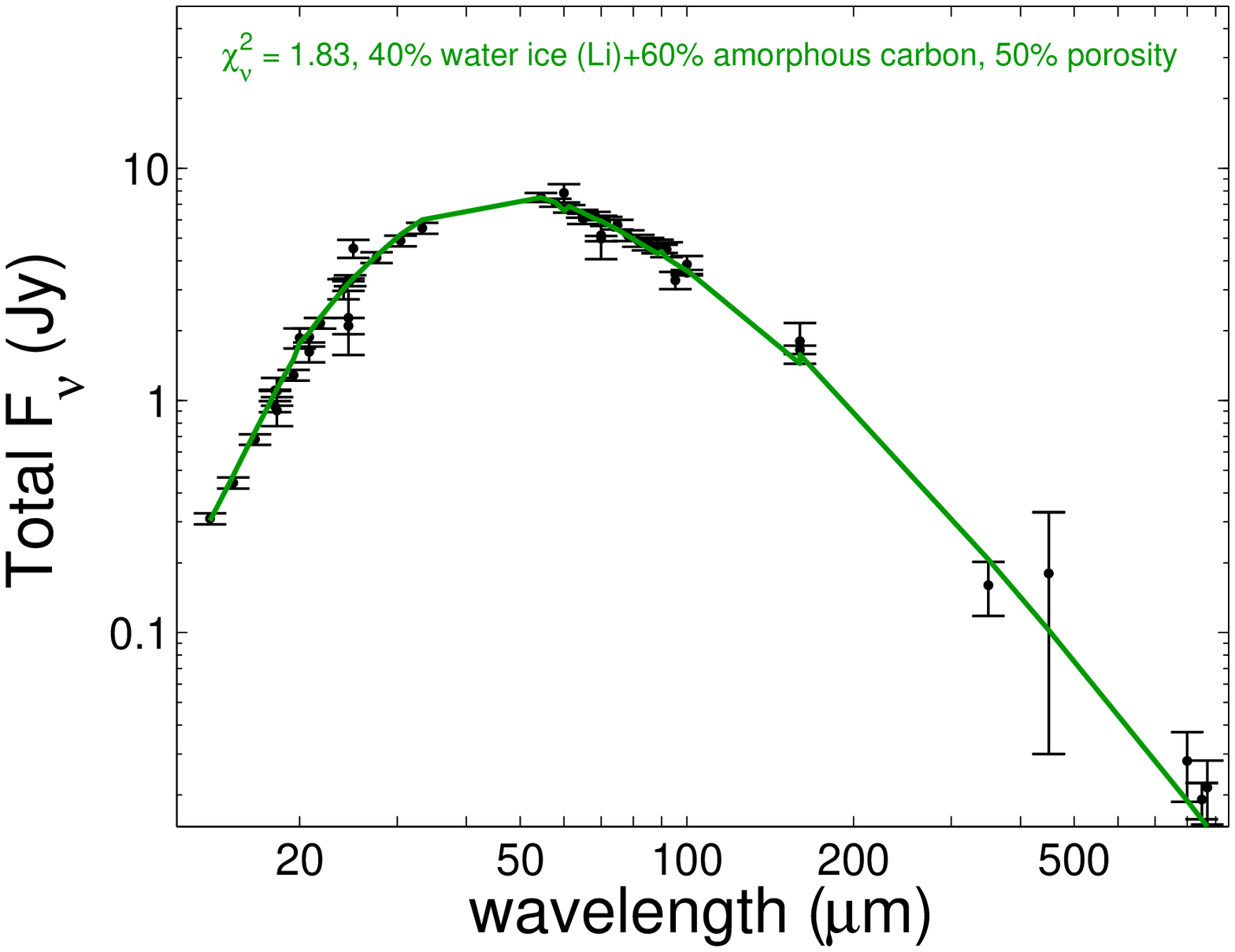}} \\
\subfloat[]{\label{fig:allmodelscattered}\includegraphics[width=0.48\textwidth]{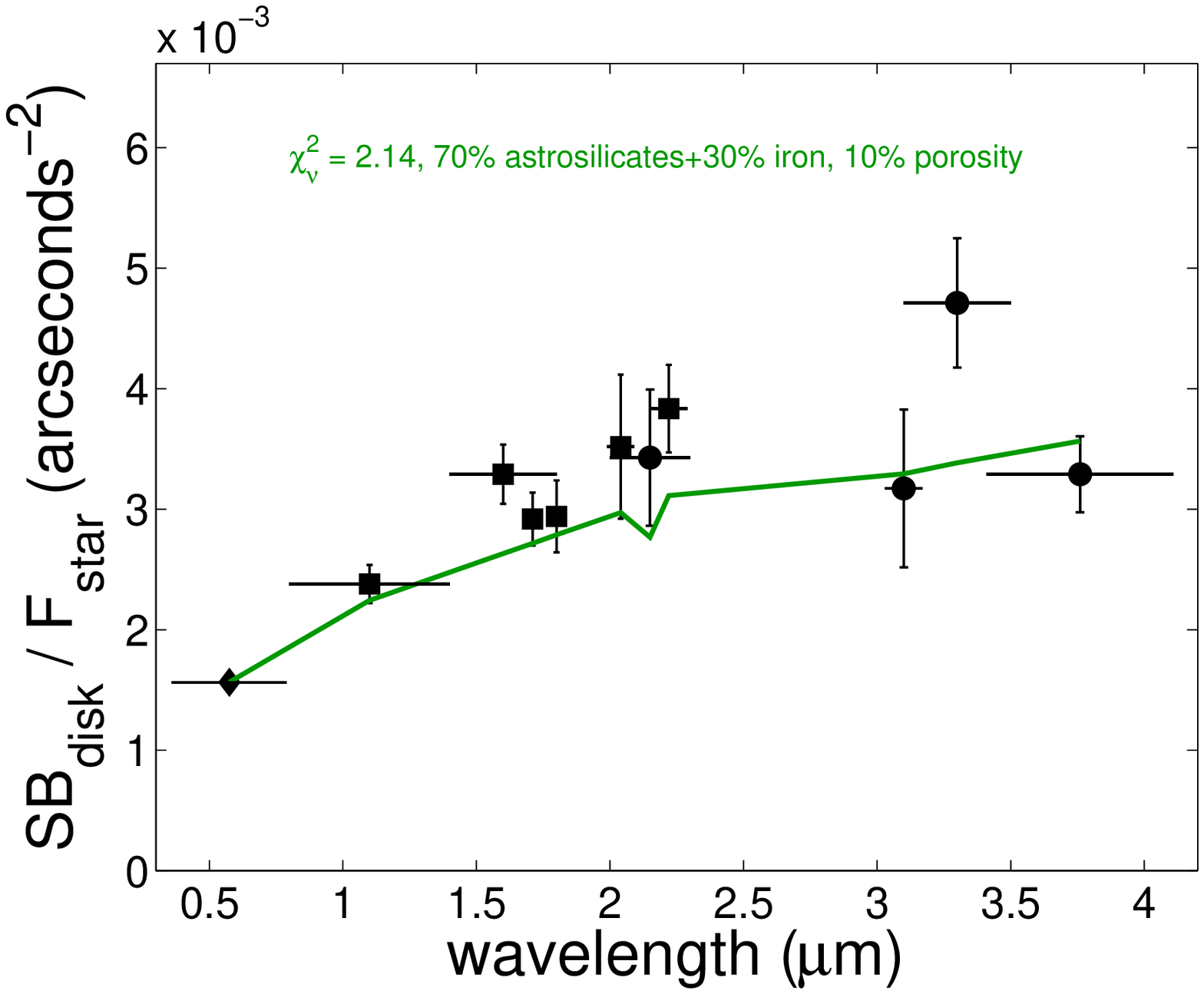}} 
\subfloat[]{\label{fig:allmodelthermal}\includegraphics[width=0.48\textwidth]{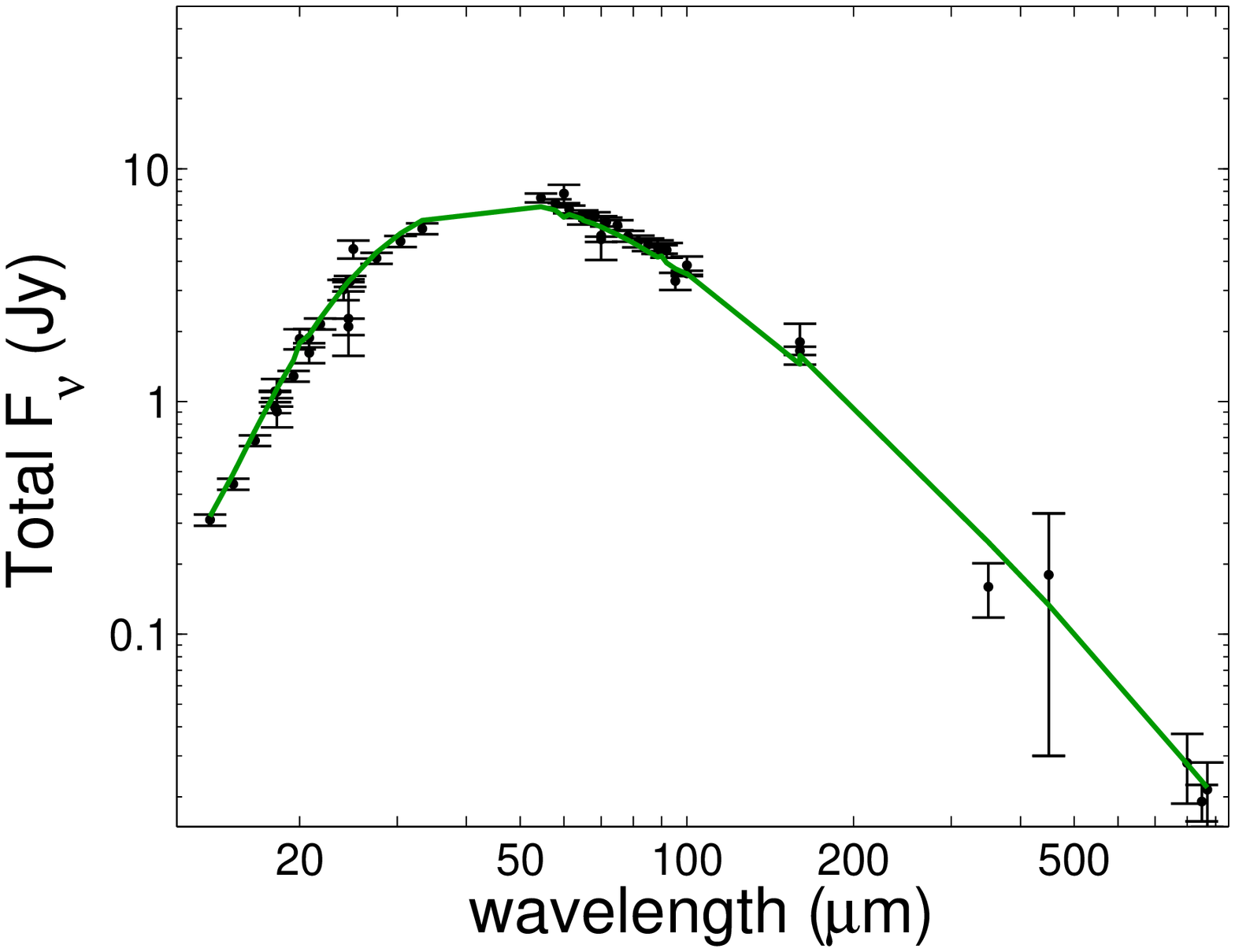}} \\
\caption{(a-b): Three ``good" model fits to the scattered light data alone compared to both the scattered light and thermal emission of the disk (thermal data bandwidths are not plotted for visual clarity). The best fit requires porous silicates (porous olivine and orthopyroxene), but good fits are also achieved with mixtures of water ice and carbon/troilite. However, these three models are poor fits to the thermal data, demonstrating the importance of the phase function and dust albedo for matching scattered light data. (c-d): Analogous to (a-b) but for the best fit to the thermal data alone. When compared to the scattered light, this model (porous water ice and amorphous carbon) is a poor fit. (e-f): The best fit to both the scattered light and thermal data compared to the both. This model (slightly porous silicates and iron) is a mediocre overall fit to both sets of data ($\chi_{\nu}^{2}$ = 2.14).}
\label{fig:scatteredmodelsfig}
\end{figure*}

\begin{figure*}[t]
\centering
\subfloat[]{\label{fig:firstchis}\includegraphics[width=0.98\textwidth]{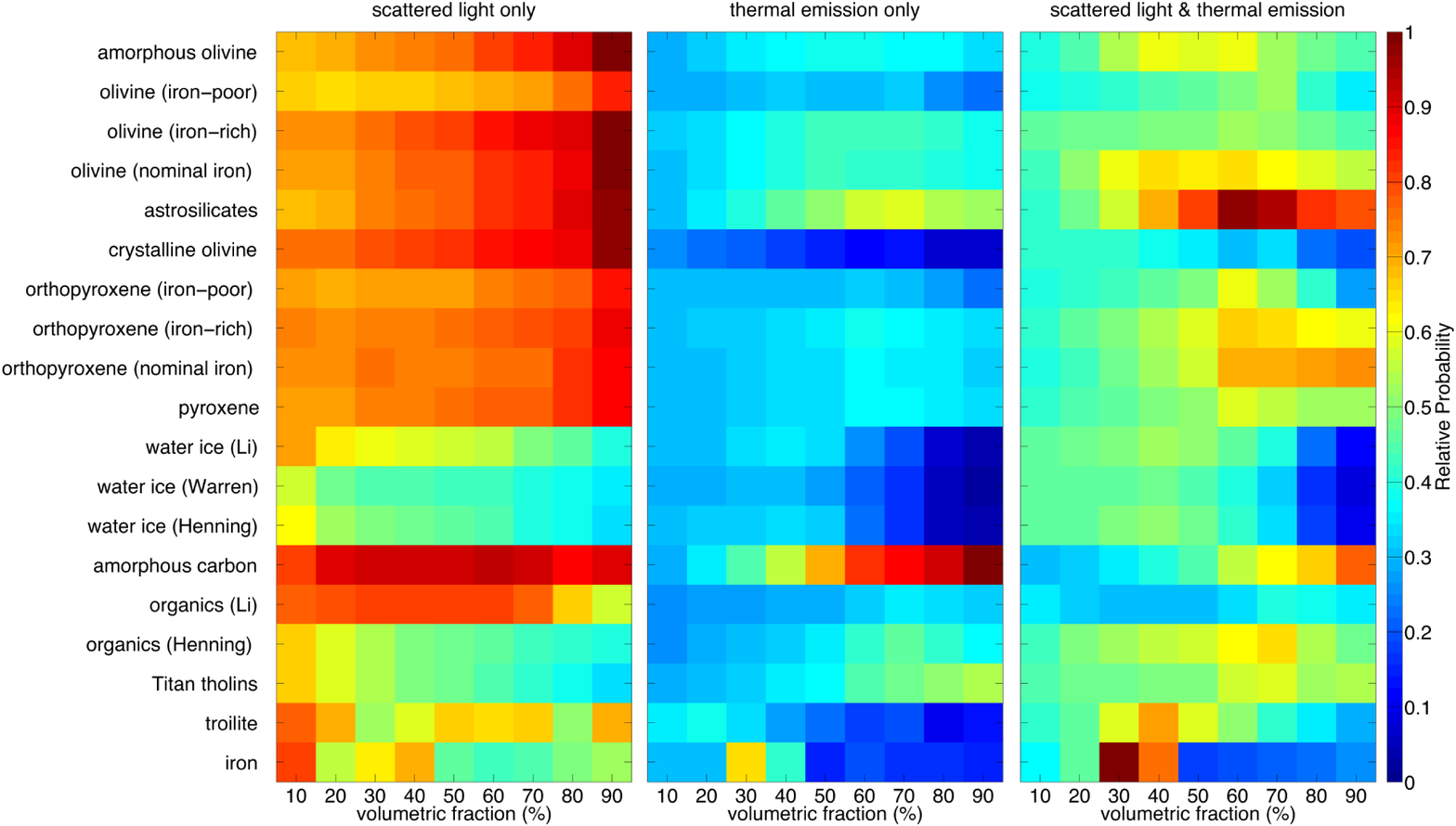}} \\
\subfloat[]{\label{fig:secondchis}\includegraphics[width=0.98\textwidth]{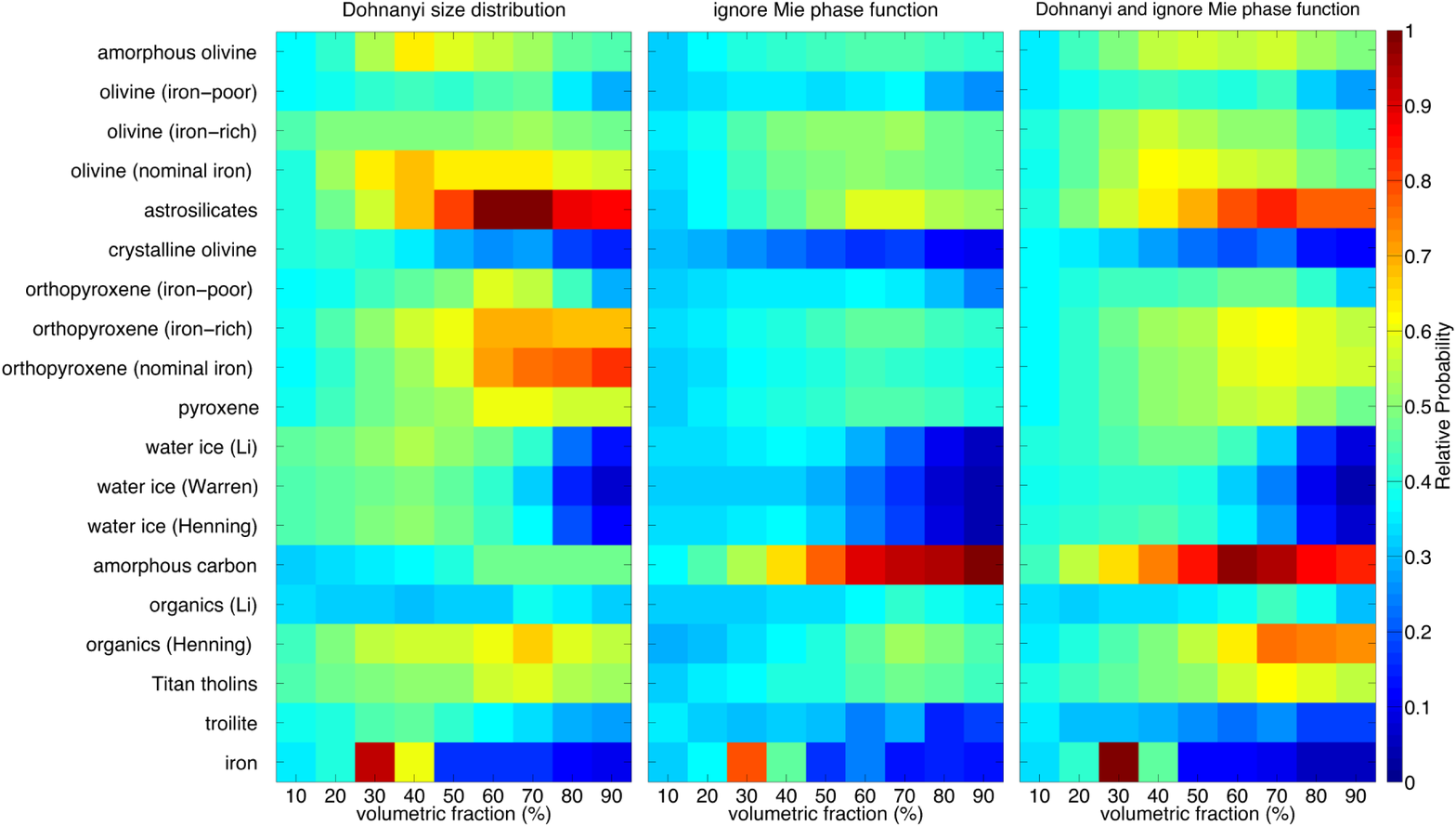}} 
\caption{Relative weighted probability that a root composition at a given volumetric fraction was used in good fits to the data. In other words, compositions with probabilities close to 1 were used more often in good fits than compositions with probabilities close to 0. (a): Preferred compositions for fits to the scattered light data only (left panel), thermal emission data only (middle panel), and scattered light and thermal emission together (right panel). (b): Fits to both the scattered light and thermal emission together while forcing a Dohnanyi size distribution (left panel), while ignoring the Mie phase function and relaxing the Dohnanyi constraint (middle panel), and while forcing a Dohnanyi size distribution and ignoring the Mie phase function (right panel). In general, silicates (olivine, pyroxene) and organics (especially amorphous carbon) are more frequently preferred than the other root compositions, and at higher fractional abundances.}
\label{fig:chis}
\end{figure*}

We first considered compositional model fits to the scattered light data alone. These data come from HST/STIS, HST/NICMOS, and MagAO/Clio-2, and are shown in Fig. \ref{fig:finalphot}. We considered only the porous 2-component mixtures, since these resulted in predominantly better fits to the data than the simpler models. The lowest reduced chi-square value achieved was 0.89, indicating a very good fit (with perhaps overestimated errorbars). However, there were more than 1000 fits with reduced chi-square values $\leq$ 2. This means that there are many reasonably good fits to the available scattered light data (see Fig. \ref{fig:scatteredmodels}). Therefore to assess which root compositions were most often favored in the fits, we weighted each root composition at a given volumetric fraction in all 6705 fits by the inverse of the corresponding reduced chi-square values, 1/$\chi_{\nu}^2$, then computed the weighted average probabilities and normalized all of these values by the highest single probability. For example, in Fig. \ref{fig:chis}, compositions with values close to 1 were favored more often in good fits than compositions with values close to 0.

From Fig. \ref{fig:firstchis}, we can immediately distinguish the compositions that were \textit{least} favored: water ice, organics (Henning), tholins, iron, and troilite. The compositions that were most often favored are the olivines and pyroxenes (i.e., silicates), amorphous carbon, and organics (Li). The discrepancy between Henning and Li's organics likely arises from their differing optical constants.  


\begin{figure}[h!]
\centering
\includegraphics[scale=0.45]{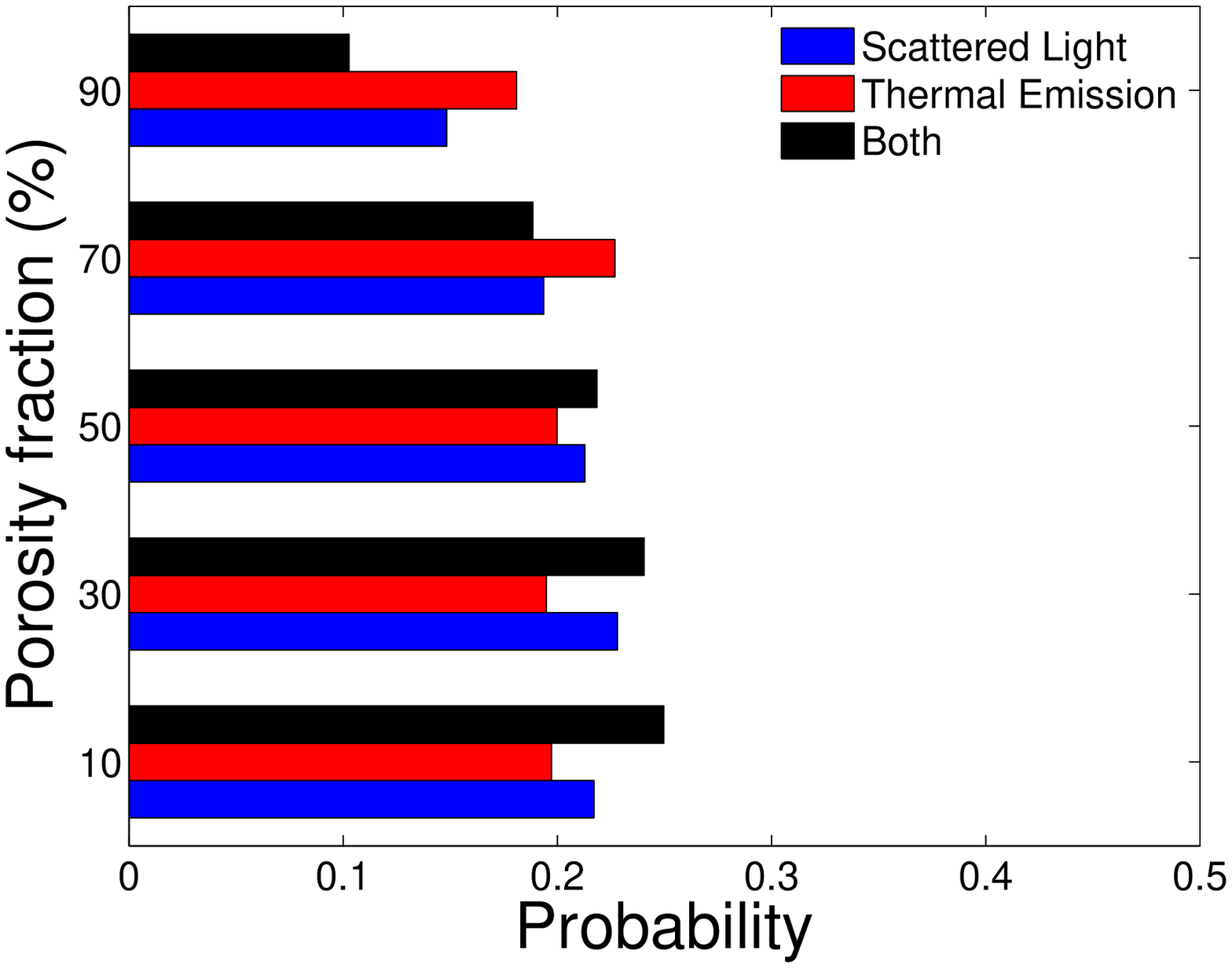}
\caption{The relative probabilities that a given porosity fraction was preferred in fits to the data. For the scattered light fits, there was a slight preference for porosity = 30$\%$. For the thermal emission fits, the most preferred porosity level was 70$\%$. For fits to all the data, low porosities were more preferred than higher porosities.}
\label{fig:allporosities}
\end{figure}

Regarding porosity, we found that there was a slight preference for porosity = 30$\%$, but this was only marginally favored over the others (Fig. \ref{fig:allporosities}). The size distribution power-laws for each family were generally shallow, varying slightly around -3, and the minimum and maximum grain size were \about 2 \microns ~and 500 \microns, respectively. The absence of very small grains and the shallow power law suggest that large grains are required to reproduce the red slope of the scattered light photometry. This is comparable to the results of \cite{4796noorganics} and \cite{hr4796li}. 

Before proceeding to fitting the disk's thermal emission alone, we examined how well the good fits to the scattered light data reproduced the disk's thermal emission. Fig. \ref{fig:scatteredmodelsthermal} shows that three good fits to the scattered light data shown in Fig. \ref{fig:scatteredmodels} are poor fits to the thermal data. This suggests that caution should be exercised when interpreting models of a disk's scattered light data alone. 

\subsubsection{Thermal emission only}
\label{sec:thermal}
We next fit the disk's thermal emission alone\footnote{We excluded five thermal emission data points when calculating the weights due to their being multiple $\sigma$ discrepant from neighboring data. These data points are marked by an $*$ in Table \ref{tab:unresolved}.}. We excluded emission at $\lambda <$ 13 \microns ~to ensure the models were testing predominantly the outer (known) disk alone. We found that the best fit composition had a reduced chi-square of 1.83, indicating a marginally good fit. Whereas there were \about 1000 fits with reduced chi-square values $< 2$ for the scattered light data, there were only 8 such fits for the thermal data. Nonetheless, it is still instructive to gauge which root compositions were the most frequently preferred. Therefore we calculated the relative probabilities for each root composition's fractional abundance in all 6705 porous 2-component fits, as before for the scattered light case.

Fig. \ref{fig:firstchis} shows that amorphous carbon and astrosilicates were the most frequently preferred root compositions. The least preferred were the water ices, iron, troilite, and crystalline olivine. It is not surprising that crystalline olivine was not preferred because its optical constants have very narrow features in the thermal infrared. To detect such features, high-resolution spectra are required (e.g., \cite{betapicolivine}).

The preferred fractional porosity was 70$\%$ (Fig. \ref{fig:allporosities}), marginally favored over the others. The average power-law for the thermal fits was \about -3.4, with the best-fitting models using \about -4. The minimum and maximum grain sizes were \about 3.5 \microns ~and 600 \microns, similar to the large grains preferred in fits to the scattered light data alone.

As was the case for the scattered light models, the best fitting thermal emission model failed to reproduce the scattered light data (Fig. \ref{fig:thermalmodel} and \ref{fig:thermalmodelscattered}). This once again shows that when fitting just the scattered light alone or just the thermal data alone, the preferred model fails to reproduce both the datasets simultaneously. 

\subsubsection{Scattered light and thermal emission together}
\label{sec:scatteredplusthermal}
Can any compositional mixture reproduce the entire SED (both scattered light and thermal emission) of the HR 4796A debris disk? To test this, we repeated the model fitting process using both the scattered light and thermal emission data simultaneously. The best-fitting model had a reduced chi-square value of 2.14, indicating a mediocre fit (Fig. \ref{fig:allmodelscattered} and \ref{fig:allmodelthermal}). As before, we computed the relative probability of each root composition's fractional abundance. Fig. \ref{fig:firstchis} shows that the results are similar to the previous cases, with silicates and organics being the most frequently preferred and water ice, iron, and cyrstalline olivine being the least preferred. 

Lower porosities were preferred more than higher porosities (Fig. \ref{fig:allporosities}). The average size distribution power law was \about -3.4, with the power-law being \about -2.5 for the best-fitting mixtures. The minimum and maximum grain sizes were \about 3 \microns ~and 500 \microns, similar to the previous cases. 


\subsubsection{Variations on fitting constraints}
Up to this point, we have tried to fit the scattered light and thermal emission of the HR 4796A debris disk using porous two-composition mixtures with few fitting constraints. For example, we tested a wide range of size distribution power-law exponents, and we used the phase function generated by Mie theory. Which compositions are preferred if these two constraints are varied? 

To test this, first we refit the data (scattered light and thermal simultaneously) while enforcing a Dohnanyi size distribution (power-law exponent = -3.5). Second, we refit the data while ignoring the phase function predicted by Mie theory. Specifically, we fit the thermal data, and then arbitrarily scaled the model disk until it best matched the scattered light data. Finally, we repeated these same fits while also enforcing a Dohnanyi size distribution. 

\begin{figure*}[t]
\centering
\subfloat[]{\label{fig:dona}\includegraphics[width=0.48\textwidth]{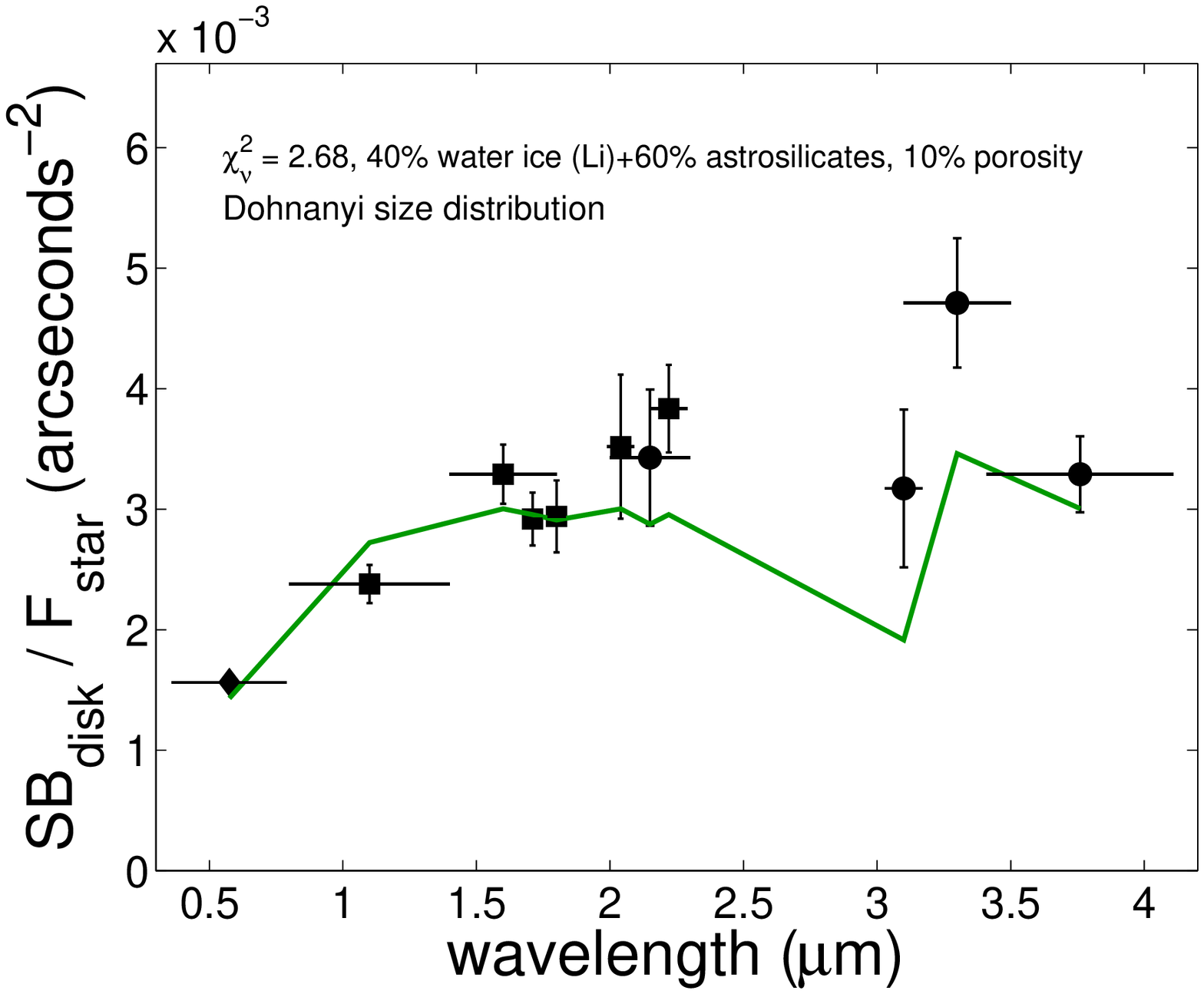}} 
\subfloat[]{\label{fig:donb}\includegraphics[width=0.48\textwidth]{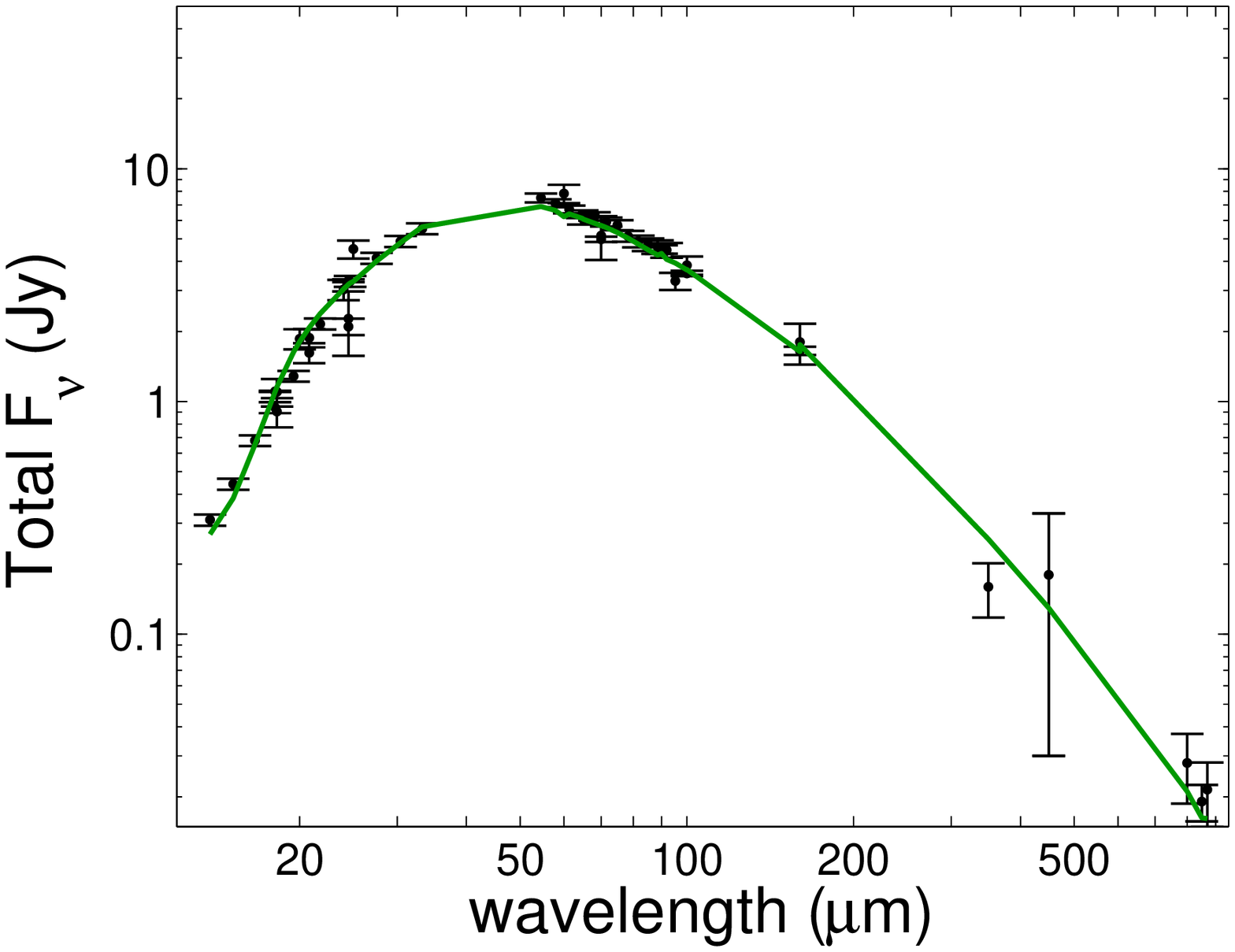}} \\
\subfloat[]{\label{fig:anya}\includegraphics[width=0.48\textwidth]{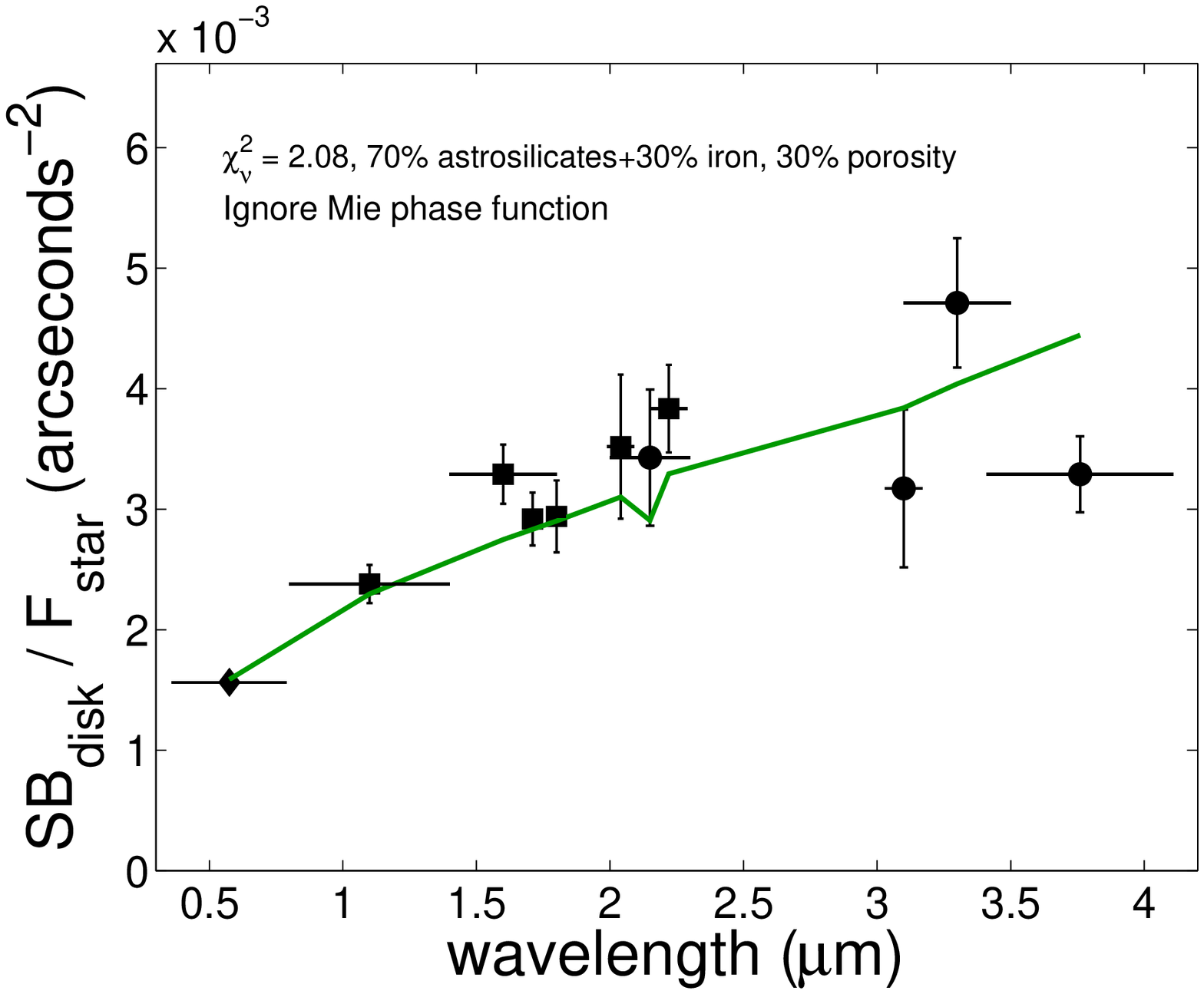}} 
\subfloat[]{\label{fig:anyb}\includegraphics[width=0.48\textwidth]{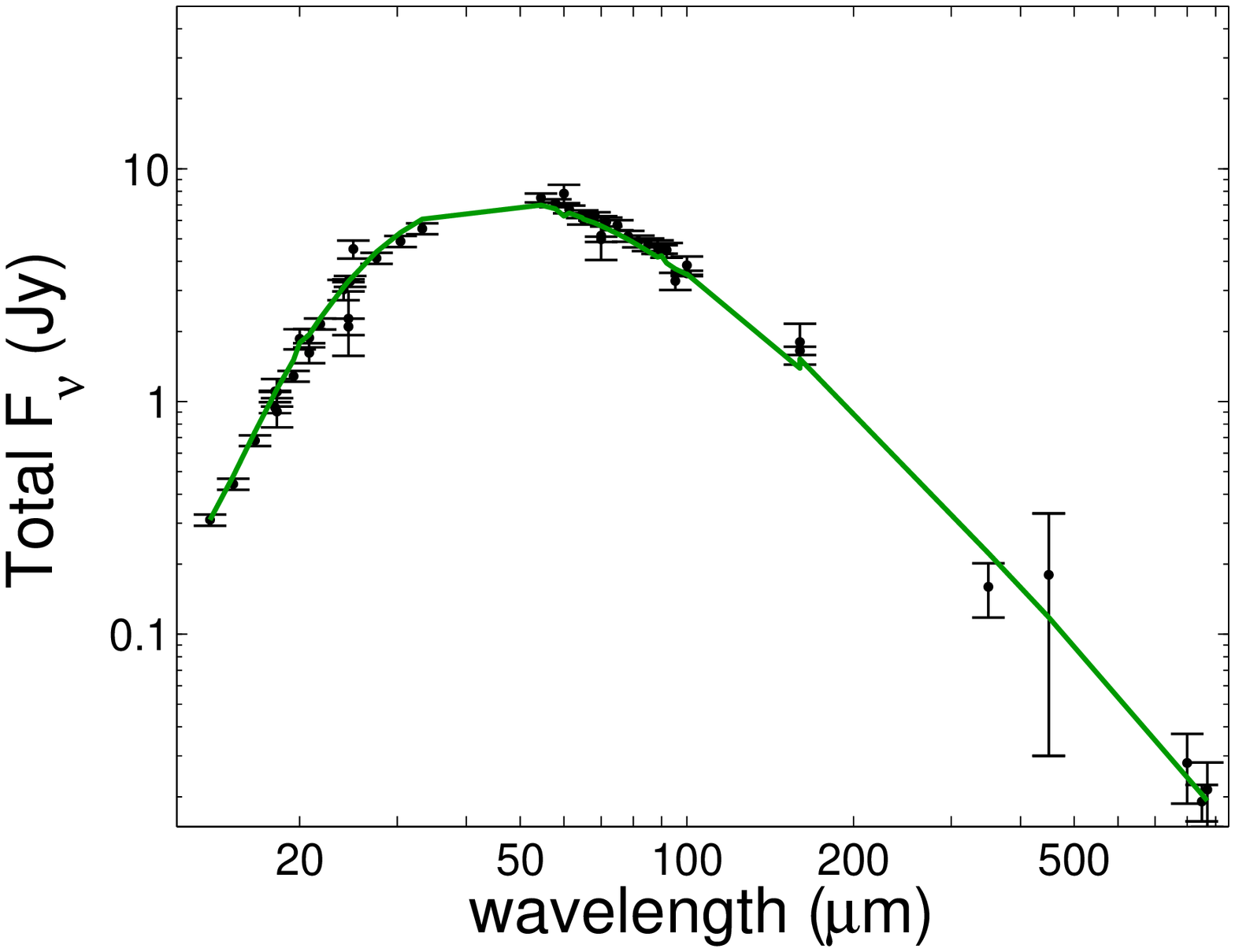}} \\
\subfloat[]{\label{fig:botha}\includegraphics[width=0.48\textwidth]{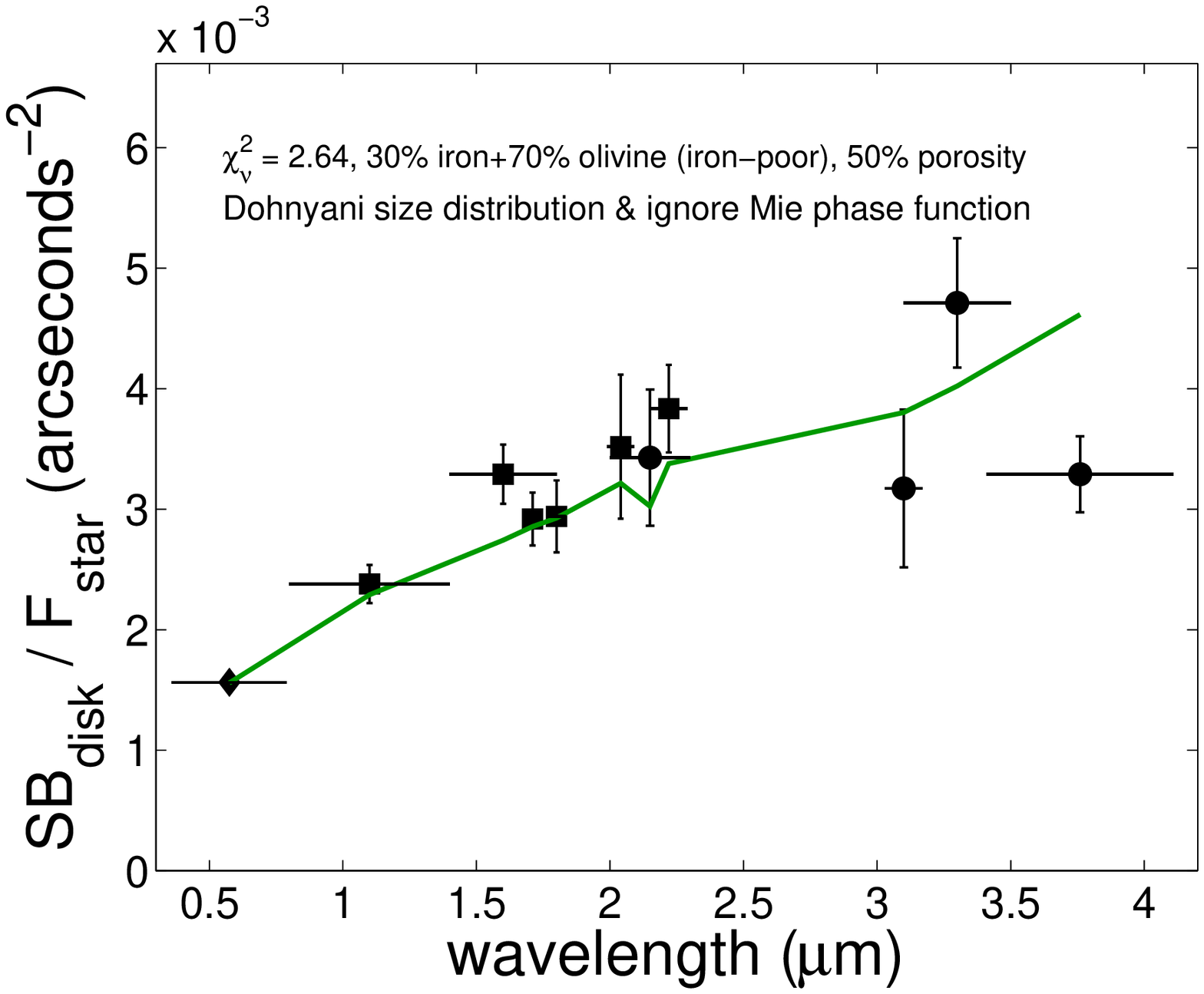}} 
\subfloat[]{\label{fig:bothb}\includegraphics[width=0.48\textwidth]{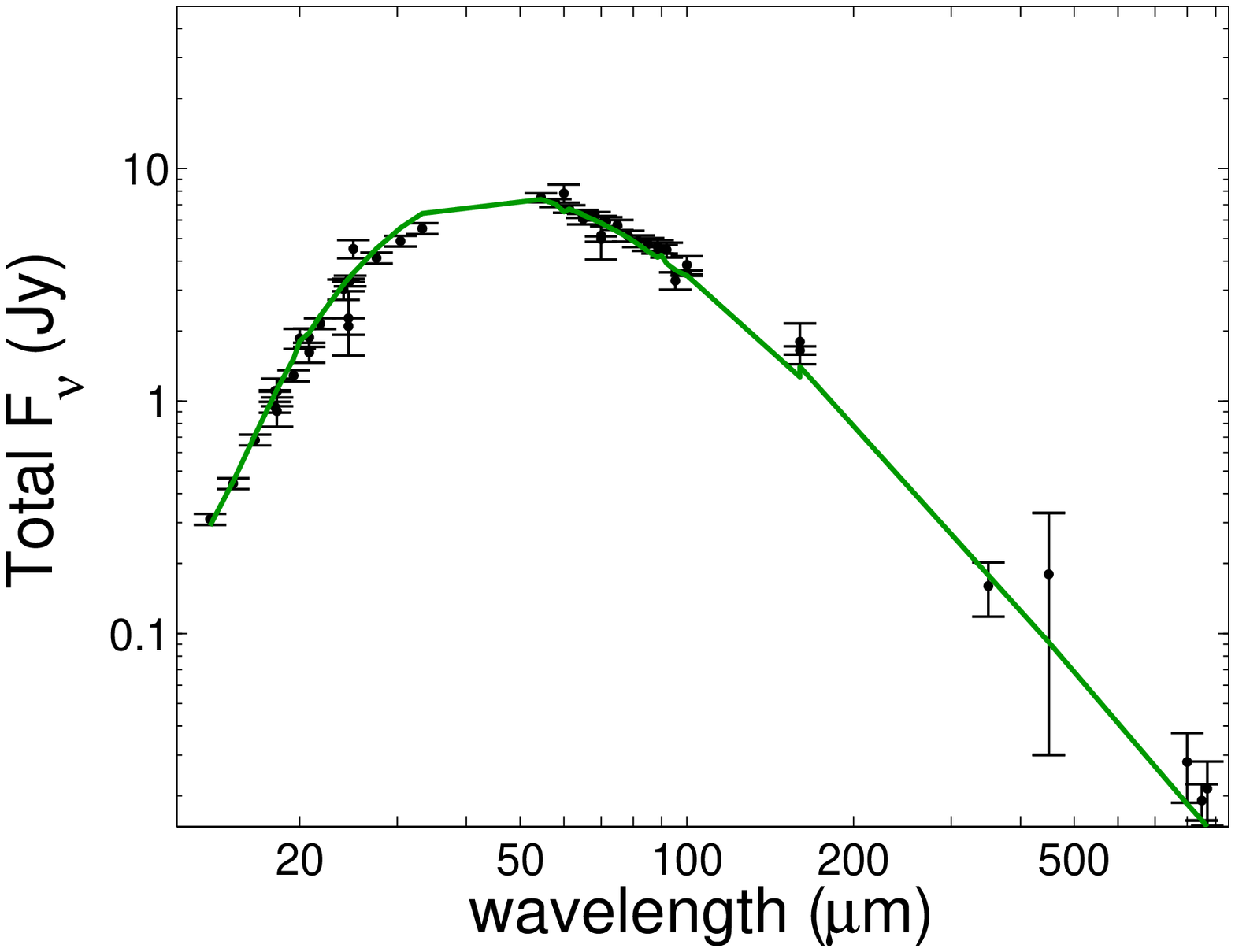}} \\
\caption{(a-b): The best-fitting composition to both the scattered light and thermal emission when a Dohnanyi size distribution is enforced. (c-d): The best-fitting composition when any size distribution is allowed, but the Mie phase function is ignored. (e-f): The best-fitting composition when a Dohnanyi size distribution is enforced \textit{and} the Mie phase function is ignored. Only mediocre fits are achieved in all three cases.}
\end{figure*}

Fig. \ref{fig:secondchis} shows that for a Dohnanyi size distribution, silicates and organics were the predominantly preferred compositions and the best overall fit (Fig. \ref{fig:dona} and \ref{fig:donb}) had a reduced chi-square of 2.68, indicating a poor fit. If we ignored the Mie phase function and relaxed the Dohnanyi constraint, organics and silicates were the most preferred compositions and the best overall fit (Fig. \ref{fig:anya} and \ref{fig:anyb}) had a reduced chi-square of 2.08, indicating a mediocre fit. If we ignored the Mie phase function and also enforced a Dohnanyi size distribution, organics and silicates were the most often preferred and the best overall fit (Fig. \ref{fig:botha} and \ref{fig:bothb}) had a reduced chi-square of 2.64, indicating a poor fit. These results are similar to our findings in the previous fitting cases, with slight variations. 

\section{Discussion}
\label{sec:discussion}
\subsection{Morphology}

How can we place the morphological features of the HR 4796A debris disk into context? Perhaps it is a clone of the disk around Fomalhaut. Ignoring the differences in stellar age, dust optical depth, and SB peak locations (\about 140 AU for Fomalhaut from \citealt{fomalhautkalas}), the two systems are strikingly similar. Both orbit A stars, both are narrow (widths $\approx$ 10-20$\%$), and both are eccentric ($e \approx 0.05-0.10$). We also know that like HR 4796A, Fomalhaut cannot have any super-Jupiters at wide separations \citep{fomalhautjanson,curriefomalhautlimits}, except that a potentially dust-enshrouded object has been observed on a very wide, eccentric, possibly ring-crossing orbit \citep{kalasnewfomalhaut,curriefomalhaut,galicherfomalhaut,fomalhautorbit,fomalhautdustcloud}. Such an object may have been stirred or scattered by a yet-undetected interior gas giant planet. But just as for any close-orbiting planet around HR 4796A, such a planet would have to be fairly low-mass \citep{currienewfomalhaut,kenworthy}, assuming the atmospheric models used to derive the masses are correct.

HR 4796A and Fomalhaut, in terms of morphology, appear in contrast to the three planetary systems that harbor imaged planets and exterior debris disks: HR 8799 \citep{marois,hr87994thplanet}, Beta Pic \citep{betapicoriginal,betapic}, and HD 95086 \citep{95086confirm}. These three systems host super-Jupiters at \about 10-80 AU with exterior \textit{broad} debris disks \citep{herschel8799,suhr8799,augereaubetapic,moor95086}. With a sample of only three, it is difficult to make strong statistical inferences. We already know that super-Jupiters rarely orbit at large separations \citep{nielsennici,wahhaj,jansonsurvey,billermovinggroups}, but perhaps they are even rarer in narrow debris disk systems. This effect, if confirmed, would not be surprising given that more massive planets have been predicted to make debris disks broader \citep{chiang,medynamics}. It might also imply we will be unlikely to image super-Jupiters in narrow debris disk systems with current or future technology. 

With regard to the halo around HR 4796A, of which we see only ``traces" due to ADI processing (Section \ref{sec:streamers}), it seems that HR 4796A is similar to several other debris disks. The disks around HD 15115 and HD 32297 have extended scattered light features that are indicative of small, blow-out grains \citep{kfg,mehd15115,glennstis}, and HD 61005 has a halo of small grains that can appear as ``halo traces" when processed with ADI \citep{moth} or high-pass filters \citep{glennstis}. Seeing evidence for halos is not surprising given our current understanding of debris disk structure \citep{strubbe}.


\subsection{Mie theory}
When fitting our composition models to just the scattered light alone or just the thermal emission alone, we found that good-fitting models could not reproduce both datasets simultaneously (Fig. \ref{fig:scatteredmodelsfig}). One likely explanation for this is that the Mie-generated product of the dust albedo and the phase function is incorrect. This likely implies that either the dust albedos are wrong or the phase function generated by Mie theory is wrong. Neither of these would be surprising. It has been suspected for some time that observed dust albedos are hard to reproduce (e.g., \citealt{hd181327ice,kristhd207129}). Furthermore \cite{perrinGPI4796} and \cite{milli4796} recently showed polarimetric images of the HR 4796A debris ring, revealing that the western side is much brighter than the eastern side in polarized light (whereas the eastern side is brighter in total intensity). This is difficult to explain unless Mie theory is not sufficient to model the properties of the dust grains \citep{milli4796}, or perhaps the disk is optically thick \citep{perrinGPI4796} (though the latter possibility needs to be rigorously tested). Another possibility is that very porous dust grains of varying compositions can sometimes produce ``polarization reversals" at high scattering angles for inclined disks \citep{polarizationreversalshen,polarizationreversal}, though no specific modeling of this phenomenon for HR 4796A has yet been carried out.

To improve the modeling of this disk (and others), the following are required. (1) The dust size distribution power-law needs to be measured (e.g., see \citealt{fomalhautq}). This would help significantly narrow down the testable parameter space, since currently there is only a theoretical motivation for forcing a Dohnanyi size distribution. (2) An alternative to Mie theory that can accurately reproduce the phase functions of observed disks (including polarized light) needs to be tested on disks with rich datasets like HR 4796A (e.g., ``distribution of hollow spheres", \citealt{dhs,milli4796}). The key to producing accurate phase functions is to actually measure the phase functions of known disks. While difficult, progress is being made (e.g., \citealt{stark181327}). Once the phase functions of a few debris disks are well-characterized, we can compare to the predictions of Mie theory and others like DHS and then explore generalizing to other debris disks. (3) It is possible that more realistic (i.e., larger) uncertainties for disk thermal emission data are required. Our generally mediocre fits to the data that included thermal emission imply that either our models are not good representations of the data (see points (1) and (2) above), or the data are incorrect/the uncertainties are underestimated.  

\subsection{Chemical composition}
What can we conclude about the chemical composition of the debris ring around HR 4796A? Below we discuss the likelihood that each composition comprises at least some fraction of the dust. Note that our inferences are based on the general trends in Fig. \ref{fig:chis} and largely ignore the fact that, other than the scattered light alone case, we could only achieve mediocre fits to the data. Given the compounding uncertainties in the optical constants, the mixing rules, the disk's phase function, and the applicability of Mie theory, it is a reasonable approach to observe which compositions succeed more often than others, even if our best-fitting models are not perfect matches to the data.


\subsubsection{Silicates: likely}
The silicate compositions (amorphous olivine, iron-poor/nominal iron/iron-rich olivine, astrosilicates, and the pyroxenes) were always generally preferred over the other compositions (Fig. \ref{fig:chis}). Furthermore, these compositions were favored at higher fractional abundances (i.e., $>50\%$). Therefore we consider it likely that the HR 4796A dust contains silicates.



\subsubsection{Crystalline Olivine: plausible at low fractional abundance}
Crystalline Olivine was only favored in the scattered light case. However, because we did not have high-resolution thermal spectra (near 69 \microns ~in particular; \citealt{betapicolivine}), its unique spectral features would not have been detected even if it was abundant. On the other hand, \cite{hr4796crystallinesilicates} found no spectral features indicative of crystalline silicates at 8-13 \microns. Therefore we conclude that if crystalline olivine is present, it is likely at low fractional abundance.


\subsubsection{Water Ice: unlikely or at low fractional abundance}
The Water Ice compositions (Li/Warren/Henning water ice) were generally not preferred. Furthermore, good fits containing water ice seemed to require smaller fractional abundances (i.e., the most probable volumetric fractions were usually $<50\%$). Therefore we conclude that water ice must either be at very low abundance or not present at all in the HR 4796A dust. The lack of water ice may not be surprising, since for stars earlier than M type, UV sputtering has been predicted to remove icy grains on very short timescales \citep{watericesurvival}. If the collisional timescale for the dust around HR 4796A is much longer than the photosputtering timescale, grains will be ice-poor for most of their lives, potentially explaining our findings. Interestingly, the lack of water ice in the dust around HR 4796A is counter to a few other debris disks that have been found to require more water ice \citep{donaldson32297,hd181327ice,chenice181327}. This could point to real differences in dust compositions, though resolved scattered light at multiple wavelengths from 1-4 \microns ~is still absent for these disks.

\subsubsection{Organics: likely}
Compositions containing carbon (amorphous carbon, Li/Henning organics, Titan tholins) were generally favored, and at high volumetric fractions. Therefore we consider it likely that the HR 4796A dust contains organics. The specific ``type" of organics is still not certain, given the discrepancies between the root compositions (e.g., the Li and Henning organics), which most likely arise due to their differing optical constants. The tholins that were proposed in \cite{4796organics} were favored less often than the other organics, suggesting that any complex organics are unlikely to be the dominant constituent of the dust. Nonetheless organics in general being favored suggests that at least some of the base building blocks of Earth-like life, which are also found in interstellar dust \citep{ismdustcomposition} and solar system comets (e.g., \citealt{stardustwaterice,cometdustcarbon}), may be present around HR 4796A.

\subsubsection{Troilite: unlikely}
Troilite was generally infrequently preferred and there was no discernible trend in volumetric fraction. Therefore we consider troilite to be unlikely to comprise the dust around HR 4796A.

\subsubsection{Iron: plausible}
Iron was generally not frequently preferred, other than an apparently special case involving 30$\%$ fractional abundance (Fig. \ref{fig:chis}). However, the iron-rich and nominal iron compositions of the olivines and orthopyroxenes were generally favored over the iron-poor cases. This might suggest that more iron is required in the silicates. Therefore we consider iron to be a plausible constituent of the dust around HR 4796A.

\section{Summary}
\label{sec:summary}
We have resolved the HR 4796A debris ring with MagAO/VisAO and MagAO/Clio-2 at seven wavelengths spanning \about 0.7-4 \microns. We compiled these data with existing archival HST/STIS and HST/NICMOS images of the ring at \about 0.5-2 \microns. We also compiled all available thermal emission data, including previously unpublished Spitzer/MIPS data. With such a rich data set, we set out to constrain the morphology of the ring and the dust grain composition. 

We found that the deprojected ring is offset by 4.76$\pm$1.6 AU and is mildly eccentric ($e = 0.06\pm$0.02), in agreement with previous studies. We measured the width of the ring at multiple wavelengths, finding that it is narrow (14$^{+3}_{-2}\%$, 11.1$^{+2.4}_{-1.6}$ AU). Using the predictions from \cite{medynamics}, this width implies that if there is a single shepherding planet orbiting interior to the ring, it must be less massive than \about 4 \mj. This limit is \about equivalent to the mass of any self-luminous planets that could have been detected in our \lprime ~data beyond \about 60 AU. 

We found that the best fits to the scattered light data alone and thermal data alone did not agree. This suggests that caution should be exercised if fitting to only scattered light data or only thermal data. A likely explanation is that Mie theory cannot reproduce the product of albedo and phase function for the observed dust grains. When we fit all of the data together simultaneously, we find only mediocre fits ($\chi_{\nu}^{2}$ \about 2), with silicates and organics generally being the most frequently preferred over the other compositions. Water ice was generally not preferred, suggesting that it is either not present in the dust or at very low abundance. These findings generally agree with previous modeling efforts that preferred mixtures of silicates, organics, and some water ice \citep{hr4796augereau,hr4796li,milli4796}. Our results suggest that some of the common constituents of interstellar dust and solar system comets may reside around this interesting young star, though improved modeling is required to determine the exact chemical composition of the dust.

\acknowledgments
We thank the referee, Christian Thalmann, for helpful comments and suggestions. We are grateful to the entire LCO observing support staff for their help preparing and operating the telescope and instruments during the observing runs. We thank the teams at the Steward Observatory Mirror Lab/CAAO (University of Arizona), Microgate (Italy), and ADS (Italy) for building the phenomenal adaptive secondary mirror (ASM) for use in the AO. The MagAO ASM was developed with support from the NSF MRI program. The MagAO pyramid wavefront sensor was developed with help from the NSF TSIP program and the Magellan partners. The Active Optics guider was developed by Carnegie Observatories with custom optics from the MagAO team. The VisAO camera and commissioning was supported with help from the NSF ATI program. C.C.S. would like to acknowledge support of this research by an appointment to the NASA Postdoctoral Program at Goddard Space Flight Center, administered by Oak Ridge Associated Universities through a contract with NASA. J.R.M and K.M.M. were supported under contract with the California Institute of Technology (Caltech) funded by NASA through the Sagan Fellowship Program. P.S.S. acknowledges support from the NASA/JPL contract 1256424 given to the University of Arizona. A.J.W. acknowledges the support of the Carnegie node of the NASA Astrobiology Institute under Cooperative Agreement NNA09DA81A.


\appendix
Obtaining accurate photometry of the HR 4796A debris ring requires a careful procedure involving several different correction factors. Previous studies of the debris disks around HD 15115 and HD 32297 have demonstrated similar procedures \citep{debes,me32297,currie32297,mehd15115}, which we largely follow here. 

In general, the true intensity $I$ at a given wavelength $\lambda$ of a debris disk measured in an aperture of a given size measured a distance $r$ from the star is calculated as follows:
\begin{equation}
I_{true}(r,\lambda) = I_{measured}(r,\lambda) \times C_{PSF} \times C_{ap} \times C_{bias},
\label{eqn:corrections}
\end{equation}
where $C$ denotes a given correction factor corresponding to PSF convolution, aperture size, or data reduction bias. For HR 4769A, we calculated all three $C$ values for each independent wavelength (image) at the two disk ansae. We then multiplied these correction factors with the measured disk SB values using Eq. \ref{eqn:corrections} to obtain the true disk SB values. In the following discussion, all apertures sizes used for the correction factors were identical to the apertures used to compute the photometry of the real disk at each wavelength as described in Section \ref{sec:scatphotometry}.

\begin{figure}[h]
\centering
\subfloat[]{\label{fig:LFake}\includegraphics[scale=0.23]{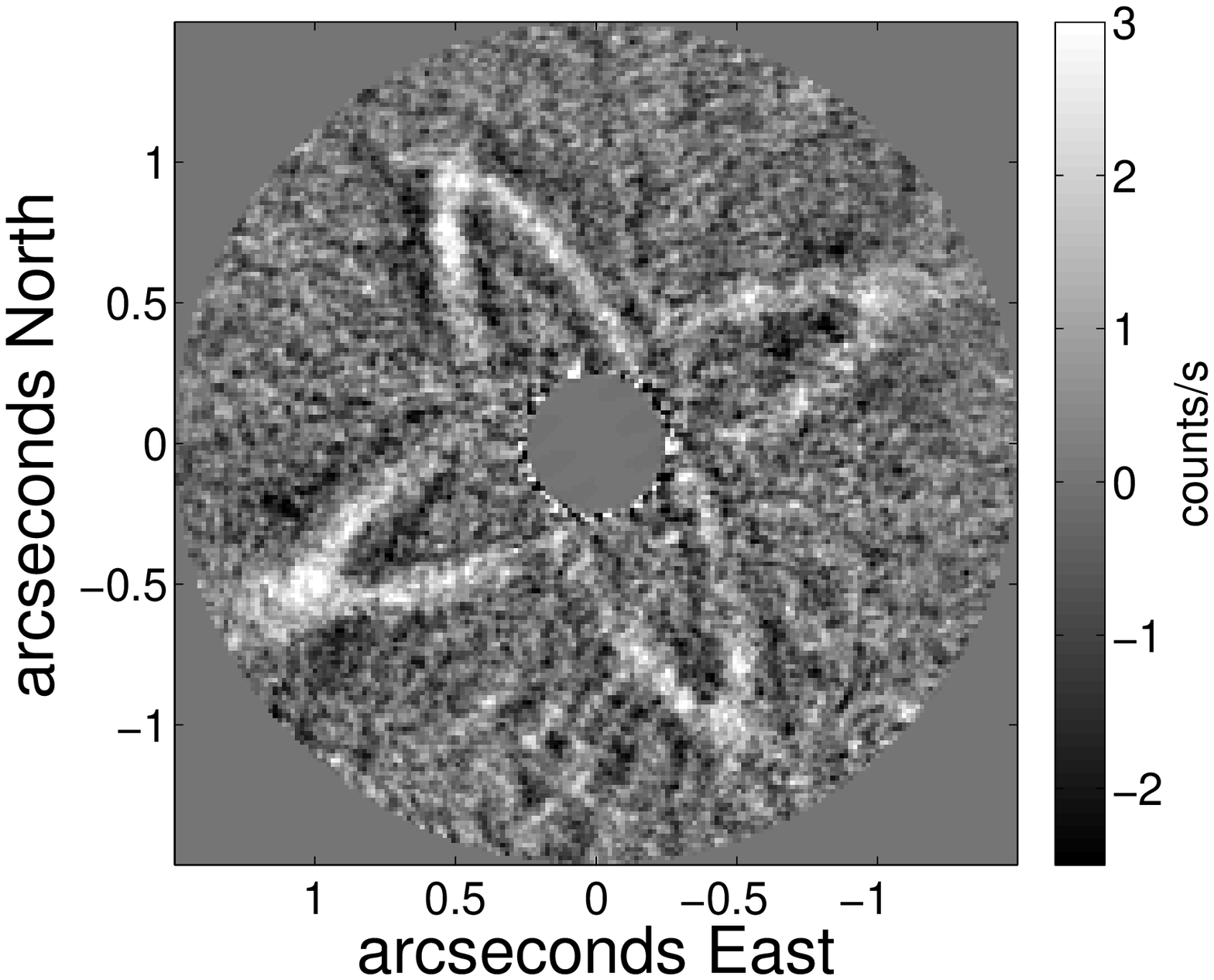}} 
\subfloat[]{\label{fig:LsFake}\includegraphics[scale=0.23]{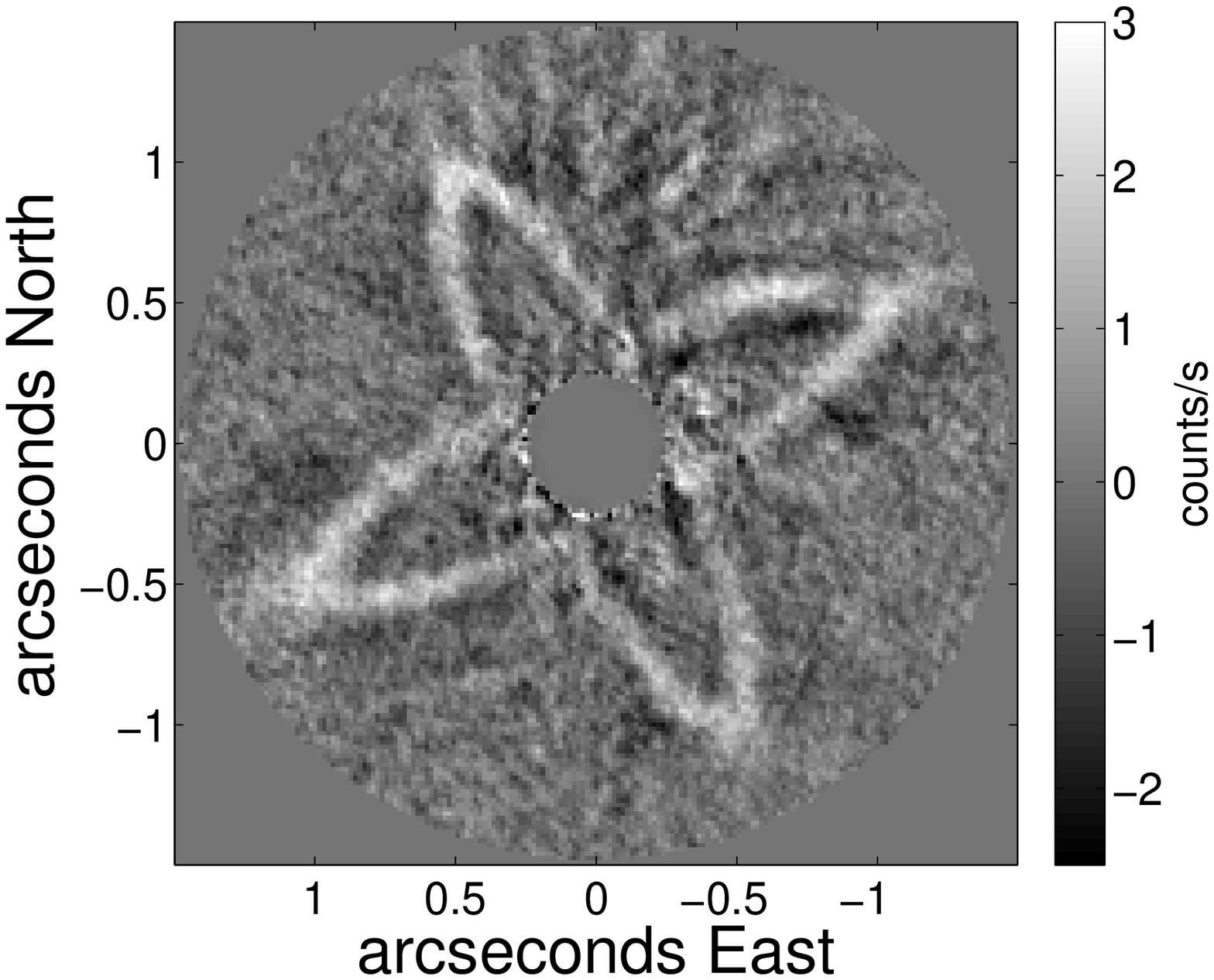}} 
\subfloat[]{\label{fig:IceFake}\includegraphics[scale=0.23]{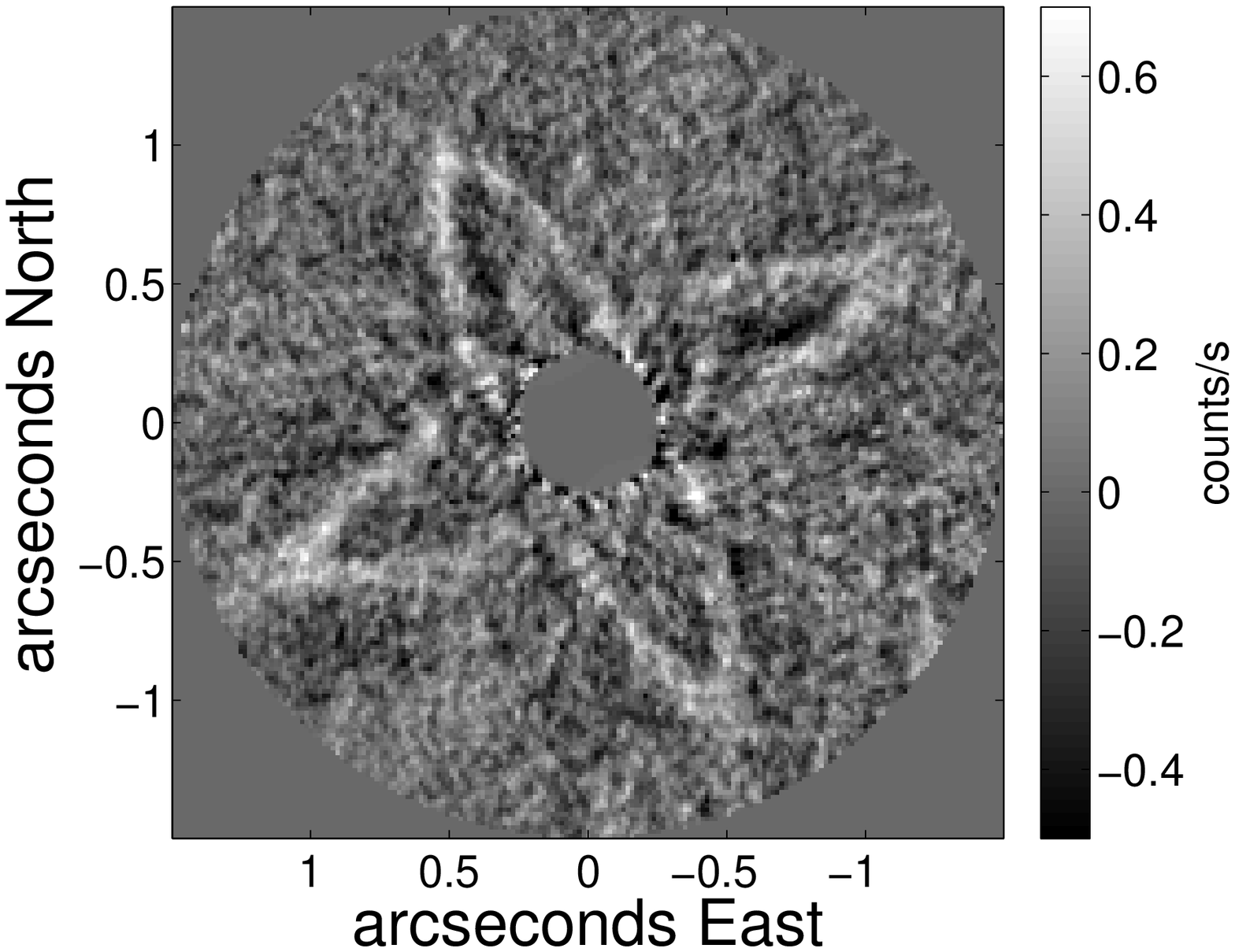}} 
\subfloat[]{\label{fig:KsFake}\includegraphics[scale=0.23]{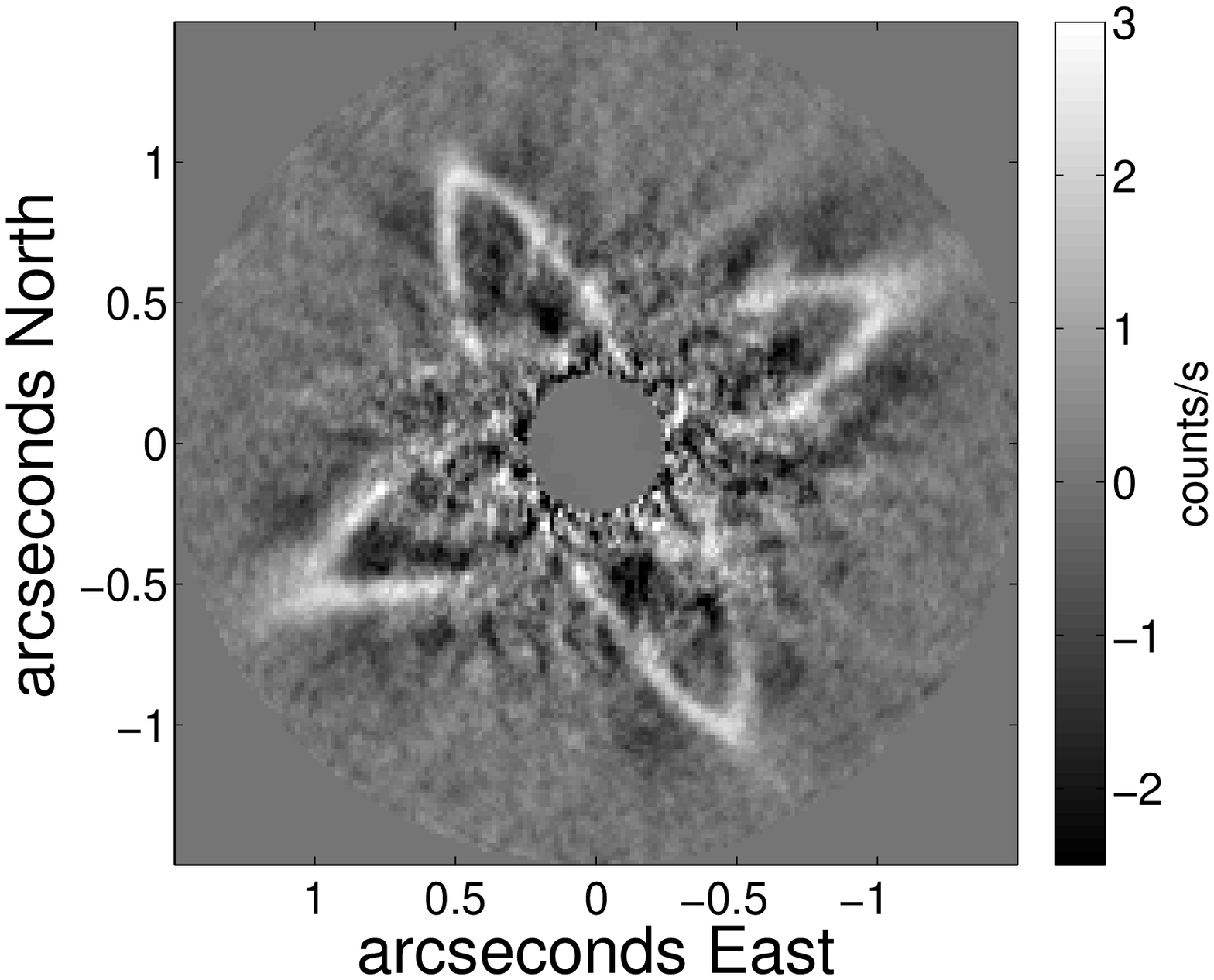}} 
\caption{Inserted and recovered model disks for the Clio-2 2-4 \microns ~data. The recovered model disk is qualitatively similar to the real disk at each wavelength.}
\label{fig:fakedisks}
\end{figure}

We computed $C_{PSF}$, the factor that corrects for the convolution of the disk with the telescope/instrument PSF, by generating an unconvolved model of the HR 4796A debris ring from the parameters given in \cite{thalmannhr4796}, which they used to reproduce their observed disk at $H$ band. This model was sampled at the respective plate scales of STIS, NICMOS, and Clio-2 and was otherwise identical across wavelength. We then convolved the model disk images with the telescope/instrument PSF. For the STIS and NICMOS data, the PSF was generated using the $TinyTim$ software\footnote{http://www.stsci.edu/hst/observatory/focus/TinyTim}. For the ground-based Clio-2 data, the unsaturated photometric images of the star were used.


We computed $C_{ap}$, the factor that corrects for the different aperture sizes used at each wavelength, in the following way: we treated the HST/NICMOS unconvolved model as the reference and compared the flux inside the aperture to the fluxes of the unconvolved model disks computed using the respective apertures at the other wavelengths/plate scales. 

For the HST data (both STIS and NICMOS), $C_{bias} \approx 1$ because the disk does not self-subtract (as is the case for ground-based/ADI data). The bias correction factors equaling \about unity was verified via insertion of model disks into the HST data. For the Clio-2 data, we computed $C_{bias}$ as follows: we scaled and inserted the convolved model disk images at each wavelength into the raw images 90\degrees ~rotated from the real disk, re-reduced the data, and measured the flux inside the appropriate apertures at the recovered model disk ansae. These values were compared with the expected values, measured in the same way on the noiseless scaled convolved images. An additional multiplicative factor was included to account for the attenuation of the real disk by the model disk and vice-versa. Fig. \ref{fig:fakedisks} shows the inserted and recovered model disks along with the real disk at 2-4 \microns. The model disk is a good qualitative match to the real disk, which is all that is necessary for the bias correction described above. 

As a sanity check, we can compare the the \ks band Clio-2 disk photometry with with the HST/NICMOS data at 2.05 \microns ~and 2.22 \microns. A priori, the photometry should be \about equal regardless of the telescope/instrument. The parameters of the model disk can affect how well the photometry agrees at 2 \microns. Therefore we varied the inner and outer power-laws of the model disk density distribution. In general, the photometry was always consistent within the respective error bars. However, the best match was obtained using an inner power-law exponent of -19.6 and an outer exponent = 6. Fig. \ref{fig:finalphot} shows the final scattered light photometry, verifying our photometric corrections and calibrations.


\bibliographystyle{apj}
\bibliography{ms}

\end{document}